\documentclass[aps,preprint,onecolumn,12pt,nofootinbib]{revtex4}
\usepackage{hyperref}
\usepackage{amsfonts}
\usepackage{bm}
\usepackage{mathrsfs}
\usepackage{soul}
\usepackage{makecell,multirow}
\usepackage{rotating}
\usepackage[utf8]{inputenc}
\usepackage{amsmath}
\usepackage{makecell}

\usepackage{amssymb}
\usepackage{tensor}
\usepackage{graphicx}
\usepackage{bigints}
\usepackage{bbm}
\usepackage{graphicx}
\usepackage{subfigure}
\usepackage{epsf}
\usepackage{bm}
\usepackage{amsmath}
\usepackage{amsfonts}
\usepackage{amssymb}
\usepackage{graphicx}
\usepackage{tabularx}
\usepackage{multirow}
\usepackage{color}%
\providecommand{\U}[1]{\protect\rule{.1in}{.1in}}

\newcommand{\ie}{\begin{equation}}
\newcommand{\fe}{\end{equation}}

\newcommand{\mincir}{\raise
-3.truept\hbox{\rlap{\hbox{$\sim$}}\raise4.truept\hbox{$<$}\ }}
\newcommand{\magcir}{\raise
-3.truept\hbox{\rlap{\hbox{$\sim$}}\raise4.truept\hbox{$>$}\ }}

\providecommand{\U}[1]{\protect\rule{.1in}{.1in}}

\usepackage{tikz,xcolor,hyperref}

\definecolor{lime}{HTML}{A6CE39}
\DeclareRobustCommand{\orcidicon}{%
	\begin{tikzpicture}
	\draw[lime, fill=lime] (0,0) 
	circle [radius=0.16] 
	node[white] {{\fontfamily{qag}\selectfont \tiny ID}};
	\draw[white, fill=white] (-0.0625,0.095) 
	circle [radius=0.007];
	\end{tikzpicture}
	\hspace{-2mm}
}

\foreach \x in {A, ..., Z}{%
	\expandafter\xdef\csname orcid\x\endcsname{\noexpand\href{https://orcid.org/\csname orcidauthor\x\endcsname}{\noexpand\orcidicon}}
}

\begin{document}

\title{\Large{Non--commutativity in Hayward spacetime}}


\author{N. Heidari\orcidA{}}
\email{heidari.n@gmail.com}

\affiliation{Center for Theoretical Physics, Khazar University, 41 Mehseti Street, Baku, AZ-1096, Azerbaijan.}
\affiliation{School of Physics, Damghan University, Damghan, 3671641167, Iran.}

\author{A. A. Ara\'{u}jo Filho\orcidB{}}
\email{dilto@fisica.ufc.br (The corresponding author)}

\affiliation{Departamento de Física, Universidade Federal da Paraíba, Caixa Postal 5008, 58051-970, João Pessoa, Paraíba, Brazil}
\affiliation{Departamento de Física, Universidade Federal de Campina Grande Caixa Postal 10071, 58429-900 Campina Grande, Paraíba, Brazil.}


\author{Iarley P. Lobo\orcidD{}}
\email{lobofisica@gmail.com}

\affiliation{Department of Chemistry and Physics, Federal University of Para\'iba, Rodovia BR 079 - km 12, 58397-000 Areia-PB,  Brazil.}


\begin{abstract}

In this work, we propose a new black hole solution, namely, a Hayward--like metric incorporating corrections due to non--commutativity { by taking into account $\partial_r\wedge\partial_\theta$ Moyal twist}. We begin by deriving this solution using the non--commutative gauge theory framework. The general properties of the metric are then analyzed, including the event horizon structure and the Kretschmann scalar. Analogous to the standard Hayward solution, the modified black hole remains regular, provided that additional { dependence on the angle $\theta$}. Next, we examine the thermodynamic properties, computing the Hawking temperature, entropy, and heat capacity. {From }the temperature profile{, we verify that there is no physical} remnant mass when $T^{(\Theta,l)} \to 0${, indicating a complete evaporation process}. Quantum radiation is analyzed by considering both bosonic and fermionic particle modes, with an estimation of the particle creation density provided for each case. The effective potential is evaluated perturbatively to accomplish the analysis of quasinormal modes and the time--domain response for scalar perturbations. The study of null geodesics is explored to enable the characterization of the photon sphere and black hole shadows. Additionally, constraints on the shadows are estimated based on EHT (Event Horizon Telescope) data. Furthermore, the Gaussian curvature is determined to assess the stability of critical orbits, followed by an analysis of gravitational lensing using the Gauss--Bonnet theorem. Finally, the constraints (bounds) on the parameters $\Theta$ (non--commutativity) and $l$ (``Hayward parameter'') are derived based on solar system tests, including the perihelion precession of Mercury, light deflection, and the Shapiro time delay effect.

\end{abstract}
\maketitle

\tableofcontents

	
\section{Introduction}

The formulation of gravity within general relativity inherently involves a geometric description of spacetime, where nonlinear effects make solving the Einstein field equations highly challenging. Even under imposed symmetries or constraints, finding exact solutions remains a difficult task \cite{misner1973gravitation,wald2010general}. To overcome these difficulties, a weak--field approximation is frequently applied, transforming the equations into a more manageable form. This method allows for the study of gravitational waves, which arise as perturbations in spacetime. These oscillations play a significant role in black hole physics, influencing their stability, the emission of Hawking radiation, and interactions with their surrounding environment.

The principles of general relativity do not inherently impose a fundamental constraint on the precision of spatial measurements. Nonetheless, various theoretical perspectives propose that a minimal length scale, frequently linked to the Planck length, may serve as a natural limit. To account for this, alternative spacetime models incorporating non--commutative properties have been formulated, drawing inspiration from string theory \cite{3,szabo2006symmetry,szabo2003quantum}. Such models have also been widely examined within the framework of supersymmetric Yang--Mills theories \cite{ferrari2003finiteness,ferrari2004superfield,ferrari2004towards}. Within gravitational scenarios, the implementation of non--commutativity is often achieved via the Seiberg--Witten map, which enables modifications to the underlying symmetry structure \cite{chamseddine2001deforming}.

The application of non--commutative geometry has played a crucial role in advancing black hole research \cite{zhao2023quasinormal,anacleto2023absorption,Anacleto:2019tdj,heidari2024exploring,1,mann2011cosmological,campos2022quasinormal,anacleto2021quasinormal,2,karimabadi2020non,lopez2006towards,modesto2010charged,nicolini2009noncommutative,heidari2023gravitational,Heidari:2025sku}. In particular, studies have focused on its implications for black hole evaporation \cite{23araujo2023thermodynamics,myung2007thermodynamics} and modifications to thermodynamic behavior \cite{nozari2007thermodynamics,lopez2006towards,banerjee2008noncommutative,nozari2006reissner,sharif2011thermodynamics}. Moreover, the thermal characteristics of field theories have been explored in various contexts \cite{araujo2023thermodynamical,furtado2023thermal}.

Fundamentally, the idea of non--commutative spacetime arises from modifying the algebra of coordinates, expressed through the commutator relation $[x^\mu, x^\nu] = \mathbbm{i} \Theta^{\mu \nu}$, where $\Theta^{\mu \nu}$ is an anti--symmetric tensor characterizing non--commutativity. This departure from conventional spacetime geometry has prompted diverse strategies for incorporating such effects into gravitational frameworks. One approach involves altering gauge symmetries by extending the de Sitter group SO(4,1) and merging it with the Poincaré group ISO(3,1) through the Seiberg--Witten map. Within this framework, Chaichian et al. \cite{chaichian2008corrections} proposed a modified version of the Schwarzschild metric that integrates non--commutative corrections.

Nicolini et al. \cite{nicolini2006noncommutative} proposed a different way to incorporate non--commutativity into general relativity by modifying the matter distribution instead of altering the Einstein tensor. Rather than treating mass as a point--like source, this approach introduces a continuous mass distribution. Two distinct profiles describe this modification: a Gaussian density function, $\rho_\Theta = M (4\pi \Theta)^{-\frac{3}{2}} e^{-\frac{r^2}{4\Theta}}$, and a Lorentzian profile, $\rho_\Theta = M \sqrt{\Theta} \pi^{-\frac{3}{2}} (r^2 + \pi \Theta)^{-2}$. Furthermore, Ref. \cite{Juric:2025kjl} recently outlined a general framework for black hole construction in non--commutative gravity, expanding the earlier established methodologies encountered in the literature \cite{chaichian2008corrections,Touati:2023cxy}.

The observation of gravitational waves by LIGO and Virgo \cite{018,017,016} has provided new directions for investigating gravitational effects, including lensing phenomena in the weak--field regime \cite{020,019}. Earlier studies on gravitational lensing primarily addressed light deflection over vast cosmological scales, frequently employing the Schwarzschild metric as a reference model \cite{021}. Subsequent research broadened this scope to incorporate more general static and spherically symmetric backgrounds \cite{022}. Nevertheless, in regions where gravitational fields are extremely strong, such as near black holes, light deflection becomes more pronounced, requiring refined analytical (or numerical) techniques to describe these deviations accurately.

The successful direct imaging of a supermassive black hole by the Event Horizon Telescope has driven further exploration of gravitational lensing phenomena \cite{024,027,023,029,026,025,028}. In earlier studies, Virbhadra and Ellis introduced a simplified lensing equation to examine black holes within asymptotically flat spacetimes \cite{031,030}. Their analysis showed that intense gravitational fields could generate multiple images, emphasizing the significance of developing accurate theoretical models for these scenarios.

Further theoretical advancements by Fritelli et al. \cite{032}, Bozza et al. \cite{033,034}, and Tsukamoto \cite{035,igata2025deflection} have enhanced the study of gravitational lensing in strong-field regimes. These refinements have been applied across various spacetime configurations, including Schwarzschild black holes \cite{metcalf2019strong,Donmez:2024lfi,Koyuncu:2014nga,virbhadra1998role,Donmez:2023wtf,virbhadra2000schwarzschild,Donmez:2023egk,Ovgun:2018tua,bisnovatyi2017gravitational,ezquiaga2021phase,Okyay:2021nnh,Ovgun:2018fnk,Li:2020dln,cunha2018shadows,Pantig:2022ely,oguri2019strong,Kuang:2022xjp,Pantig:2022gih,grespan2023strong,virbhadra2002gravitational}, axissymmetric black holes \cite{hsieh2021gravitational,hsieh2021strong,37.4,37.1,37.6,37.3,37.2,37.5,jusufi2018gravitational}, and wormholes \cite{38.2,38.1,ovgun2019exact,38.5,38.4,38.3}. Lensing studies have also been extended to modified gravity frameworks \cite{chakraborty2017strong,40,nascimento2024gravitational} and charged Reissner--Nordström backgrounds \cite{036.1,tsukamoto2023gravitational,036,zhang2024strong,036.2}. Moreover, investigations into gravitational distortions have examined the influence of extreme fields on light trajectories \cite{virbhadra2022distortions,virbhadra2024conservation}.

Hawking’s contributions established a direct relationship between black hole thermodynamics and quantum field theory \cite{o111,o11,o1}. His findings indicated that black holes emit thermal radiation, which gradually depletes their mass --- a phenomenon now referred to as Hawking radiation \cite{gibbons1977cosmological,eeeOvgun:2015box,eeeKuang:2018goo,eeeOvgun:2019ygw,eeeKuang:2017sqa,eeeOvgun:2019jdo}. This breakthrough significantly advanced the study of quantum effects in curved spacetime \cite{o9,araujo2023analysis,o8,sedaghatnia2023thermodynamical,araujo2024dark,o6,o3,o4,aa2024implications,o7}. Building on this, Kraus and Wilczek \cite{o10}, along with Parikh and Wilczek \cite{o12,o13,011}, later reinterpreted this radiation as a consequence of quantum tunneling. Their semi--classical perspective has since been applied to numerous black hole models, further refining the theoretical understanding of energy emission in strong gravitational fields \cite{zhang2005new,senjaya2024bocharova,anacleto2015quantum,calmet2023quantum,giavoni2020quantum,hawking1975particle,johnson2020hawking,hollands2015quantum,silva2013quantum,mirekhtiary2024tunneling,medved2002radiation,vanzo2011tunnelling,del2024tunneling,touati2024quantum,mitra2007hawking}.

Furthermore, the gravitational waves offer a fundamental tool to investigate astrophysical and cosmological events, such as binary mergers and stellar oscillations \cite{heuvel2011compact, dziembowski1992effects,pretorius2005evolution,unno1979nonradial,kjeldsen1994amplitudes,hurley2002evolution,yakut2005evolution}. Their spectral features are shaped by the nature of their sources, with perturbed black holes generating gravitational radiation at distinct frequencies, commonly referred to as quasinormal modes \cite{blazquez2018scalar, oliveira2019quasinormal,konoplya2011quasinormal,heidari2024impact,roy2020revisiting,Berti:2022xfj,horowitz2000quasinormal,london2014modeling,flachi2013quasinormal,kokkotas1999quasi,rincon2020greybody,nollert1999quasinormal,ferrari1984new,jusufi2024charged,santos2016quasinormal,araujo2024dark,ovgun2018quasinormal,maggiore2008physical,berti2009quasinormal,herceg2025noncommutative}.

{A recent work, Ref. \cite{Juric:2025kjl}, introduced alternative approaches for constructing black hole solutions within the framework of non--commutative gauge theory. In particular, the authors pointed out that the seminal paper \cite{chaichian2008corrections}, published several years earlier, did not present the complete form of the non--commutative corrections. In other words, they identified a missing term in the original equations, namely,
\ie
 - \frac{1}{16} \Theta^{\nu \rho} \Theta^{\lambda\tau} \Big[ \tilde{\omega}^{\mathfrak{A}\mathfrak{C}}_{\nu} \Tilde{\omega}^{\mathfrak{C}\mathfrak{D}}_{\lambda} \Big(  D_{\tau}R^{\mathfrak{D}5}_{\rho\mu} + \partial_{\tau}R^{\mathfrak{D}5}_{\rho\mu}   \Big)   \Big],
\nonumber
\fe
which leads to substantial corrections in the tetrad fields, and consequently affects the metric components themselves.

}

In general lines, this work presents a new black hole solution that integrates non--commutative effects within a Hayward--like metric. The analysis encompasses its geometric structure, thermodynamic behavior, quasinormal spectra, and observational features. Model parameters are constrained through solar system tests, considering perihelion precession, light bending, and Shapiro time delay.


\section{The new non--commutative black hole}

This part establishes the foundational approach for analyzing gravity within a non--commutative gauge theory framework. The gauge group under consideration, as mentioned earlier, corresponds to the de Sitter group, $\mathrm{SO}(4,1)$. To begin, attention is directed toward the construction of the $\mathrm{SO}(4,1)$ gauge theory in a commutative $(3+1)$--dimensional Minkowski spacetime. In this context, the metric, when expressed in spherical coordinates, is given by
\ie
\mathrm{d}s^{2}=\mathrm{d}r^{2}+r^{2}\mathrm{d}\Omega^{2}_{2}-c^{2}\mathrm{d}t^{2}.
\fe

The angular component of the metric is represented as $\mathrm{d}\Omega^{2}_{2} = \mathrm{d}\theta^{2} + \sin^{2}\theta \,\mathrm{d}\varphi^{2}$. The symmetry group under consideration, $\mathrm{SO}(4,1)$, is generated by a set of ten elements, denoted $\mathfrak{M}_{\mathfrak{A}\mathfrak{B}}$, which adhere to the antisymmetric property $\mathfrak{M}_{\mathfrak{A}\mathfrak{B}} = -\mathfrak{M}_{\mathfrak{B}\mathfrak{A}}$. In addition, the indices $\mathfrak{A}, \mathfrak{B}$ take values in $\{{\mathfrak{A}}, 5\}$, where ${\mathfrak{A}}, {\mathfrak{B}}$ run from $0$ to $3$. It is worth mentioning that these generators naturally split into two groups: the set $\mathfrak{M}_{{\mathfrak{A}\mathfrak{B}}} = -\mathfrak{M}_{{\mathfrak{B}\mathfrak{A}}}$, which governs rotational transformations, and the set $\mathfrak{P}_{{\mathfrak{A}}} = \mathfrak{M}_{{\mathfrak{A}} 5}$, which encodes translational operations.

The gauge potentials in their undeformed configuration, represented by $\Tilde{\omega}^{\mathfrak{A}\mathfrak{B}}_{\mu}(x)$, satisfy the antisymmetric relation $\Tilde{\omega}^{\mathfrak{A}\mathfrak{B}}_{\mu}(x) = -\Tilde{\omega}^{\mathfrak{B}\mathfrak{A}}_{\mu}(x)$. Unlike the spin connection, which adheres to $\Tilde{\omega}^{{\mathfrak{A}\mathfrak{B}}}_{\mu}(x) = -\Tilde{\omega}^{{\mathfrak{B}\mathfrak{A}}}_{\mu}(x)$, and the tetrad fields, denoted as $\mathfrak{e}^{{\mathfrak{A}}}_{\mu}(x)$, these gauge potentials play a distinct role. The components $\hat{\Tilde{\omega}}^{{\mathfrak{A}} \, 5}_{\mu}(x)$ establish a connection to the tetrads through the proportionality $\hat{\Tilde{\omega}}^{{\mathfrak{A}}\, 5}_{\mu}(x) = \mathfrak{K} \, \hat{\mathfrak{e}}^{{\mathfrak{A}}}_{\mu}(x)$, where $\mathfrak{K}$ acts as a contraction parameter. Furthermore, an extra gauge field, expressed as $\hat{\Tilde{\omega}}^{55}_{\mu}(x) = \mathfrak{K} \hat{\phi}_{\mu}(x,\Theta)$, emerges, with the function $\hat{\phi}_{\mu}(x,\Theta)$ vanishing when $\mathfrak{K} \to 0$. Taking this limit effectively collapses the gauge structure to that of the Poincaré group, $\mathrm{ISO}(3,1)$ \cite{2,1}. The associated field strength tensor for the gauge potential ${\Tilde{\omega}}^{\mathfrak{A}\mathfrak{B}}_{\mu}(x)$ reads
\ie
F^{\mathfrak{A}\mathfrak{B}}_{\mu} = \partial_{\mu}\omega^{\mathfrak{A}\mathfrak{B}}_{\nu} - \partial_{\nu}\omega^{\mathfrak{A}\mathfrak{B}}_{\mu} + \left(\Tilde{\omega}^{\mathfrak{A}\mathfrak{C}}_{\mu}\Tilde{\omega}^{\mathfrak{D}\mathfrak{B}}_{\nu}-\Tilde{\omega}^{\mathfrak{A}\mathfrak{C}}_{\nu}\Tilde{\omega}^{\mathfrak{D}\mathfrak{B}}_{\mu}\right)\mathfrak{n}_{\mathfrak{C}\mathfrak{D}},
\fe 
in which the spacetime indices range over $\mu, \nu = 0,1,2,3$, and the metric tensor is given by $\mathfrak{n}_{\mathfrak{A}\mathfrak{B}} = \mathrm{diag}(+,+,+,-,+)$. This formulation can also be rewritten in an equivalent form, leading to an alternative expression
\begin{subequations}
	\begin{align}
&F^{{\mathfrak{A}}5}_{\mu\nu}=\mathfrak{K}\left[\partial_{\mu}\mathfrak{e}^{{\mathfrak{A}}}_{\nu}-\partial_{\nu}\mathfrak{e}^{{\mathfrak{A}}}_{\mu}+\left(\Tilde{\omega}^{{\mathfrak{A}\mathfrak{B}}}_{\mu}\mathfrak{e}^{a}_{\nu}-\Tilde{\omega}^{{\mathfrak{A}\mathfrak{B}}}_{\nu}\mathfrak{e}^{{\mathfrak{C}}}_{\mu}\right)\mathfrak{n}_{{\mathfrak{B}\mathfrak{C}}}\right]=\mathfrak{K}T^{{\mathfrak{A}}}_{\mu\nu},\label{torsion}\\
&F^{{\mathfrak{A}\mathfrak{B}}}_{\mu\nu} = \partial_{\mu} \Tilde{\omega}^{{\mathfrak{A}\mathfrak{B}}}_{\nu}-\partial_{\nu}\Tilde{\omega}^{{\mathfrak{A}\mathfrak{B}}}_{\mu}+\left(\Tilde{\omega}^{{\mathfrak{A}\mathfrak{C}}}_{\mu}\Tilde{\omega}^{{\mathfrak{D}\mathfrak{B}}}_{\nu}-\Tilde{\omega}^{{\mathfrak{A}\mathfrak{C}}}_{\nu}\Tilde{\omega}^{{\mathfrak{D}\mathfrak{B}}}_{\mu}\right)\mathfrak{n}_{{\mathfrak{C}\mathfrak{D}}}+\mathfrak{K}\left(\mathfrak{e}^{{\mathfrak{A}}}_{\mu}\mathfrak{e}^{{\mathfrak{B}}}_{\nu}-\mathfrak{e}^{{\mathfrak{A}}}_{\nu}\mathfrak{e}^{{\mathfrak{B}}}_{\mu}\right)=R^{{\mathfrak{A}\mathfrak{B}}}_{\mu\nu},
\end{align}
\end{subequations}
with the metric components are given by $\mathfrak{n}_{ab} = \mathrm{diag}(+,+,+,-)$. Notably, the Poincaré gauge group under consideration is directly associated with the geometric framework of Riemann--Cartan spacetime, which accommodates both curvature and torsion effects \cite{1,6}. The torsion tensor is introduced as $T^{{\mathfrak{A}}}_{\mu\nu} \equiv F^{{\mathfrak{A}}5}_{\mu\nu}/\mathfrak{K}$, while the curvature tensor is defined as $R^{{\mathfrak{A}\mathfrak{B}}}_{\mu\nu} \equiv F^{{\mathfrak{A}\mathfrak{B}}}_{\mu\nu}$. Both quantities are formulated in terms of the tetrad fields $\mathfrak{e}^{{\mathfrak{A}}}_{\mu}(x)$ and the spin connection $\Tilde{\omega}^{{\mathfrak{A}\mathfrak{B}}}_{\mu}(x)$. In the particular case where the torsion field vanishes, as expressed in Eq. \eqref{torsion}, the spin connection can entirely be addressed by the tetrad fields alone.

The focus now shifts to examining a possible structure for gauge fields exhibiting spherical symmetry within the framework of the $\mathrm{SO}(4,1)$ group \cite{1,6}
\ie
\label{tetrads1}
	\mathfrak{e}^{1}_{\mu} = \left(\frac{1}{\Tilde{\mathcal{A}}}, 0,0,0\right), \quad \mathfrak{e}^{2}_{\mu} = \left(0, r,0,0\right), \quad \mathfrak{e}^{3}_{\mu} = \left(0,0,r\, \mathrm{sin}\theta,0\right), \quad \mathfrak{e}^{0}_{\mu} = \left(,0,0,0, \Tilde{\mathcal{A}}\right),
\fe
and 
\ie
\label{omega}
	\begin{split}
		& \Tilde{\omega}^{12}_{\mu} = \left(0, \Tilde{\mathcal{W}},0,0\right), \quad 
		\Tilde{\omega}^{13}_{\mu} = \left(0,0, \Tilde{\mathcal{Z}}\, \sin \theta,0\right),  \quad \Tilde{\omega}^{10}_{\mu} = \left(0,0,0,\Tilde{\mathcal{U}}\right),\\
		& \Tilde{\omega} ^{23}_{\mu} = \left(0,0,-\cos \theta, \Tilde{\mathcal{V}}\right), \quad \Tilde{\omega}^{20}_{\mu} = \Tilde{\omega}^{30}_{\mu} = \left(0,0,0,0\right),
	\end{split}
\fe
where the quantities $\Tilde{\mathcal{A}}$, $\Tilde{\mathcal{U}}$, $\Tilde{\mathcal{V}}$, $\Tilde{\mathcal{W}}$, and $\Tilde{\mathcal{Z}}$ are functions that depend only on the radial coordinate in three--dimensional space. In addition, the torsion tensor features nonvanishing components, which can be represented as \cite{2}
\ie
\label{torsion2}
	\begin{split}
		&T^{0}_{01} = -\frac{\Tilde{\mathcal{A}}\Tilde{\mathcal{A}}'+\Tilde{\mathcal{U}}}{\mathcal{A}}, \qquad 
		T^{2}_{03} = r\, \Tilde{\mathcal{V}} \sin \theta \,T^{2}_{12} = \frac{\Tilde{\mathcal{A}} + \Tilde{\mathcal{W}}}{\Tilde{\mathcal{A}}},\\
		& T^{3}_{02} = -r\, \Tilde{\mathcal{V}}, \,\,\,\qquad\qquad T^{3}_{13} = \frac{\left(\Tilde{\mathcal{A}} + \Tilde{\mathcal{Z}}\right)\sin \theta}{\Tilde{\mathcal{A}}}.
	\end{split}
\fe
Consequently, the curvature tensor is expressed as \cite{2}
\ie
\label{curvature}
\begin{split}
&R^{01}_{01} = \Tilde{\mathcal{U}}', \quad R^{23}_{01} = - \Tilde{\mathcal{V}}', \quad R^{13}_{23} = \left(\Tilde{\mathcal{Z}} -\Tilde{\mathcal{W}}\right) \cos\theta,\quad R^{01}_{01} = -\Tilde{\mathcal{U}}\Tilde{\mathcal{W}}, \quad R^{13}_{01} = - \Tilde{\mathcal{V}}\Tilde{\mathcal{W}},\\
& R^{03}_{03} = -\Tilde{\mathcal{U}} \Tilde{\mathcal{Z}} \sin \theta, \quad R^{12}_{03} = \Tilde{\mathcal{V}} \Tilde{\mathcal{Z}} \sin \theta \, R^{12}_{12} = \Tilde{\mathcal{W}}',\quad R^{23}_{23} = \left(1-\Tilde{\mathcal{Z}} \Tilde{\mathcal{W}}\right)\sin\theta, \quad R^{13}_{13} = \Tilde{\mathcal{Z}}' \sin\theta.	\end{split}
\fe
The notation $\Tilde{\mathcal{A}}'$, $\Tilde{\mathcal{U}}'$, $\Tilde{\mathcal{V}}'$, $\Tilde{\mathcal{W}}'$, and $\Tilde{\mathcal{Z}}'$ represents derivatives with respect to the radial coordinate. In order to eliminate the torsion field, as specified in Eq. \eqref{torsion2}, the following constraints must be satisfied
\ie
\Tilde{\mathcal{V}} = 0, \qquad \Tilde{\mathcal{U}} = - \Tilde{\mathcal{A}}\Tilde{\mathcal{A}}', \qquad \Tilde{\mathcal{W}} =  - \Tilde{\mathcal{A}} = \Tilde{\mathcal{Z}}.
\fe
Here, by considering the corresponding field equation, the following relations must be imposed
\ie
\label{fieldequation}
R^{{\mathfrak{A}}}_{\mu} - \frac{1}{2} R\, \mathfrak{e}^{{\mathfrak{A}}}_{\mu} = 0,
\fe
formulated in terms of the tetrad fields $\mathfrak{e}^{{\mathfrak{A}}}_{\mu}(x)$, with the curvature components defined as $R^{{\mathfrak{A}}}_{\mu} = R^{{\mathfrak{A}\mathfrak{B}}}_{\mu\nu} \mathfrak{e}^{\nu}_{{\mathfrak{B}}}$ and the scalar curvature given by $R = R^{{\mathfrak{A}\mathfrak{B}}}_{\mu\nu} \mathfrak{e}^{\mu}_{{\mathfrak{A}}} \mathfrak{e}^{\nu}_{{\mathfrak{B}}}$, the obtained solution reads
\ie
\Tilde{\mathcal{A}}(r) = \sqrt{1 - \frac{2M r^2}{r^3 + 2M l^2}}.
\fe

In this context, $M$ denotes the black hole mass, while $l$ corresponds to ``Hayward parameter''. The formulation of the modified metric, expressed as $\mathrm{d}s^{2} = g^{(\Theta,l)}_{\mu\nu}(x,\Theta)\mathrm{d}x^{\mu}\mathrm{d}x^{\nu}$, requires working in spherical coordinates, where $x^{\mu} = (t, r, \theta, \varphi)$ defines the coordinate system for a $(3+1)$--dimensional non--commutative Schwarzschild spacetime. To achieve this, the deformed tetrad fields $\hat{\mathfrak{e}}^{{\mathfrak{A}}}_{\mu}(x,\Theta)$ must be determined. These tetrads emerge through the contraction of the non--commutative gauge group $\mathrm{SO}(4,1)$ with the Poincaré group $\mathrm{ISO}(3,1)$, employing the Seiberg--Witten map framework \cite{3,4,5}. The nature of the resulting non--commutative spacetime is then characterized by the following conditions:
\ie
\label{NonCommSTcond1}
\left[x^{\mu},x^{\nu}\right]=i\Theta^{\mu\nu}.
\fe

The parameters $\Theta^{\mu\nu}$ are taken to be real and obey the antisymmetry condition $\Theta^{\mu\nu} = -\Theta^{\nu\mu}$. Consequently, the gravitational sector in a non--commutative framework, particularly the deformed tetrad fields $\hat{\mathfrak{e}}^{{\mathfrak{A}}}_{\mu}(x,\Theta)$ and the gauge connection $\hat{\Tilde{\omega}}^{\mathfrak{A}\mathfrak{B}}_{\mu}(x,\Theta)$, can be systematically expanded in terms of the non--commutativity parameter $\Theta$ as a power series \cite{1,2,3,4} 
\ie
\begin{split}
		&\hat{\mathfrak{e}}^{{\mathfrak{A}}}_{\mu}(x,\Theta) = \mathfrak{e}^{{\mathfrak{A}}}_{\mu}(x) - i \Theta^{\nu\rho}\mathfrak{e}^{{\mathfrak{A}}}_{\mu\nu\rho}(x) + \Theta^{\nu\rho}\Theta^{\lambda\tau}e^{{\mathfrak{A}}}_{\mu\nu\rho\lambda\tau}(x)\dots,\\	
		&\hat{\Tilde{\omega}}^{\mathfrak{A}\mathfrak{B}}_{\mu}(x,\Theta) = \Tilde{\omega}^{\mathfrak{A}\mathfrak{B}}_{\mu}(x)-i \Theta^{\nu\rho} \Tilde{\omega}^{\mathfrak{A}\mathfrak{B}}_{\mu\nu\rho}(x) + \Theta^{\nu\rho}\Theta^{\lambda\tau}\Tilde{\omega}^{\mathfrak{A}\mathfrak{B}}_{\mu\nu\rho\lambda\tau}(x)\dots  \,\,. \label{omeganoncom}
	\end{split}
\fe

The deformed tetrad fields $\hat{\mathfrak{e}}^{{\mathfrak{A}}}_{\mu}(x,\Theta)$ emerge as a consequence of introducing non-commutative corrections to the gauge connection $\hat{\Tilde{\omega}}^{\mathfrak{A}\mathfrak{B}}_{\mu}(x,\Theta)$. Their expansion follows from the formulation provided in Eq. \eqref{omeganoncom} truncated to second-order terms in the parameter $\Theta$ \cite{Juric:2025kjl} 
\ie
\omega^{\mathfrak{A}\mathfrak{B}}_{\mu\nu\rho} (x) = \frac{1}{4} \left\{\Tilde{\omega}_{\nu},\partial_{\rho}\Tilde{\omega}{\mu}+R_{\rho\mu}\right\}^{\mathfrak{A}\mathfrak{B}},\label{noncommcorr-tetrad}
\fe
{
\begin{equation}
	\begin{split}
		\Tilde{\omega}_{\mu\nu\rho\lambda\tau}^{\mathfrak{A}\mathfrak{B}} = &\frac{1}{16}\biggl[-\left\{\left\{\Tilde{\omega}_{\lambda},\left(\partial_\tau \Tilde{\omega}_{\nu} + R_{\tau\nu}\right)\right\},\left(\partial_\rho  \Tilde{\omega}_{\mu} + R_{\rho\mu}\right)\right\}^{\mathfrak{A}\mathfrak{B}} \\
  & - \left\{\Tilde{\omega}_{\nu}, \partial_\rho \left\{\Tilde{\omega}_{\lambda},\left(\partial_{\tau}\Tilde{\omega}_{\mu} + R_{\tau\mu}\right)\right\} \right\}^{\mathfrak{A}\mathfrak{B}}  +2 \left[\partial_{\lambda} \Tilde{\omega}_{\nu}, \partial_{\tau} \left(\partial_{\rho} \Tilde{\omega}_{\mu} + R_{\rho\mu}\right)\right]^{\mathfrak{A}\mathfrak{B}}  \\
  & + \left\{\Tilde{\omega}_{\nu},2\left\{R_{\rho\lambda},R_{\mu\tau}\right\}\right\}^{\mathfrak{A}\mathfrak{B}} - \left\{\Tilde{\omega}_{\nu},\left\{\Tilde{\omega}_{\lambda}, D_\tau R_{\rho\mu} + \partial_\tau R_{\rho \mu}\right\}\right\}^{\mathfrak{A}\mathfrak{B}}   \biggr].
	\end{split}
\label{noncommcorr-omega}
\end{equation}
}

Derived using the Seiberg--Witten map, Eqs. \eqref{noncommcorr-tetrad} and \eqref{noncommcorr-omega} must satisfy 
\ie
\left[\alpha,\beta\right]^{\mathfrak{A}\mathfrak{B}} =  \alpha^{\mathfrak{A}\mathfrak{C}}\beta^{\mathfrak{B}}_{\mathfrak{C}}-\beta^{\mathfrak{A}\mathfrak{C}}\alpha^{\mathfrak{B}}_{\mathfrak{C}},\qquad
\left\{\alpha,\beta\right\}^{\mathfrak{A}\mathfrak{B}} = \alpha^{\mathfrak{A}\mathfrak{C}}\beta^{\mathfrak{B}}_{\mathfrak{C}}+\beta^{\mathfrak{A}\mathfrak{C}}\alpha^{\mathfrak{B}}_{\mathfrak{C}},
\fe
and
\ie
	D_{\mu}R^{\mathfrak{A}\mathfrak{B}}_{\rho\sigma} = \partial_{\mu}R^{\mathfrak{A}\mathfrak{B}}_{\rho\sigma} +\left(\Tilde{\omega}^{\mathfrak{A}\mathfrak{C}}_{\mu}R^{\mathfrak{D}\mathfrak{B}}_{\rho\sigma}+\Tilde{\omega}^{\mathfrak{B}\mathfrak{C}}_{\mu}R^{\mathfrak{D}\mathfrak{A}}_{\rho\sigma}\right)\mathfrak{n}_{\mathfrak{C}\mathfrak{D}}.
\fe

Certain restrictions must be highlighted regarding the gauge connection $\hat{\Tilde{\omega}}^{\mathfrak{A}\mathfrak{B}}_{\mu}(x,\Theta)$, which adhere to the following constraints 
\ie
\label{CondDefOmega}
\hat{\Tilde{\omega}}^{\mathfrak{A}\mathfrak{B}\star}_{\mu}(x,\Theta) = -\hat{\Tilde{\omega}}^{\mathfrak{A}\mathfrak{B}}_{\mu}(x,\Theta), \quad 
\hat{\Tilde{\omega}}^{\mathfrak{A}\mathfrak{B}}_{\mu}(x,\Theta) ^{r} \equiv \hat{\Tilde{\omega}}^{\mathfrak{A}\mathfrak{B}}_{\mu}(x,-\Theta) = -\hat{\omega}^{\mathfrak{B}\mathfrak{A}}_{\mu}(x,\Theta).
\fe

It is important to mention that the notation ${}^\star$ represents the complex conjugate operation. Furthermore, the non--commutative modifications derived from the conditions imposed in Eq. \eqref{CondDefOmega} can be expressed as
\ie
\Tilde{\omega}^{\mathfrak{A}\mathfrak{B}}_{\mu} (x) = - \Tilde{\omega}^{\mathfrak{B}\mathfrak{A}}_{\mu} (x), \quad \Tilde{\omega}^{\mathfrak{A}\mathfrak{B}}_{\mu\nu\rho} (x) = \Tilde{\omega}^{\mathfrak{B}\mathfrak{A}}_{\mu\nu\rho} (x), \quad \Tilde{\omega}^{\mathfrak{A}\mathfrak{B}}_{\mu\nu\rho\lambda\tau} (x) = -\Tilde{\omega}^{\mathfrak{B}\mathfrak{A}}_{\mu\nu\rho\lambda\tau} (x).
\fe

The above expressions follow from the application of Eqs. \eqref{noncommcorr-tetrad} and \eqref{noncommcorr-omega}, considering the constraints of a vanishing torsion field $T^{{\mathfrak{A}}}_{\mu\nu}$ and taking the limit $\mathfrak{K} \to 0$. Under these conditions, we have
\ie
\label{ComConjDefTetrads}
\hat{\mathfrak{e}}^{{\mathfrak{A}}\star}_{\mu}(x,\Theta) = \mathfrak{e}^{{\mathfrak{A}}}_{\mu}(x)+i \Theta^{\nu\rho}\mathfrak{e}^{{\mathfrak{A}}}_{\mu\nu\rho}(x)+\Theta^{\nu\rho}\Theta^{\lambda\tau}\mathfrak{e}^{{\mathfrak{A}}}_{\mu\nu\rho\lambda\tau}(x)\dots,
\fe
where
\ie
\begin{split}
\mathfrak{e}^{{\mathfrak{A}}}_{\mu\nu\rho} &= \frac14	\left[\Tilde{\omega}^{{\mathfrak{A}\mathfrak{C}}}_{\nu}\partial_{\rho} \mathfrak{e}^{{\mathfrak{D}}}_{\mu} + \left(\partial_{\rho}\Tilde{\omega}^{{\mathfrak{A}\mathfrak{C}}}_{\mu} + R^{{\mathfrak{A}\mathfrak{C}}}_{\rho\mu}\right)\mathfrak{e}^{{\mathfrak{D}}}_{\nu}\right]\mathfrak{n}_{{\mathfrak{C}\mathfrak{D}}},
\end{split}
\fe
in which \cite{Juric:2025kjl}
{
\begin{align}
&\mathfrak{e}_{\mu \nu \rho \lambda \tau }^{\mathfrak{A}} = \\
&= \frac{1}{16}\Bigl[
   2\,\Bigl\{R_{\tau \nu},\,R_{\mu \rho}\Bigr\}^{\mathfrak{A}\mathfrak{B}}\,\mathfrak{e}_{\lambda}^{\mathfrak{C}}
   \;-\;\Tilde{\omega}_{\lambda}^{\mathfrak{A}\,\mathfrak{B}}\,\Bigl(D_{\rho}\,R_{\tau \mu}^{\mathfrak{C}\,\mathfrak{D}}
     \;+\;\partial_{\rho}\,R_{\tau \mu}^{\mathfrak{C}\,\mathfrak{D}}\Bigr)\,\mathfrak{e}_{\nu}^{m}\,\mathfrak{n}_{\mathfrak{D}\,\mathfrak{M}}
   -\,\Bigl\{\Tilde{\omega}_{\nu},\,\bigl(D_{\rho}\,R_{\tau \mu}
     + \partial_{\rho}\,R_{\tau \mu}\bigr)\Bigr\}^{ab}\,\mathfrak{e}_{\lambda}^{c}
     \nonumber \\
&\quad
   \;-\;\partial_{\tau}\,\Bigl\{\Tilde{\omega}_{\nu},\,\bigl(\partial_{\rho}\,\Tilde{\omega}_{\mu}
     + R_{\rho \mu}\bigr)\Bigr\}^{\mathfrak{A}\,\mathfrak{B}}\,\mathfrak{e}_{\lambda}^{\mathfrak{C}}
   -\,\Tilde{\omega}_{\lambda}^{\mathfrak{A}\,\mathfrak{B}}\,\partial_{\tau}\Bigl(
       \Tilde{\omega}_{\nu}^{\mathfrak{C}\,\mathfrak{D}}\,\partial_{\rho}\,\mathfrak{e}_{\mu}^{\mathfrak{M}}
       + \bigl(\partial_{\rho}\,\Tilde{\omega}_{\mu}^{\mathfrak{C}\,\mathfrak{D}}
         + R_{\rho \mu}^{\mathfrak{C}\,\mathfrak{D}}\bigr)\,\mathfrak{e}_{\nu}^{m}
     \Bigr)\,\mathfrak{n}_{\mathfrak{D}\,\mathfrak{M}}
     \nonumber \\
&\quad
   \;+\;2\,\partial_{\nu}\,\Tilde{\omega}_{\lambda}^{\mathfrak{A}\,\mathfrak{B}}\,
        \partial_{\rho}\partial_{\tau}\,\mathfrak{e}_{\mu}^{\mathfrak{C}}
   -\,2\,\partial_{\rho}\Bigl(\partial_{\tau}\,\Tilde{\omega}_{\mu}^{\mathfrak{A}\,\mathfrak{B}}
     + R_{\tau \mu}^{\mathfrak{A}\,\mathfrak{B}}\Bigr)\,\partial_{\nu}\,\mathfrak{e}_{\lambda}^{\mathfrak{C}}
   \;-\;\Bigl\{\Tilde{\omega}_{\nu},\,\bigl(\partial_{\rho}\,\Tilde{\omega}_{\lambda}
     + R_{\rho \lambda}\bigr)\Bigr\}^{\mathfrak{A}\,\mathfrak{B}}\,\partial_{\tau}\,\mathfrak{e}_{\mu}^{\mathfrak{C}}
\nonumber \\
&\quad
   \;-\,\Bigl(\partial_{\tau}\,\Tilde{\omega}_{\mu}^{\mathfrak{A}\,\mathfrak{B}}
     + R_{\tau \mu}^{\mathfrak{A}\,\mathfrak{B}}\Bigr)\,\Bigl(
       \Tilde{\omega}_{\nu}^{\mathfrak{C}\,\mathfrak{D}}\,\partial_{\rho}\,\mathfrak{e}_{\lambda}^{\mathfrak{M}}
       + \bigl(\partial_{\rho}\,\Tilde{\omega}_{\lambda}^{\mathfrak{C}\,\mathfrak{D}}
         + R_{\rho \lambda}^{\mathfrak{C}\,\mathfrak{D}}\bigr)\,\mathfrak{e}_{\nu}^{\mathfrak{M}}\,\eta_{\mathfrak{D}\,\mathfrak{M}}
     \Bigr)
\Bigr]\;\mathfrak{n}_{\mathfrak{B}\,\mathfrak{C}}
\nonumber \\
&\quad
\; - \frac{1}{16}\,\Tilde{\omega}_{\lambda}^{\mathfrak{A}\,\mathfrak{C}}\,\Tilde{\omega}_{\nu}^{\mathfrak{D}\,\mathfrak{B}}\,\mathfrak{e}_{\rho}^{\mathfrak{F}}\,
   R_{\tau \mu}^{\mathfrak{G}\,\mathfrak{M}}\,\mathfrak{n}_{\mathfrak{C}\,\mathfrak{D}}\,\mathfrak{n}_{\mathfrak{F}\,\mathfrak{G}}\,\mathfrak{n}_{\mathfrak{B}\,\mathfrak{M}}\,.
\label{3.12}
\end{align}

It is worth noting that this final term was overlooked in the seminal work \cite{chaichian2008corrections}, as well as in the subsequent studies that built upon its framework.
} As a result, the modified metric tensor takes the form
\ie
\label{DefMetTensor}
g^{(\Theta)}_{\mu\nu}\left(x,\Theta\right) = \frac{1}{2} \mathfrak{n}_{{\mathfrak{A}}{\mathfrak{B}}}\Bigg[\hat{\mathfrak{e}}^{{\mathfrak{A}}}_{\mu}(x,\Theta)\ast\hat{\mathfrak{e}}^{{\mathfrak{B}}\star}_{\nu}(x,\Theta)+\hat{\mathfrak{e}}^{{\mathfrak{B}}}_{\mu}(x,\Theta)\ast\hat{\mathfrak{e}}^{{\mathfrak{A}}\star}_{\nu}(x,\Theta)\Bigg],
\fe
with the symbol $\ast$ denotes the standard star product. For the subsequent calculations, natural units will be employed, fixing $\hbar = c = G = 1$. Under this convention, the following expression holds
\ie \label{ggtt}
\begin{split}
& g_{tt}^{(\Theta,l)}  = -1 + \frac{2 M r^2}{2 l^2 M+r^3} - \frac{1}{{2}}\left\{ \frac{1}{2} \left(\frac{6 M r^4}{\left(2 l^2 M+r^3\right)^2}-\frac{4 M r}{2 l^2 M+r^3}\right)^2 \right. \\
&\left.  +\frac{r \left(\frac{6 M r^4}{\left(2 l^2 M+r^3\right)^2}-\frac{4 M r}{2 l^2 M+r^3}\right)^3}{4 \left(1-\frac{2 M r^2}{2 l^2 M+r^3}\right)} +r \left(1-\frac{2 M r^2}{2 l^2 M+r^3}\right)^{3/2}  \right.  \\
& \left.   \times \left[ \frac{3 \left(\frac{6 M r^4}{\left(2 l^2 M+r^3\right)^2}-\frac{4 M r}{2 l^2 M+r^3}\right)^3}{8 \left(1-\frac{2 M r^2}{2 l^2 M+r^3}\right)^{5/2}} +\frac{\frac{324 M r^8}{\left(2 l^2 M+r^3\right)^4}-\frac{432 M r^5}{\left(2 l^2 M+r^3\right)^3}+\frac{120 M r^2}{\left(2 l^2 M+r^3\right)^2}}{2 \sqrt{1-\frac{2 M r^2}{2 l^2 M+r^3}}} \right. \right. \\
& \left. \left.  -\frac{3 \left(\frac{36 M r^3}{\left(2 l^2 M+r^3\right)^2}-\frac{4 M}{2 l^2 M+r^3}-\frac{36 M r^6}{\left(2 l^2 M+r^3\right)^3}\right) \left(\frac{6 M r^4}{\left(2 l^2 M+r^3\right)^2}-\frac{4 M r}{2 l^2 M+r^3}\right)}{4 \left(1-\frac{2 M r^2}{2 l^2 M+r^3}\right)^{3/2}}     \right] \right. \\
& \left. + \frac{5}{2} r \left(\frac{6 M r^4}{\left(2 l^2 M+r^3\right)^2}-\frac{4 M r}{2 l^2 M+r^3}\right) \sqrt{1-\frac{2 M r^2}{2 l^2 M+r^3}}   \right. \\
& \left. \times  \left[ \frac{\frac{36 M r^3}{\left(2 l^2 M+r^3\right)^2}-\frac{4 M}{2 l^2 M+r^3}-\frac{36 M r^6}{\left(2 l^2 M+r^3\right)^3}}{2 \sqrt{1-\frac{2 M r^2}{2 l^2 M+r^3}}}-\frac{\left(\frac{6 M r^4}{\left(2 {l}^2 M+r^3\right)^2}-\frac{4 M r}{2 l^2 M+r^3}\right)^2}{4 \left(1-\frac{2 M r^2}{2 l^2 M+r^3}\right)^{3/2}}   \right] \right. \\
& \left. + \left(1-\frac{2 M r^2}{2 l^2 M+r^3}\right)^{3/2}    \times  \left[ \frac{\frac{36 M r^3}{\left(2 l^2 M+r^3\right)^2}-\frac{4 M}{2 l^2 M+r^3}-\frac{36 M r^6}{\left(2 l^2 M+r^3\right)^3}}{2 \sqrt{1-\frac{2 M r^2}{2 l^2 M+r^3}}}  \right. \right. \\
& \left. \left. -\frac{\left(\frac{6 M r^4}{\left(2 l^2 M+r^3\right)^2}-\frac{4 M r}{2 l^2 M+r^3}\right)^2}{4 \left(1-\frac{2 M r^2}{2 l^2 M+r^3}\right)^{3/2}}  \right]     \right\}\Theta^{2} ,
\end{split}
\fe

\ie \label{grr}
\begin{split}
 g_{rr}^{(\Theta,l)} & = \frac{1}{1-\frac{2 M r^2}{2 l^2 M+r^3}}+\frac{ \left(\frac{\frac{36 M r^3}{\left(2 l^2 M+r^3\right)^2}-\frac{4 M}{2 l^2 M+r^3}-\frac{36 M r^6}{\left(2 l^2 M+r^3\right)^3}}{2 \sqrt{1-\frac{2 M r^2}{2 l^2 M+r^3}}}-\frac{\left(\frac{6 M r^4}{\left(2 l^2 M+r^3\right)^2}-\frac{4 M r}{2 l^2 M+r^3}\right)^2}{4 \left(1-\frac{2 M r^2}{2 l^2 M+r^3}\right)^{3/2}}\right) \Theta ^2}{{2} \sqrt{1-\frac{2 M r^2}{2 l^2 M+r^3}}},
\end{split}
\fe

\ie
\begin{split} \label{gtheta}
 g^{(\Theta,l)}_{\theta\theta} = & \, \, r^{2} {+ \frac{1}{16 \left(2 l^2 M+r^3\right)^3 \left(2 l^2 M+r^2 (r-2 M)\right)} }  \\
 & {\times \Bigg\{ 16 l^8 M^4+32 l^6 M^3 r^2 (r-23 M)+r^{10} \left(64 M^2-32 M r+r^2\right)   } \\
 & {  8 l^4 M^2 r^4 \left(98 M^2+108 M r+3 r^2\right)++8 l^2 M r^7 \left(-160 M^2+69 M r+r^2\right)  \Bigg\}\Theta^{2}}
\end{split}
\fe

\ie\label{gphi}
\begin{split}
g^{(\Theta,l)}_{\varphi\varphi} = & \,\, r^{2} \sin^{2}\theta + { \frac{1}{16} \Bigg\{ 5 \cos^2(\theta ) + \frac{1}{\left(2 l^2 M+r^3\right)^3 \left(2 l^2 M+r^2 (r-2 M)\right)} \times }\\
&  {\times \Bigg[ 4 \sin ^2(\theta ) \Bigg(16 l^8 M^4+16 l^6 M^3 r^2 (2 r-5 M)+8 l^4 M^2 r^4 \left(16 M^2+6 M r+3 r^2\right)  }  \\
& { +4 l^2 M r^7 \left(-28 M^2+9 M r+2 r^2\right)+r^{10} \left(2 M^2-4 M r+r^2\right)\Bigg)  \Bigg]    \Bigg\} }.
\end{split}
\fe

\begin{figure}
    \centering
    \includegraphics[scale=0.7]{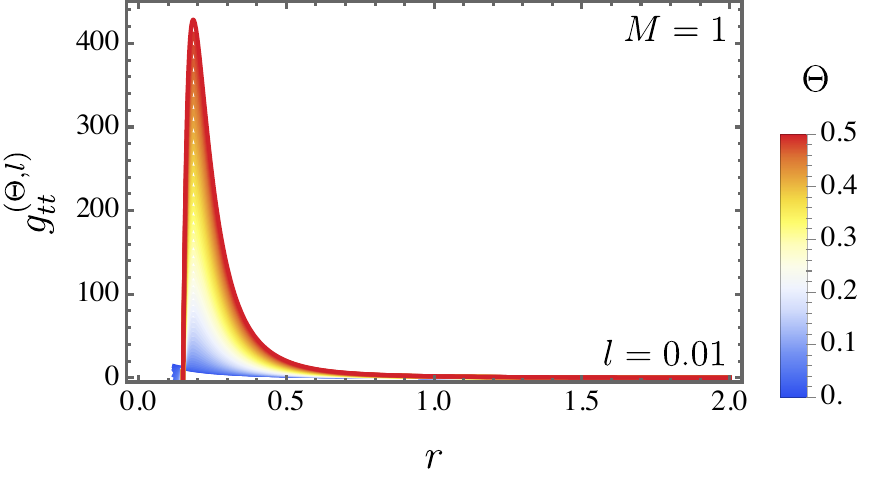}
    \caption{The metric component $g^{(\Theta,l)}_{tt}$ as a function of $r$ for different values of the non--commutative parameter $\Theta$, with $l$ fixed at $0.01$.}
    \label{metricgtt}
\end{figure}

\begin{figure}
    \centering
    \includegraphics[scale=0.7]{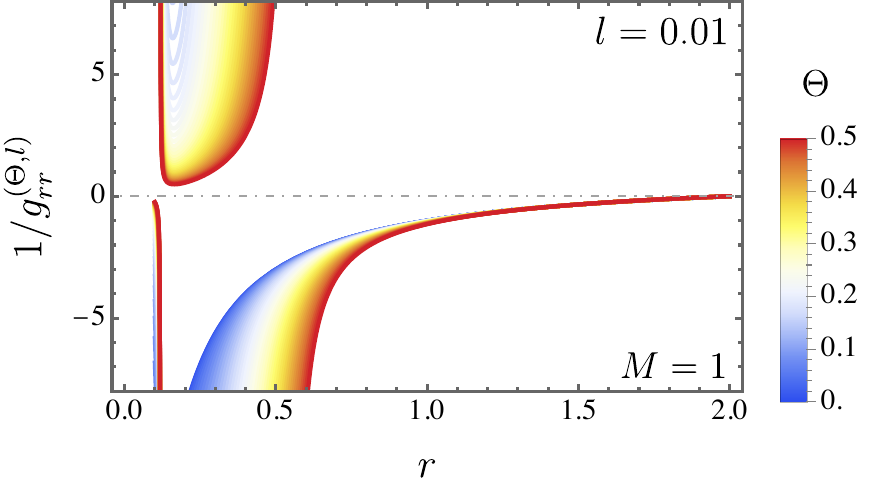}
    \caption{The metric component $1/g^{(\Theta,l)}_{rr}$ as a function of $r$ for different values of the non--commutative parameter $\Theta$, with $l$ fixed at $0.01$.}
    \label{metricgrr}
\end{figure}

\begin{figure}
    \centering
    \includegraphics[scale=0.7]{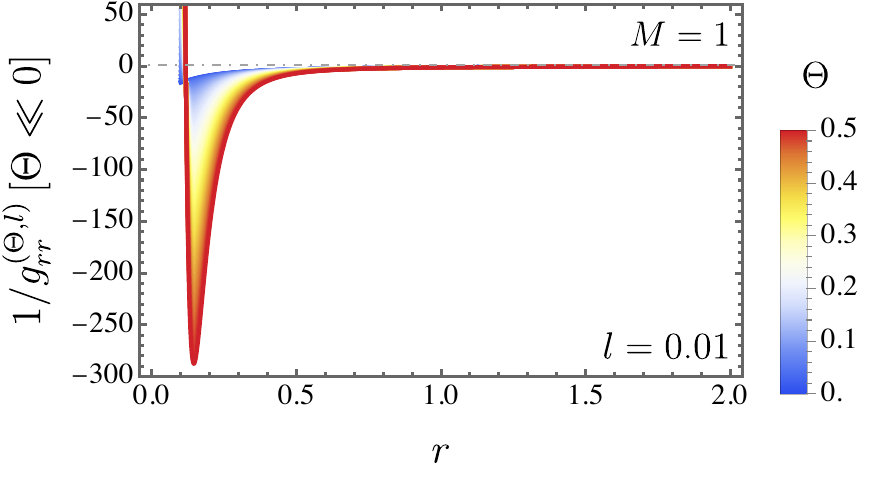}
    \caption{The metric component $1/g^{(\Theta,l)}_{rr}$, in the regime of small $\Theta$, as a function of $r$ for different values of the non--commutative parameter $\Theta$, with $l$ fixed at $0.01$.}
    \label{expansiongrr}
\end{figure}

{
Based on the newly derived metric components, several observations are in order. First, when comparing the two approaches for constructing non--commutative black holes—namely, the original method in \cite{chaichian2008corrections} and the recent corrections proposed in Ref.~\cite{Juric:2025kjl}—the differences in the $g^{(\Theta,l)}_{tt}$ and $g^{(\Theta,l)}_{rr}$ components are minimal, amounting to a mere factor of $1/2$ in each. A similar feature also arises in the Schwarzschild case, at least for the same Moyal twist employed here, namely $\partial_r\wedge\partial_\theta$. Also, in order to verify the behavior of $ g_{tt}^{(\Theta,l)}$ and $ 1/g_{rr}^{(\Theta,l)}$ we present Figs. \ref{metricgtt} and \ref{metricgrr}. 

However, for the remaining components, the differences become more significant. In particular, for $g^{(\Theta,l)}_{\theta\theta}$, the deviation between the two constructions is considerably more pronounced
\ie
g^{(\Theta,l) [ \text{Chaichian}]}_{\theta\theta} - g^{(\Theta,l)  [\text{Jurić}]}_{\theta\theta}  =  \frac{3 \Theta ^2 M r^2 \left(56 l^4 M^2 - 106 l^2 M r^3+5 r^6\right)}{16 \left(2 l^2 M+r^3\right)^3}.
\fe
In addition, for the $g^{(\Theta,l)}_{\varphi\varphi}$ component, the discrepancy between the two methods becomes even more pronounced, as we can see below
\ie
\begin{split}
& g^{(\Theta,l)  [\text{Chaichian}]}_{\varphi\varphi} - g^{(\Theta,l)  [\text{Jurić}]}_{\varphi\varphi} \\
& = -\frac{\Theta^{2}}{8 \left(2 l^2 M+r^3\right)^3 \left(2 l^2 M+r^2 (r-2 M)\right)} \\
& \times \Bigg\{ 32 l^8 M^4+64 l^6 M^3 r^2 (r-M)+8 l^2 M r^7 \left(2 r^2-7 M^2\right)+r^{10} \left(M^2-5 M r+2 r^2\right)      \\
& +4 l^4 M^2 r^4 \left(16 M^2-3 M r+12 r^2\right) \\
&   +M r^2 \cos 2 \theta  \Big[32 l^6 M^3-4 l^4 M^2 r^2 (16 M+9 r)+8 l^2 M r^5 (7 M-3 r)+r^8 (r-M)\Big]    \Bigg\} 
\end{split}
\fe
Therefore, when focusing on thermodynamics, Hawking radiation, and particle production—along the lines of Refs.~\cite{12araujo2024particle,12araujo2025does,12aa2025particle}—the impact of the correction remains almost negligible, as these analyses primarily depend on the $g_{tt}(r)$ and $g_{rr}(r)$ components. In contrast, quantities such as quasinormal modes, time--domain profiles, greybody factors, absorption cross--sections, shadows, and evaporation dynamics will exhibit substantial modifications in their respective formulations.

Furthermore, from a straightforward inspection of $1/g_{rr}^{(\Theta,l)}$, it becomes evident that non--commutativity does not affect the location of the event horizon in comparison with the Hayward black hole, namely,
\ie
\begin{split}
\label{eventhorizonhay}
r_{h} = &  \, \frac{1}{3} \Bigg( 2 M+\sqrt[3]{-27 l^2 M+3 \sqrt{81 l^4 M^2-48 l^2 M^4}+8 M^3}  \\
&  +\frac{4 M^2}{\sqrt[3]{-27 l^2 M+3 \sqrt{81 l^4 M^2-48 l^2 M^4}+8 M^3}}\Bigg)\\
& \approx \, \, 2 M-\frac{l^2}{2 M},
\end{split}
\fe
by taking in account $l$ small. On the other hand, if we solve it for $M$, we obtain
\ie
\label{masss}
M = \frac{r_{h}^3}{2 \left(r_{h}^2-l^2\right)} \approx \, \, \frac{r_{h}}{2} + \frac{l^2}{2 \, r_{h}}.
\fe

Nevertheless, an estimate can be obtained by expanding $1/g_{rr}^{(\Theta,l)}$ up to second order and solving the resulting expression. In other words, the corresponding expanded form of Eq.~(\ref{grr}) reads:
\ie
\label{exxppss}
\frac{1}{g_{rr}^{(\Theta,l)}} \approx  \frac{1}{\frac{2 M r^2}{1+ 2 l^2 M-2 M r^2+r^3}}+\frac{\Theta ^2 M \left(16 l^6 M^3-48 l^4 M^2 r^3+48 l^2 M^2 r^5-24 l^2 M r^6-3 M r^8+2 r^9\right)}{2 \left(2 l^2 M+r^3\right)^4},
\fe
and solving it for $r$ yields a single real and positive solution: $r = 1.99994$ (for $\Theta = 0.01$ and $l = 0.01$), which is slightly smaller than the Schwarzschild event horizon. Also, its corresponding behavior is shown in Fig. \ref{expansiongrr}. However, in our forthcoming analysis, we shall consider the full version of Eq. (\ref{grr}) (instead of the expansion encountered in Eq. (\ref{exxppss})), which leads to no corrections to the event horizon due to non--commutativity, as demonstrated in Ref. \cite{Juric:2025kjl}. 

}

\begin{figure}
    \centering
    \includegraphics[scale=0.55]{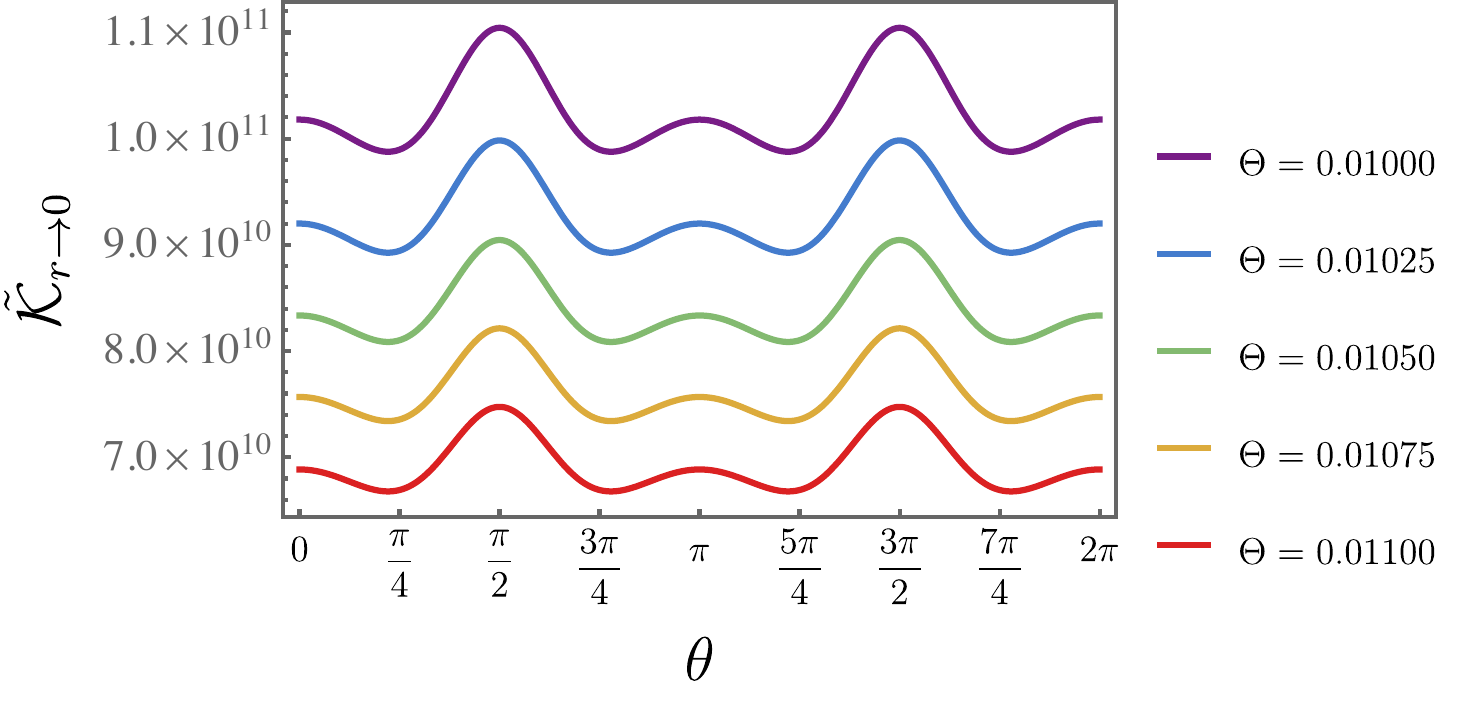}
    \caption{The behavior of $\tilde{\mathcal{K}}_{r\to 0}$ as a function of $\theta$ for a fixed value of $l = 0.1$.}
    \label{kkk}
\end{figure}

It is worth commenting that the Kretschmann scalar contains {152.282} terms, and its expression will be omitted for obvious reasons. In the limit $r \to 0$, it becomes {
\ie
\begin{split}
\tilde{\mathcal{K}}_{r\to 0} = &  - \frac{185664 l^2 \Theta ^2}{\left(\Theta ^2-2 l^2\right)^4}+\frac{49152 l^8}{\Theta ^4 \left(\Theta ^2-2 l^2\right)^4}-\frac{190464 l^6}{\Theta ^2 \left(\Theta ^2-2 l^2\right)^4}+\frac{303744 l^4}{\left(\Theta ^2-2 l^2\right)^4} \\
& +\frac{37776 \Theta ^4}{\left(\Theta ^2-2 l^2\right)^4}-\frac{983040 l^8}{\Theta ^4 \left(\Theta ^2-2 l^2\right)^4 (\cos (2 \theta )+9)^2}+\frac{1392640 l^4}{\left(\Theta ^2-2 l^2\right)^4 (\cos (2 \theta )+9)^2}\\
& +\frac{6553600}{\Theta ^4 (\cos (2 \theta )+9)^4} +\frac{327680 l^6}{\Theta ^2 \left(\Theta ^2-2 l^2\right)^4 (\cos (2 \theta )+9)^2}-\frac{1146880 l^2 \Theta ^2}{\left(\Theta ^2-2 l^2\right)^4 (\cos (2 \theta )+9)^2}\\
& +\frac{245760 \Theta ^4}{\left(\Theta ^2-2 l^2\right)^4 (\cos (2 \theta )+9)^2}-\frac{327680 l^8}{\Theta ^4 \left(\Theta ^2-2 l^2\right)^4 (\cos (2 \theta )+9)}-\frac{1064960 l^4}{\left(\Theta ^2-2 l^2\right)^4 (\cos (2 \theta )+9)} \\
& + \frac{491520 l^2 \Theta ^2}{\left(\Theta ^2-2 l^2\right)^4 (\cos (2 \theta )+9)}-\frac{81920 \Theta ^4}{\left(\Theta ^2-2 l^2\right)^4 (\cos (2 \theta )+9)}+\frac{983040 l^6}{\Theta ^2 \left(\Theta ^2-2 l^2\right)^4 (\cos (2 \theta )+9)},
\end{split}
\fe
or, retaining only the leading--order contribution in $\Theta$, namely $\Theta^{2}$, we obtain
\ie
\begin{split}
\tilde{\mathcal{K}}^{\text{leading-order}}_{r\to 0} & \approx \, \frac{4284 \Theta ^2}{l^6}+\frac{5120}{l^4 (\cos (2 \theta )+9)}-\frac{25600}{l^4 (\cos (2 \theta )+9)^2}+\frac{2856}{l^4}\\
& -\frac{5760 \cos ^2(2 \theta )}{l^2 \Theta ^2 (\cos (2 \theta )+9)^2}-\frac{83200 \cos (2 \theta )}{l^2 \Theta ^2 (\cos (2 \theta )+9)^2}-\frac{384640}{l^2 \Theta ^2 (\cos (2 \theta )+9)^2}.
\end{split}
\fe

}
In other words, the black hole may remain regular under the enhancement of non--commutativity introduced via the non--commutative gauge theory. Moreover, an unusual dependence on the angle $\theta$ emerges, in contrast to the standard Hayward solution without non--commutativity, where $\tilde{\mathcal{K}}^{Hay}_{r\to 0} = 24/l^4$. Also, even in the implementation of non--commutativity through the non--commutative gauge theory applied to the Schwarzschild black hole, which yielded a regular solution $\tilde{\mathcal{K}}^{Schw ( \Theta)}_{r\to 0} = \frac{6208}{3 \Theta^4}$ \cite{chaichian2008corrections}, no dependence on $\theta$ was found. In addition, notice that this angular dependence implies that the expression for $\tilde{\mathcal{K}}_{r\to 0}$ must be evaluated within an appropriate domain to ensure regularity. In Fig. \ref{kkk}, the behavior of $\tilde{\mathcal{K}}_{r\to 0}$ is shown as a function of $\theta$ for a fixed value of $l = 0.1$. Therefore, we conclude that our proposed solution remains regular {with a dependence on the angle $\theta$}.

\section{Thermodynamics}

This section examines the thermodynamic properties of the system, beginning with the Hawking temperature, $T^{(\Theta,l)}$, derived from the surface gravity. The analysis then extends to the entropy, $S^{(\Theta,l)}$, and heat capacity, $C_{V}^{(\Theta,l)}$. Due to their extensive length, the explicit expressions for these quantities, apart from the Hawking temperature, will be omitted.

Furthermore, all thermodynamic state quantities will be plotted as functions of the mass $M$. As will be shown, the behavior of $T^{(\Theta,l)}$ suggests the existence of a remnant mass, characterized by the condition $T^{(\Theta,l)} \to 0$.


\subsection{Hawking temperature}

\begin{figure}
    \centering
    \includegraphics[scale=0.7]{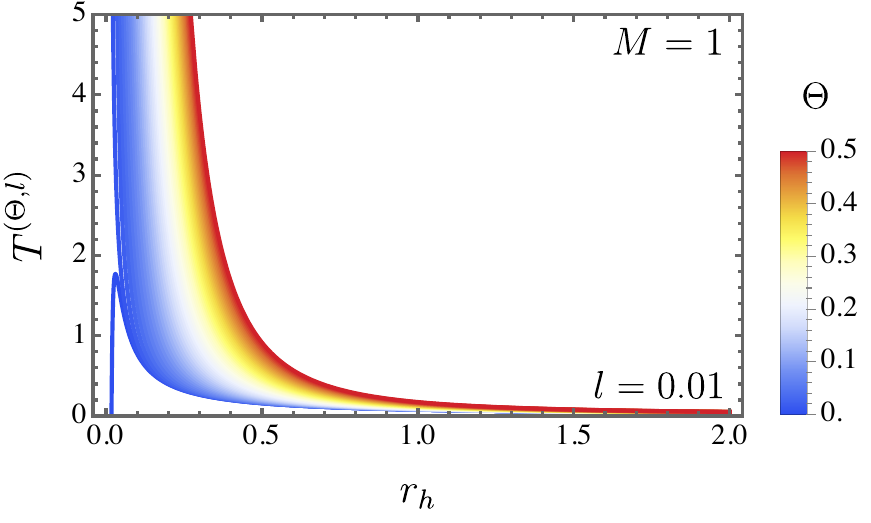}
    \caption{The Hawking temperature $T^{(\Theta,l)}$ is plotted as a function of the event horizon $r_{h}$ for different values of $\Theta$, with $l$ fixed at $0.01$.}
    \label{hawkingtemppr}
\end{figure}

\begin{figure}
    \centering
    \includegraphics[scale=0.7]{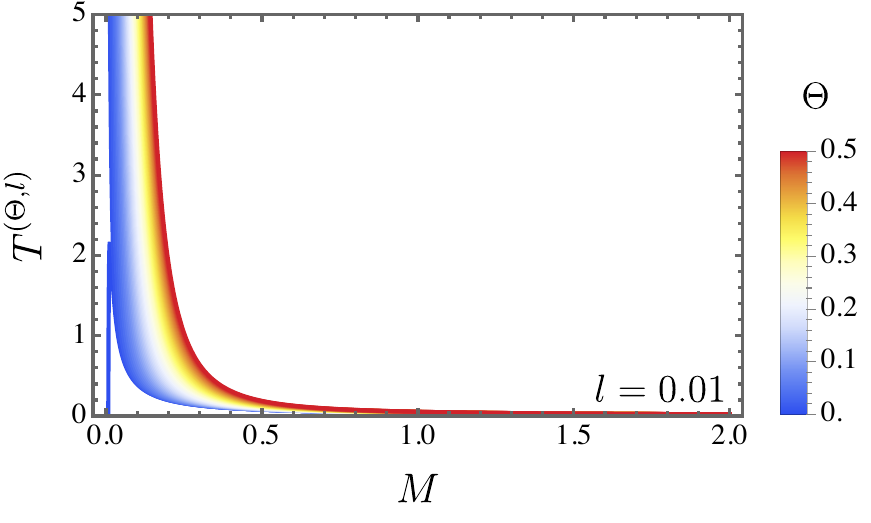}
    \caption{The Hawking temperature $T^{(\Theta,l)}$ is plotted as a function of mass $M$ for different values of $\Theta$, with $l$ fixed at $0.01$.}
    \label{hawkingtempp}
\end{figure}

By using the surface gravity approach, the Hawking temperature reads 
\ie
\begin{split}
\label{temppppe}
T^{(\Theta,l)} & =   \frac{1}{{4\pi \sqrt {{{|g^{(\Theta,l)}_{tt}|}}{{g^{(\Theta,l)}_{rr}}}} }}{\left. {\frac{{\mathrm{d}{{|g^{(\Theta,l)}_{tt}|}}}}{{\mathrm{d}r}}} \right|_{r = {r_{h}}}}  \\
  {\approx} & \, \, { \frac{M}{2 \pi  r_{h}^2}-\frac{4 l^2 M^2}{\pi  r_{h}^5}    -\frac{1670 l^2 \Theta ^2 M^5}{\pi  r_{h}^8 (2 M-r_{h})^2}+\frac{2286 l^2 \Theta ^2 M^4}{\pi  r_{h}^7 (2 M-r_{h})^2}-\frac{1026 l^2 \Theta ^2 M^3}{\pi  r_{h}^6 (2 M-r_{h})^2} } \\
  & {+\frac{150 l^2 \Theta ^2 M^2}{\pi  r_{h}^5 (2 M-r_{h})^2}+\frac{10 \Theta ^2 M^3}{\pi  r_{h}^5 (2 M-r_{h})}-\frac{33 \Theta ^2 M^2}{4 \pi  r_{h}^4 (2 M-r_{h})}+\frac{3 \Theta ^2 M}{2 \pi  r_{h}^3 (2 M-r_{h})} }        ,
\end{split}
\fe
{where the expansion has been carried out up to second order in $\Theta$, and $l$. In contrast to the Schwarzschild case (with the same Moyal product considered here), the surface gravity is well defined. Accordingly, by substituting Eq. (\ref{masss}) into Eq.~(\ref{temppppe}), we obtain
\ie
\begin{split}
\label{refMhawking}
T^{(\Theta,l)} = & \, \frac{1}{4 \pi  r_{h}} -\frac{835 l^8 \Theta ^2}{16 \pi  r_{h}^{11}}-\frac{1889 l^6 \Theta ^2}{16 \pi  r_{h}^9}-\frac{l^6}{\pi  r_{h}^7}-\frac{619 l^4 \Theta ^2}{8 \pi  r_{h}^7}\\
& -\frac{2 l^4}{\pi  r_{h}^5}-\frac{163 l^2 \Theta ^2}{16 \pi  r_{h}^5} -\frac{3 l^2}{4 \pi  r_{h}^3}-\frac{\Theta ^2}{8 \pi  l^2 r_{h}}+\frac{19 \Theta ^2}{16 \pi  r_{h}^3}.
\end{split}
\fe
Since both $\Theta$ and $l$ are assumed to be small (as we shall be verifying in the next sections -- bounds based on the solar system tests), we shall retain only the leading--order terms of the above expression, namely,
\ie
\label{hhasds}
T \approx \, \frac{1}{4 \pi  r_{h}} - \frac{3 l^2}{4 \pi  r_{h}^3} +\frac{11 \Theta ^2}{8 \pi  r_{h}^3} .
\fe

Here, we observe that the first and second terms on the right--hand side correspond to the Schwarzschild and Hayward black holes, respectively. As expected, the third term arises from the non--commutative corrections introduced in the present work.

To examine how the Hawking temperature varies with the event horizon, we display its behavior in Fig.~\ref{hawkingtemppr}, which shows that increasing $\Theta$ enhances the magnitude of $T^{(\Theta,l)}$. On the other hand, we can write it as a function of the mass $M$. For this purpose, we we the expression of Eq. (\ref{eventhorizonhay}) and substitute this into Eq.~(\ref{hhasds}), yielding:
\ie
\label{masshawww}
T^{(\Theta,l)} \approx \, \frac{1}{8 \pi  M} -\frac{l^2}{16 \left(\pi  M^3\right)}+\frac{11 \Theta ^2}{64 \pi  M^3},
\fe
where also leading terms are taken into account only.

Expressing it in this form is essential for identifying the existence of a remnant mass ($M_{\text{rem}}$) and for analyzing the black hole evaporation process through the Stefan--Boltzmann law. Although this latter aspect is indeed significant, it falls outside the scope of the present work. A comprehensive investigation, including the analysis of particle creation via greybody factors, emission rates, absorption cross sections, and the evaporation lifetime, will be presented in a forthcoming study, as outlined in the conclusion section.

By setting $T^{(\Theta,l)} \to 0$ in Eq.~(\ref{masshawww}), one obtains two distinct solutions for $M$. However, none of them correspond to real and physically defined quantities, namely:
\ie
M^{(1)}_{\text{rem}} = \frac{1}{2} \sqrt{2 l^2-\frac{11 \Theta ^2}{2}},
\fe
\ie
M^{(2)}_{\text{rem}} = -\frac{1}{2} \sqrt{2 l^2-\frac{11 \Theta ^2}{2}}.
\fe
As is straightforward to verify, the dominant term, $-11\Theta^2/2$, is negative, resulting in imaginary values for both quantities. In this case, our interpretation is that there is no physically meaningful remnant mass, implying that the black hole considered here undergoes complete evaporation. This conclusion is further supported by Fig.~\ref{hawkingtempp}, which reinforces the absence of a remnant mass—except in the limiting case $\Theta = 0$, where the pure Hayward solution does predict a remnant, i.e., $\sqrt{2} l/2 $.

}

{As is standard in thermodynamic analyses, entropy represents a fundamental quantity of interest. However, in the present case, if one explicitly adopts the relation $S = \pi \, r_{h}^{2}$, no corrections arising from the non--commutative parameter $\Theta$ appear. For this reason, we shall omit its plots in this work.

}


\subsection{Heat capacity}

Finally, for finishing the thermodynamic behavior analysis in this study, we present the heat capacity, which is given by 
\ie
\begin{split}
C^{(\Theta,l)}_{V} & = {\frac{2}{3} \pi  r_{h}^2 \left(\frac{8 r_{h}^2}{36 l^2-57 \Theta ^2-4 r_{h}^2}-1\right) } \\
& { \approx \,  - 12 \pi  l^2-2 \pi  r_{h}^2 + \left( \frac{342 \pi  l^2}{r_{h}^2}+19 \pi \right)\Theta^{2}  . }
\end{split}
\fe

\begin{figure}
    \centering
    \includegraphics[scale=0.7]{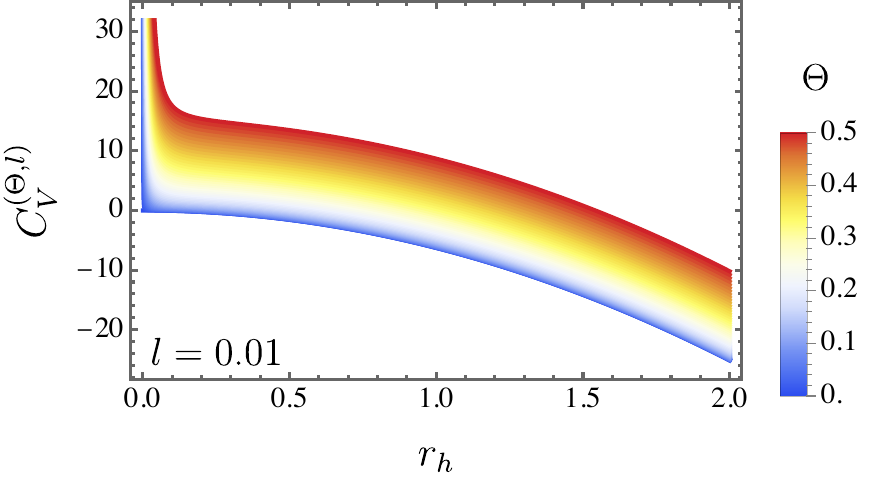}
    \caption{The heat capacity plotted as a function of mass $M$ for different values of $\Theta$, with $l$ fixed at $0.01$.}
    \label{heattt}
\end{figure}

In Fig. \ref{heattt}, we illustrate the behavior of the heat capacity, highlighting the modifications introduced by the non--commutative parameter $\Theta$. Indeed, the heat capacity exhibits two distinct regions, characterized by the presence of both positive and negative values.

\section{Quantum radiation}

In this section, building on the previous analysis, where we verified that the black hole under consideration emits thermal radiation through the Hawking temperature calculation, we now examine this quantum radiation --- the Hawking radiation. Specifically, we consider both particle modes: fermions and bosons. The particle creation densities for each case are computed, followed by a comparison to determine which mode exhibits a higher particle creation rate. As we shall see, for a given frequency $\omega$, bosons are emitted more abundantly than fermions.


\subsection{Bosonic particle modes}

To accomplish our analysis, let us consider following configuration to the metric tensor 
\ie
\label{eqwerer}
\mathrm{d}s^{2} =  g_{tt}^{(\Theta,l)}(r) \mathrm{d}t^{2} +  g_{rr}^{(\Theta,l)} (r) \mathrm{d}r^{2} + g_{\theta\theta}^{(\Theta,l)}(r,\theta)\mathrm{d}\theta^{2} + g_{\varphi\varphi}^{(\Theta,l)}(r,\theta) \mathrm{d}\varphi^{2}.
\fe
Within the Hamilton--Jacobi framework, the equation describing radial motion for a massless particle is given by \cite{Filho:2023qxu,Filho:2023voz,vanzo2011tunnelling}
\ie
\frac{1}{|g_{tt}^{(\Theta,l)}(r)|}(\partial_t I)^2 + \frac{1}{g_{rr}^{(\Theta,l)}(r)}(\partial_r I)^2=0\,.
\label{m2}
\fe

As it can be seen below, the classical action can be written with the plus and minus signs representing outgoing and ingoing particles, respectively
\ie
I_{\pm}=-\omega t\pm\int \omega \frac{\mathrm{d}r}{\sqrt{\frac{|g_{tt}^{(\Theta,l)}(r)|}{g_{rr}^{(\Theta,l)}(r)}}}\,,
\label{m3}
\fe
where $\omega = - \partial_t I$ denotes the Killing energy. Taking into account the near--horizon expansion, we have
\ie
|g_{tt}^{(\Theta,l)}(r)| =  \left.
\frac{\mathrm{d}}{\mathrm{d}r}\Bigl(|g_{tt}^{(\Theta,l)}(r)|\Bigr)
\right|_{r=r_{h}}  (r-r_{ h})+ \dots \;,\quad  \frac{1}{g_{rr}^{(\Theta,l)}(r)} 
= 
\left.
\frac{\mathrm{d}}{\mathrm{d}r}\Bigl(\tfrac{1}{g_{rr}^{(\Theta,l)}(r)}\Bigr)
\right|_{r=r_{h}}
\, (r-r_{h}) 
+ \dots
\;,
\fe
and applying Feynman's method, we directly obtain
\ie
\mbox{Im}\!\int\!\mathrm{d}I_+-\mbox{ Im}\!\int\! \mathrm{d}I_-=\frac{\pi\omega}{\kappa},
\fe
with
\ie\label{kvw}
\kappa=\frac{1}{2}\sqrt{\left.
\frac{\mathrm{d}}{\mathrm{d}r}\Bigl(|g_{tt}^{(\Theta,l)}(r)|\Bigr)
\right|_{r=r_{h}}   \left.
\frac{\mathrm{d}}{\mathrm{d}r}\Bigl(\tfrac{1}{g_{rr}^{(\Theta,l)}(r)}\Bigr)
\right|_{r=r_{h}}},
\fe
is, therefore, the so--called surface gravity. By defining 
\ie
\Gamma = e^{-\frac{2 \pi  \omega }{k}},
\fe
we directly obtain the particle creation density $n$ for fermions, which can be determined as
\ie
\begin{split}
n^{{(\Theta,l)}} & = \frac{\Gamma}{1-\Gamma} \\
& = {\frac{1}{\exp \left(\frac{\pi  \omega  \left(l^2-4 M^2\right)^8}{2 M^3 \left(l^2-4 M^2\right)^3 \left(l^6-12 l^4 M^2+112 l^2 M^4-64 M^6\right)+2 \Theta ^2 M^5 \left(-9 l^8+88 l^6 M^2-5952 l^4 M^4-10048 l^2 M^6+1280 M^8\right)}\right)-1}},
\end{split}
\fe
{where we have considered $\Theta$ and $l$ small.}

\begin{figure}
    \centering
    \includegraphics[scale=0.7]{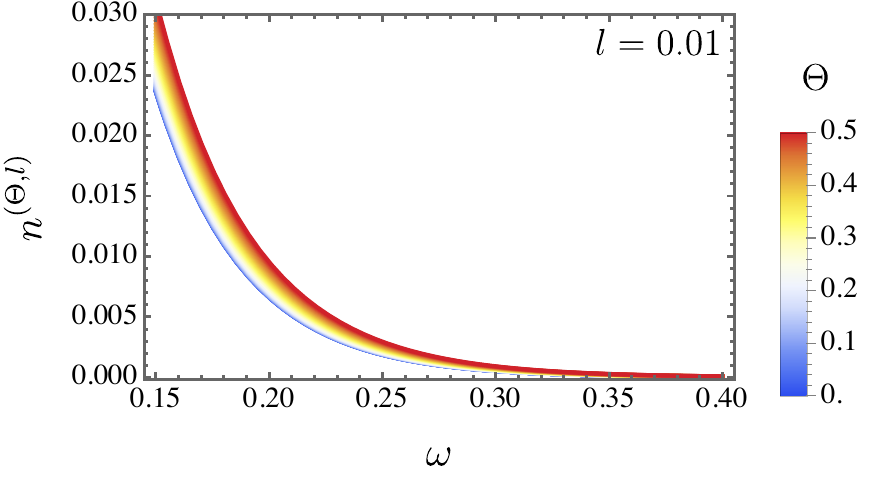}
    \caption{The particle creation density $n$ is presented as a function of the frequency $\omega$ for a fixed value of $l = 0.01$ and different values of $\Theta$.}
    \label{bosonparticles}
\end{figure}

Fig. \ref{bosonparticles} illustrates the particle creation for bosonic modes, represented by $n$. Overall, the non--commutative parameter enhances the magnitude of $n$ for a fixed value of $l = 0.01$.

\subsection{Fermionic particle modes}

In this section, it is crucial to note that the upcoming calculations disregard backreaction effects. Within the quantum tunneling framework, particles can escape a black hole by passing through the event horizon. The probability associated with this process can be derived using the methods described in Refs. \cite{angheben2005hawking,kerner2006tunnelling,kerner2008fermions} (and the references therein).


Black hole radiation, akin to black body radiation, emerges due to their intrinsic temperature. However, this emission spectrum is influenced by greybody factors, which modify the outgoing radiation. The spectrum is expected to include particles of different spins, such as fermions. Studies initiated by Kerner and Mann \cite{o69}, followed by further research \cite{o75,o72,o71,o74,o73,o70}, indicate that massless bosons and fermions radiate at the same temperature. Moreover, analyses of spin--$1$ bosons suggest that the Hawking temperature remains unchanged even when higher--order quantum corrections in $\hbar$ beyond the semiclassical regime are considered \cite{o77,o76}.

The behavior of fermions is often described through the phase of their spinor wave function, which follows the Hamilton--Jacobi equation. An alternative expression for the action, as presented in \cite{o83,o84,vanzo2011tunnelling}, is given by 
\ie
\mathcal{S}^{(\Psi)} = S^{(0)} + \psi^{(\uparrow \downarrow)},
\fe
where $S_0$ corresponds to the classical action for scalar particles and $\psi^{(\uparrow \downarrow)}$ accounts for the spin corrections. The additional spin--dependent terms account for the interaction between the particle’s spin and the background spin connection, ensuring no singularities arise at the event horizon. Since these corrections primarily influence spin precession and remain small, they are not considered in this study. Furthermore, the spin of the emitted particles has an insignificant effect on the black hole’s angular momentum, especially in the case of non--rotating black holes with masses far exceeding the Planck scale \cite{vanzo2011tunnelling}. On average, particles with opposite spins are emitted in a balanced manner, maintaining the black hole’s total angular momentum effectively unchanged.

This analysis focuses on the tunneling of fermionic particles as they cross the event horizon of a particular black hole solution. Alternative methods employing generalized Painlevé--Gullstrand or Kruskal--Szekeres coordinates are discussed in the foundational work \cite{o69}. The investigation begins with a general spacetime metric, as expressed in Eq. (\ref{eqwerer}). In curved spacetime, fermionic dynamics are dictated by the Dirac equation, which is formulated as:
\ie
\Big(\Tilde{\gamma}^\mu \nabla_\mu + m \Big) \Psi(x) = 0
\fe
with
\ie
\nabla_\mu = \partial_\mu + \frac{\mathbbm{i}}{2} {\Gamma^\alpha_{\;\mu}}^{\;\beta} \,\Tilde{\Sigma}_{\alpha\beta}\fe
and 
\ie
\Tilde{\Sigma}_{\alpha\beta} = \frac{\mathbbm{i}}{4} [\Tilde{\gamma}_\alpha,  \Tilde{\gamma}_\beta].
\fe
It is important to note that the coordinates are denoted as $x \equiv t, r, \theta, \varphi$. The matrices $\Tilde{\gamma}^\mu$ accounts for the fundamental properties of the Clifford algebra, which are expressed through the relation:
\ie
\{\Tilde{\gamma}_\alpha,\Tilde{\gamma}_\beta\} = 2 g_{\alpha\beta} \mathbbm{1},
\fe
with $\mathbbm{1}$ denotes the $4 \times 4$ identity matrix. Within this formulation, the $\Tilde{\gamma}$ matrices are specified as given below: 
\begin{eqnarray*}
 \Tilde{\gamma}^{t} &=&\frac{\mathbbm{i}}{\sqrt{{|g_{tt}^{(\Theta,l)}(r)|}}}\left( \begin{array}{cc}
\bf{1}& \bf{ 0} \\ 
\bf{ 0} & -\bf{ 1}%
\end{array}%
\right), \;\;
\Tilde{\gamma}^{r} =\sqrt{\frac{1}{g_{rr}^{(\Theta,l)}{(r)}}}\left( 
\begin{array}{cc}
\bf{0} &  \Tilde{\sigma}_{3} \\ 
 \Tilde{\sigma}_{3} & \bf{0}%
\end{array}%
\right), \\
\Tilde{\gamma}^{\theta } &=&\frac{1}{\sqrt{{g_{\theta\theta}^{(\Theta,l)}(r,\theta)}}}\left( 
\begin{array}{cc}
\bf{0} &  \Tilde{\sigma}_{1} \\ 
 \Tilde{\sigma}_{1} & \bf{0}%
\end{array}%
\right), \;\;
\Tilde{\gamma}^{\varphi } =\frac{1}{{g_{\varphi\varphi}^{(\Theta,l)}(r,\theta)} }\left( 
\begin{array}{cc}
\bf{0} &  \Tilde{\sigma}_{2} \\ 
 \Tilde{\sigma}_{2} & \bf{0}%
\end{array}%
\right).
\end{eqnarray*}%
In this context, $\Tilde{\sigma}$ corresponds to the Pauli matrices, which obey the conventional commutation relation:  $\Tilde{\sigma}_i \Tilde{\sigma}_j = \mathbf{1} \delta_{ij} + \mathbbm{i} \varepsilon_{ijk} \Tilde{\sigma}_k, \quad \text{where} \quad i,j,k =1,2,3$. Additionally, the matrix associated with $\Tilde{\gamma}^5$ can be equivalent to 
\begin{equation*}
\Tilde{\gamma}^{5} = \mathbbm{i} \Tilde{\gamma}^{t}\Tilde{\gamma}^{r}\Tilde{\gamma}^{\theta }\Tilde{\gamma}^{\varphi } = \mathbbm{i}\sqrt{\frac{1}{{|g_{tt}^{(\Theta,l)}(r)| \, g_{rr}^{(\Theta,l)}(r) \, g_{\theta\theta}^{(\Theta,l)}(r,\theta) \, g_{\varphi\varphi}^{(\Theta,l)}(r,\theta) }}}\left( 
\begin{array}{cc}
\bf{ 0} & - \bf{ 1} \\ 
\bf{ 1} & \bf{ 0}%
\end{array}%
\right)\:.
\end{equation*}

In order to describe a Dirac field with its spin aligned upward along the positive $r$--axis, the adopted ansatz is given by \cite{vagnozzi2022horizon}:
\begin{equation}
\Psi^{(+)}(x) = \left( \begin{array}{c}
\Tilde{H}(x) \\ 
0 \\ 
\Tilde{Y}(x) \\ 
0
\end{array}
\right) \exp \left[ \mathbbm{i} \, \psi^{(+)}(x)\right]\;.
\label{spinupbh} 
\end{equation}

This study focuses on the spin--up $(+)$ configuration, while the spin--down $(-)$ case, oriented along the negative $r$--axis, follows a similar treatment. Substituting the ansatz (\ref{spinupbh}) into the Dirac equation leads to  {
\begin{align}
&\frac{\mathbbm{i}}{\sqrt{|g_{tt}^{(\Theta,l)}(r)|}}\left(\partial_{t}\Tilde{H}(x) + \frac{\mathbbm{i}}{\hbar}\Tilde{H}(x)\,\partial_{t}\psi^{(+)}\right)
+\sqrt{\frac{1}{g_{rr}^{(\Theta,l)}(r)}}\left(\partial_{r}\Tilde{Y}(x) + \frac{\mathbbm{i}}{\hbar}\Tilde{Y}(x)\,\partial_{r}\psi^{(+)}\right)
+\frac{m}{\hbar}\Tilde{H}(x) = 0, \label{eq01} \\[10pt]
&\frac{\mathbbm{i}}{\sqrt{g_{\theta\theta}^{(\Theta,l)}(r,\theta)}}\left(\partial_{\theta}\Tilde{Y}(x) + \frac{\mathbbm{i}}{\hbar}\Tilde{Y}(x)\,\partial_{\theta}\psi^{(+)}\right)
-\frac{1}{\sqrt{g_{\varphi\varphi}^{(\Theta,l)}(r,\theta)}}\left(\partial_{\phi}\Tilde{Y}(x) + \frac{\mathbbm{i}}{\hbar}\Tilde{Y}(x)\,\partial_{\phi}\psi^{(+)}\right) = 0, \label{eq02} \\[10pt]
&-\frac{\mathbbm{i}}{\sqrt{|g_{tt}^{(\Theta,l)}(r)|}}\left(\partial_{t}\Tilde{Y}(x) + \frac{\mathbbm{i}}{\hbar}\Tilde{Y}(x)\,\partial_{t}\psi^{(+)}\right)
+\sqrt{\frac{1}{g_{rr}^{(\Theta,l)}(r)}}\left(\partial_{r}\Tilde{H}(x) + \frac{\mathbbm{i}}{\hbar}\Tilde{H}(x)\,\partial_{r}\psi^{(+)}\right)
+\frac{m}{\hbar}\Tilde{Y}(x) = 0, \label{eq03} \\[10pt]
&\frac{\mathbbm{i}}{\sqrt{g_{\theta\theta}^{(\Theta,l)}(r,\theta)}}\left(\partial_{\theta}\Tilde{H}(x) + \frac{\mathbbm{i}}{\hbar}\Tilde{H}(x)\,\partial_{\theta}\psi^{(+)}\right)
-\frac{1}{\sqrt{g_{\varphi\varphi}^{(\Theta,l)}(r,\theta)}}\left(\partial_{\phi}\Tilde{H}(x) + \frac{\mathbbm{i}}{\hbar}\Tilde{H}(x)\,\partial_{\phi}\psi^{(+)}\right) = 0. \label{eq04}
\end{align}
Following Vanzo et al. \cite{vanzo2011tunnelling}, we shall retain only the leading--order terms in $\hbar$
\begin{align}
&-\frac{\mathbbm{i}}{\sqrt{|g_{tt}^{(\Theta,l)}(r)|}}\left( \Tilde{H}(x)\,\partial_{t}\psi^{(+)}\right)
-\sqrt{\frac{1}{g_{rr}^{(\Theta,l)}(r)}}\left( \Tilde{Y}(x)\,\partial_{r}\psi^{(+)}\right)
+m \mathbbm{i} \Tilde{H}(x) = 0, \label{eq1} \\[10pt]
& - \frac{1}{\sqrt{g_{\theta\theta}^{(\Theta,l)}(r,\theta)}}\left(\Tilde{Y}(x)\,\partial_{\theta}\psi^{(+)}\right)
-\frac{1}{\sqrt{g_{\varphi\varphi}^{(\Theta,l)}(r,\theta)}}\left( \mathbbm{i}\Tilde{Y}(x)\,\partial_{\phi}\psi^{(+)}\right) = 0, \label{eq2} \\[10pt]
&    \frac{\mathbbm{i}}{\sqrt{|g_{tt}^{(\Theta,l)}(r)|}}\left(\Tilde{Y}(x)\,\partial_{t}\psi^{(+)}\right)
-\sqrt{\frac{1}{g_{rr}^{(\Theta,l)}(r)}}\left(\Tilde{H}(x)\,\partial_{r}\psi^{(+)}\right)
+ m \mathbbm{i} \Tilde{Y}(x) = 0, \label{eq3} \\[10pt]
& - \frac{1}{\sqrt{g_{\theta\theta}^{(\Theta,l)}(r,\theta)}}\left( \Tilde{H}(x)\,\partial_{\theta}\psi^{(+)}\right)
-\frac{\mathbbm{i}}{\sqrt{g_{\varphi\varphi}^{(\Theta,l)}(r,\theta)}}\left( \Tilde{H}(x)\,\partial_{\phi}\psi^{(+)}\right) = 0. \label{eq4}
\end{align}

}

and considering the action expressed as
\ie
\psi^{(+)}= - \omega\, t + \Tilde{\chi}(r) + L(\theta ,\varphi ) 
\fe
which leads to
\cite{vanzo2011tunnelling} 
{
\begin{align}
&+\frac{\mathbbm{i}\, \omega \Tilde{H}(x)}{\sqrt{|g_{tt}^{(\Theta,l)}(r)|}} 
-\sqrt{\frac{1}{g_{rr}^{(\Theta,l)}(r)}} \Tilde{Y}(x)\,\Tilde{\chi}^{\prime}(r)
+m \mathbbm{i} \Tilde{H}(x) = 0, \label{eq11} \\[10pt]
& - \Tilde{Y}(x) \left(  \frac{\partial_{\theta}L(\theta,\varphi)}{\sqrt{g_{\theta\theta}^{(\Theta,l)}(r,\theta)}}
+\frac{\mathbbm{i}\,\partial_{\phi} L(\theta,\varphi)}{\sqrt{g_{\varphi\varphi}^{(\Theta,l)}(r,\theta)}} \right) = 0, \label{eq21} \\[10pt]
&    -\frac{\mathbbm{i} \, \omega \Tilde{Y}(x)}{\sqrt{|g_{tt}^{(\Theta,l)}(r)|}}
-\sqrt{\frac{1}{g_{rr}^{(\Theta,l)}(r)}}\Tilde{H}(x)\Tilde{\chi}^{\prime}(r)
+ m \mathbbm{i} \Tilde{Y}(x) = 0, \label{eq31} \\[10pt]
& - \Tilde{H}(x) \left(  \frac{\partial_{\theta}L(\theta,\varphi)}{\sqrt{g_{\theta\theta}^{(\Theta,l)}(r,\theta)}}
+\frac{\mathbbm{i}\,\partial_{\phi} L(\theta,\varphi)}{\sqrt{g_{\varphi\varphi}^{(\Theta,l)}(r,\theta)}} \right) = 0. \label{eq41}
\end{align}

}

 The specific forms of $\Tilde{H}(x)$ and $\Tilde{Y}(x)$ do not affect the conclusion that Eqs. (\ref{eq21}) and (\ref{eq41}) impose the constraint {
$$\frac{\partial_{\theta}L(\theta,\varphi)}{\sqrt{g_{\theta\theta}^{(\Theta,l)}(r,\theta)}}
+\frac{\mathbbm{i}\,\partial_{\phi} L(\theta,\varphi)}{\sqrt{g_{\varphi\varphi}^{(\Theta,l)}(r,\theta)}} = 0,$$}
implying that $L(\theta, \varphi)$ must be a complex function. This condition holds for both outgoing and ingoing scenarios. Consequently, when computing the ratio of outgoing to ingoing probabilities, the terms involving $L(\theta, \varphi)$ cancel out, allowing it to be omitted from further analysis, as pointed out in Ref. \cite{vanzo2011tunnelling}. In the case of massless particles ($m=0$), Eqs. (\ref{eq11}) and (\ref{eq31}) lead to two independent solutions:
\ie
\Tilde{H}(x) = - \mathbbm{i} \Tilde{Y}, \qquad \Tilde{\chi}^{\prime }(r) = \Tilde{\chi}_{\text{out}}' {(r)} = \frac{\omega}{\sqrt{\frac{|g_{tt}^{(\Theta,l)}(r)|}{g_{rr}^{(\Theta,l)}(r)}}},
\fe
\ie
\Tilde{H}(x) = \mathbbm{i} \Tilde{Y}(x), \qquad \Tilde{\chi}^{\prime }(r) = \Tilde{\chi}_{\text{in}}' {(r)} = - \frac{\omega}{\sqrt{\frac{|g_{tt}^{(\Theta,l)}(r)|}{g_{rr}^{(\Theta,l)}(r)}}}.
\fe
In this context, $\Tilde{\chi}_{\text{out}} {(r)}$ and $\Tilde{\chi}_{\text{in}} {(r)}$ represent the solutions associated with outgoing and incoming particles, respectively \cite{vanzo2011tunnelling}. Consequently, the overall tunneling probability takes the form $\Tilde{\Gamma}_{\psi} \sim e^{-2 \, \text{Im} \, (\Tilde{\chi}_{\text{out}}{(r)} - \Tilde{\chi}_{\text{in}}{(r)})}$. Therefore,
\ie
 \Tilde{\chi}_{ \text{out}}(r)= -  \Tilde{\chi}_{ \text{in}} (r) = \int \mathrm{d} r \,\frac{\omega}{\sqrt{\frac{|g_{tt}^{(\Theta,l)}(r)|}{g_{rr}^{(\Theta,l)}(r)}}}\:.
\fe

{Furthermore, one additional comment must be made: note that, since both $\Theta$ and $l$ are extremely small—as will be confirmed in the bounds section based on solar system tests—the metric components near $r = r_h$ exhibit an approximately linear behavior. This allows us to consider the presence of a simple pole to carry out our calculations. In this sense, we may write} {
\ie
|g_{tt}^{(\Theta,l)}(r)| \frac{1}{g_{rr}^{(\Theta,l)}(r)} \approx \, |g_{tt}^{(\Theta,l)\prime}(r)| \frac{1}{g_{rr}^{(\Theta,l)\prime}(r)}(r - r_{h})^{2} + ... \, .
\fe
}

By employing Feynman’s approach, the following expression is obtained:
\ie
2\mbox{Im}\;\left[  \Tilde{\chi}_{ \text{out}}{(r)} -  \Tilde{\chi}_{ \text{in}} {(r)}\right] =\mbox{Im}\int \mathrm{d} r \,\frac{4\omega}{\sqrt{\frac{|g_{tt}^{(\Theta,l)}(r)|}{g_{rr}^{(\Theta,l)}(r)}}}=\frac{2\pi\omega}{\kappa},
\fe
{where $\kappa$ is
\ie
\kappa \approx \, \, \frac{M}{r_{h}^2}-\frac{8 l^2 M^2}{r_{h}^5} + \left[ \frac{M (28 M-9 r_{h})}{4 r_{h}^5}-\frac{l^2 M^2 (511 M-180 r_{h})}{r_{h}^8} \right]\Theta^{2}.
\fe

}

In this setting, the particle density $n_{\psi}$ corresponding to the given black hole solution follows the relation $\Tilde{\Gamma}_{\psi} \sim e^{-\frac{2 \pi \omega}{\kappa}}$ {and replacing the expression for $r_{h}$ derived from Eq. (\ref{eventhorizonhay}), we obtain
  }
\ie
\begin{split}
n_{\psi} & = {\frac{\Tilde{\Gamma}_{\psi}}{1 + \Tilde{\Gamma}_{\psi}}} \\
& { = \, \, \frac{1}{\exp \left(\frac{\pi  \omega  \left(l^2-4 M^2\right)^8}{2 M^3 \left(l^2-4 M^2\right)^3 \left(l^6-12 l^4 M^2+112 l^2 M^4-64 M^6\right)+2 \Theta ^2 M^5 \left(-9 l^8+88 l^6 M^2-5952 l^4 M^4-10048 l^2 M^6+1280 M^8\right)}\right)+1}}.
\end{split}
\fe

\begin{figure}
    \centering
    \includegraphics[scale=0.7]{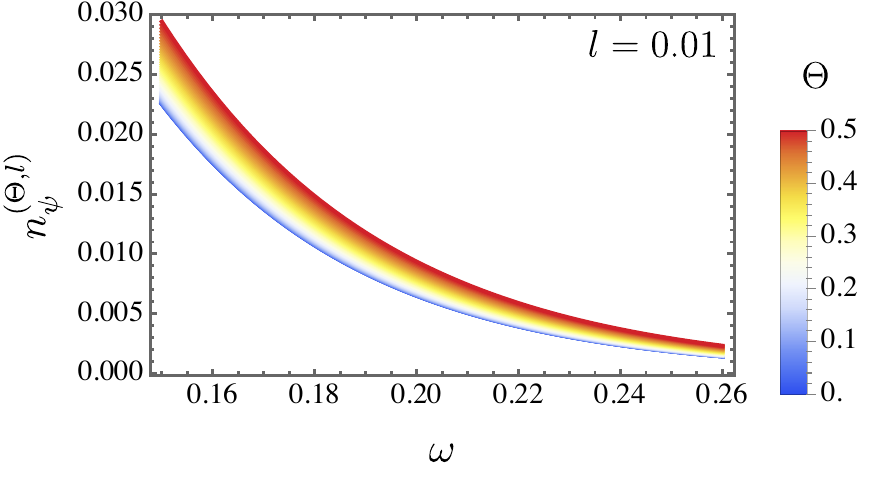}
    \caption{The particle creation density $n_{\psi}^{{(\Theta,l)}}$ is presented as a function of the frequency $\omega$ for different values of the non--commutative parameter $\Theta$, with a fixed value of $l = 0.01$.}
    \label{fermionicparticles}
\end{figure}

\begin{figure}
    \centering
    \includegraphics[scale=0.75]{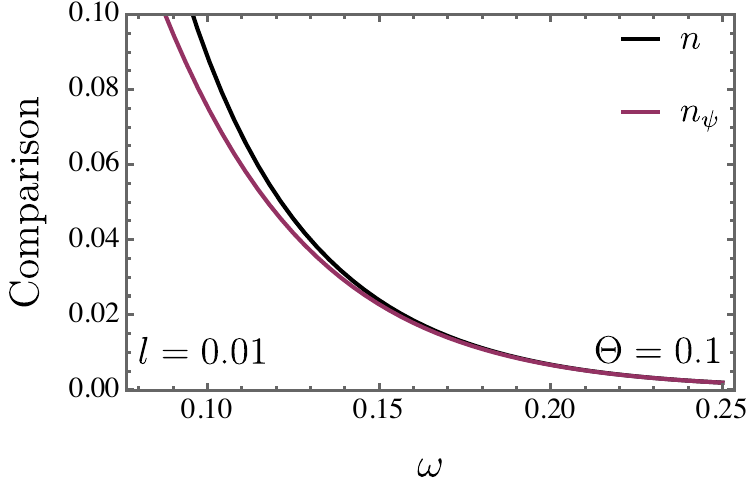}
    \caption{The {comparison of} particle creation {densities} is depicted as a function of the frequency $\omega$ for {$\Theta = 0.1$ and } $l = 0.01$ fixed.}
    \label{comparisonparticles}
\end{figure}

Fig. \ref{fermionicparticles} illustrates the particle creation density for fermions, $n_{\psi}$. Similar to the bosonic scenario, the non--commutative parameter $\Theta$ contributes to an increase in the magnitude of the particle creation density.
Additionally, Fig. \ref{comparisonparticles} presents a comparison of the particle creation densities for bosons and fermions. Overall, the bosonic case exhibits a greater magnitude at lower frequencies compared to the fermionic case.

\section{The scalar perturbation}
 
In this section, we examine the evolution of a massless scalar field. To achieve this, first, we explore the non--commutative modifications to the Hayward black hole spacetime by incorporating correction terms arising from the non--commutative geometry framework expressed in Eq. (\ref{ggtt}--\ref{gphi}).  The modified metric accounting for these corrections can be expressed as \( g_{ij}^{NC} = g_{ij} + \Theta^2 \mathcal{G}_{ij}^{NC} \), where \( g_{ij} \) represents the original Hayward black hole metric, while \( \mathcal{G}_{ij}^{NC} \) contains the coefficients associated with the non--commutative corrections, as discussed in Refs. \cite{chaichian2008corrections, chamseddine2001deforming, zet2003desitter}.  

We follow the method introduced in Ref. \cite{chen2022eikonal}. In this intriguing approach, a deformed metric remains both stationary and axisymmetric in the modified approach. This deformation is governed by a small dimensionless parameter \( \epsilon \), leading to an alternative description of the black hole geometry. Applying this approach, the initial metric represented in Eq. (\ref{ggtt}--\ref{gphi}) are reformed in a $\cos$ series as
	
	\begin{align}\label{g00}
		{g_{tt}^{(\Theta,l)}} &= - f(r)(1 + \epsilon{A_j}{{\mathop{\rm cos}\nolimits} ^j}\theta ),\\
		g_{rr}^{(\Theta,l)} &= { f(r)}^{-1}(1 + \epsilon{B_j}{{\mathop{\rm cos}\nolimits} ^j}\theta ),\\
		g_{\theta\theta}^{(\Theta,l)}& = {r^2}(1 + \epsilon{C_j}{{\mathop{\rm cos}\nolimits} ^j}\theta ),\\
		g_{\phi\phi}^{(\Theta,l)}&= {r^2}{{\mathop{\rm sin}\nolimits} ^2}\theta (1 + \epsilon{D_j}{{\mathop{\rm cos}\nolimits} ^j}\theta ),\\
		g_{tr}^{(\Theta,l)}& = \epsilon{a_j}(r){{\mathop{\rm cos}\nolimits} ^j}\theta ,\quad
		g_{r\theta}^{(\Theta,l)} = \epsilon{c_j}(r){{\mathop{\rm cos}\nolimits} ^j}\theta ,\quad\
		g_{\theta \phi}^{(\Theta,l)} = \epsilon{e_j}(r){{\mathop{\rm cos}\nolimits} ^j}\theta,\\
		g_{t\theta}^{(\Theta,l)}& = \epsilon{b_j}(r){{\mathop{\rm cos}\nolimits} ^j}\theta ,\quad
		g_{r\phi}^{(\Theta,l)} = \epsilon{d_j}(r){{\mathop{\rm cos}\nolimits} ^j}\theta. \\ \label{g13}
	\end{align}
    where $f(r)= \left(1 - \frac{2M r^2}{r^3 + 2M l^2}\right)$, is the lapse function of the Hayward black hole and the small deformed parameter is equivalent to the NC parameter $(\epsilon=\Theta^2)$. Consequently, by following Ref. \cite{chen2022eikonal,zhao2023quasinormal}, the deformed coefficients of the metric are derived as 
\begin{align}\label{A}
		{A_0} =&{-\frac{M}{2 \left(2 l^2 M+r^3\right)^4 \left(2 l^2 M+r^2 (r-2 M)\right)}\Big( 32 l^8 M^4-16 l^6 M^3 r^2 (8 M+35 r)}\\ \nonumber
        &{+48 l^4 M^2 r^5 (17 M+r)+2 l^2 M r^8 (80 r-201 M)+r^{11} (11 M-4 r)\Big)},
		\\
        {B_0} =&{\frac{M \left(-16 l^6 M^3+48 l^4 M^2 r^3+24 l^2 M r^5 (r-2 M)+r^8 (3 M-2 r)\right)}{2 \left(2 l^2 M+r^3\right)^3 \left(2 l^2 M+r^2 (r-2 M)\right)}},\\ 
		{C_0} =&{ \frac{1}{16 r^2 \left(2 l^2 M+r^3\right)^3 \left(2 l^2 M+r^2 (r-2 M)\right)}\Big(16 l^8 M^4+32 l^6 M^3 r^2 (r-23 M)}\\ \nonumber
        &{+8 l^4 M^2 r^4 \left(98 M^2+108 M r+3 r^2\right)+8 l^2 M r^7 \left(-160 M^2+69 M r+r^2\right)}\\ \nonumber
        &{+r^{10} \left(64 M^2-32 M r+r^2\right)\Big)},\\
		{D_0} = &{\frac{1}{4 r^2 \left(2 l^2 M+r^3\right)^3 \left(2 l^2 M+r^2 (r-2 M)\right)}\Big(16 l^8 M^4+16 l^6 M^3 r^2 (2 r-5 M)}\\ \nonumber
        &{+8 l^4 M^2 r^4 \left(16 M^2+6 M r+3 r^2\right)+4 l^2 M r^7 \left(-28 M^2+9 M r+2 r^2\right)}\\ \nonumber
        &{+r^{10} \left(2 M^2-4 M r+r^2\right)\Big)},\\
		&{A_j}={B_j}={C_j}=0 \quad \text{and} \quad
		{D_j}= {\frac{5({1 + {{\left( { - 1} \right)}^j}})}{{16{r^2}}}} \quad \text{for}\quad j>0,\\ \label{abcd}
		&{a_j}(r)= {b_j}(r) = {c_j}(r) = {d_j}(r) = 0.
\end{align}

Now, we start by expressing the Klein--Gordon equation in curved spacetime as follow:
\begin{equation}
    \frac{1}{\sqrt{-g}} \partial_\mu \left( \sqrt{-g} g^{\mu \nu} \partial_\nu \psi \right) = 0.
\end{equation}
Next, we decompose the scalar field \( \psi \) in terms of the two Killing vectors \( \partial_t \) and \( \partial_\phi \) that generate the symmetries of the spacetime. The wave function can be written as a sum over angular and radial modes
\begin{equation}
    \psi = \int_{-\infty}^{\infty} \mathrm{d} \omega \sum_{m = -\infty}^{\infty} e^{im\varphi} D_{m, \omega}^2 \psi_{m,\omega}(r,\theta) e^{-i\omega t},
\end{equation}
where \( D_{m,\omega}^2 \psi_{m,\omega}(r, \theta) = 0 \) and \( m \) and \( \omega \) are the azimuthal number and mode frequency, respectively. To account for non--commutative corrections up to first order in \( \Theta^2 \), we expand the operator \( D_{m,\omega}^2 \psi_{m,\omega} \) as follows \cite{chen2022eikonal, zhao2023quasinormal}
\begin{equation}
    D_{m,\omega}^2 = D_{(0)m,\omega}^2 + \Theta^2 D_{(1)m,\omega}^2.
\end{equation}

Substituting the metric coefficients from Eqs. (\eqref{g00} -- \eqref{g13}) into the Klein--Gordon equation, we obtain the following expressions for the operators
\begin{align}
    D_{(0)m,\omega}^2 &= - \left( \omega^2 - \frac{m^2 f(r)}{r^2 \sin^2 \theta} \right) - \frac{f(r)}{r^2} \partial_r \left( r^2 f(r) \partial_r \right) - \cos \theta \, \partial_r \left( r^2 f(r) \partial_r \right) \\
    &- \frac{f(r)}{r^2 \sin^2 \theta} \partial_\theta \left( \sin \theta \, \partial_\theta \right), \\
    D_{(1)m,\omega}^2 &= \frac{m^2 f(r)}{r^2 \sin^2 \theta} (A_j - D_j) \cos \theta - \frac{f(r)}{r^2} (A_j - B_j) \cos \theta \, \partial_r \left( r^2 f(r) \partial_r \right) \\
    &- \frac{f(r)^2}{r^2} (A'_j - B'_j + C'_j + D'_j) \cos \theta \, \partial_r - \frac{f(r)}{r^2} (A_j - C_j) \cos \theta \left( \cot \theta \, \partial_\theta + \partial_\theta^2 \right) \\
    &- \frac{f(r)}{2r^2} (A_j + B_j - C_j + D_j) \partial_\theta \cos \theta \, \partial_\theta - \frac{2i\omega f(r)}{r} a_j \cos \theta (r \partial_r + 1).
\end{align}

We also introduce the tortoise coordinate \( r^* \), which is defined by
\begin{equation}
    \frac{\mathrm{d}r^*}{\mathrm{d}r} = \frac{1}{f(r) \left( 1 + \frac{1}{2} \Theta^2 b_{lm}^j (A_j - B_j) \right)}.
\end{equation}
Expanding \( \psi_{m,\omega} \) in terms of Legendre functions \( P_{lm}(\cos\theta) \) and radial wave functions \( R_{l,m}(r) \), we write the field as
\begin{equation}
    \psi_{m,\omega} = \sum_{l' = |m|}^\infty P_{l'}^m (\cos\theta) R_{l',m}(r),
\end{equation}
and the radial wave function satisfies the Schr\"{o}dinger--like equation

\begin{equation}
    \partial_{r^*}^2 \Psi_{lm} + \omega^2 \Psi_{lm} = V_{\rm{eff}}(r) \Psi_{lm},
\end{equation}
and the effective potential \( V_{\rm{eff}} \) is then given by
\begin{equation}
    V_{\rm{eff}} = V_{\rm{H}} + \Theta^2 V_{\rm{NC}},
\end{equation}
where \( V_{\rm{H}} \) is the effective potential for the original black hole in the commutative scenario
\begin{equation}
    V_{\rm{H}} = f(r) \left( \frac{\ell (\ell + 1)}{r^2} + \frac{1}{r}\frac{\mathrm{d}f(r)}{\mathrm{d}r}  \right),
\end{equation}
and \( V_{\rm{NC}} \) represents the non--commutative correction to the potential which can be obtained after performing the necessary algebraic manipulations

\ie
    \begin{split}
		& V_{\rm {NC}}= \frac{f(r)}{r}\frac{{\mathrm{d}f(r)}}{{\mathrm{d}r}}b_{lm}^0\left( {{A_0} - {B_0}} \right) +  \left[\frac{f(r)}{{{r^2}}}(a_{lm}^0\left( {{A_0} - {D_0}} \right) - c_{lm}^0\left( {{A_0} - {C_0}}\right) \right. \\ 
		& \left. -\frac{{d_{lm}^0}}{2}({A_0} + {B_0} - {C_0} + {D_0})
		+ \frac{1}{{4{r^2}}}\frac{\mathrm{d}}{{\mathrm{d}{r^*}}}\left(b_{lm}^0{r^2}\frac{\mathrm{d}}{{\mathrm{d}{r^*}}}({A_0} - {B_0} + {C_0} + {D_0})\right) - \frac{{b_{lm}^0}}{4}\frac{{{\mathrm{d}^2}}}{{\mathrm{d}{r^*}^2}}({A_0} - {B_0})) \right]\\ 
		& - \frac{f(r)}{{{r^2}}}\sum\limits_{j = 1}^\infty  {\left(a_{lm}^j + \frac{1}{2}d_{lm}^j \right){D_j}}  + \sum\limits_{j = 1}^\infty  {\frac{1}{{4{r^2}}}\frac{\mathrm{d}}{{\mathrm{d}{r^*}}}\left(b_{lm}^j{r^2}\frac{\mathrm{d}}{{\mathrm{d}{r^*}}}\right){D_j}}. \nonumber
	\end{split}
    \fe

\begin{figure}[ht]
		\centering
		\includegraphics[width=80mm]{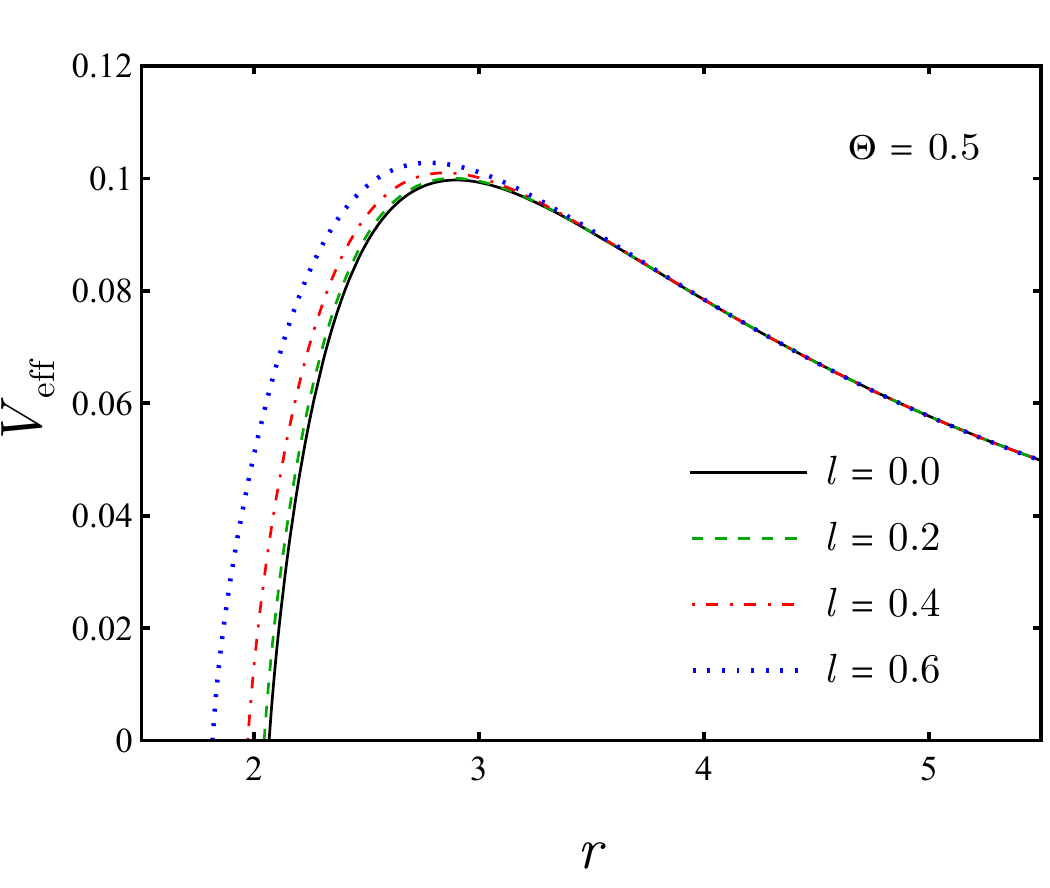} 
		\hfill
	   \caption{{Effective potential for scalar field with $M = 1$, $\ell = 1  (m=\pm 1) $ for both Schwarzschild case $(l = 0)$ and Hayward spacetime with $l = 0.2,0.4$ and $0.6$ in non--commutative framework with parameter $\Theta = 0.5$ }}
		\label{fig:VeffSH}
	\end{figure}

\begin{figure}[ht]
		\centering
		\includegraphics[width=54mm]{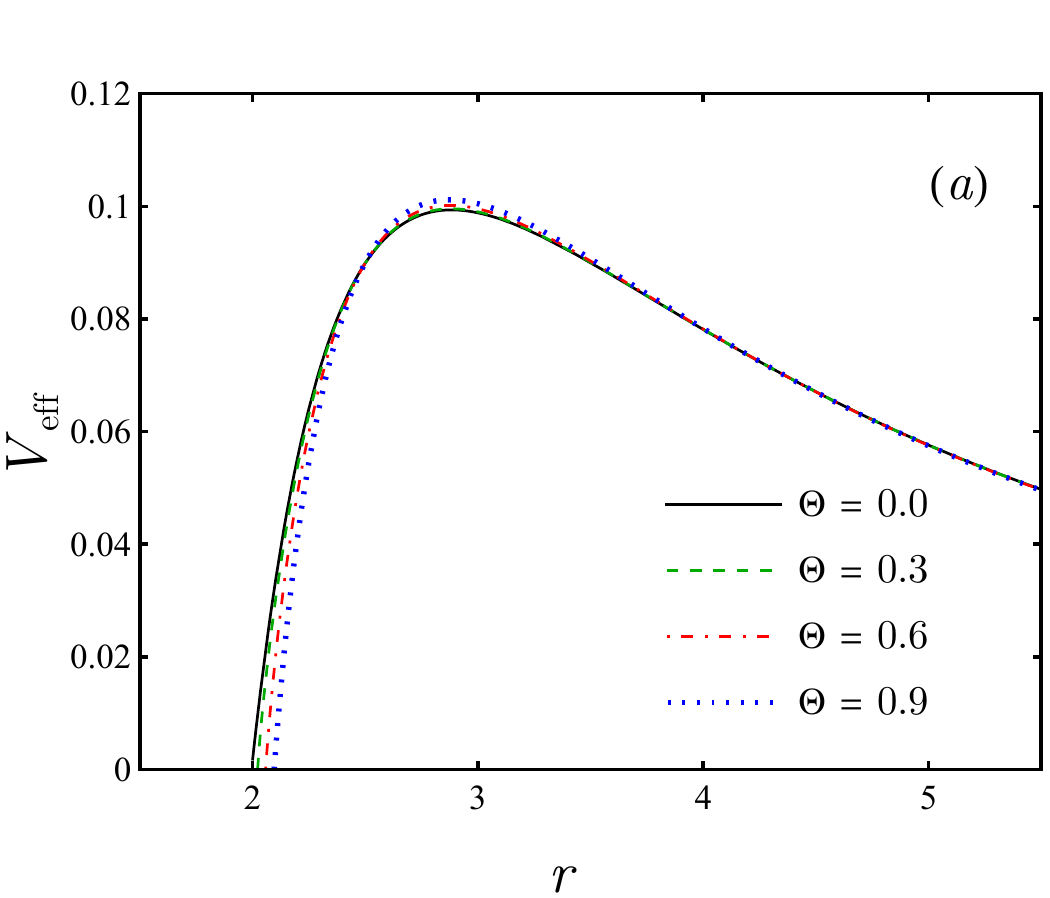} 
		\hfill
		\includegraphics[width=54mm]{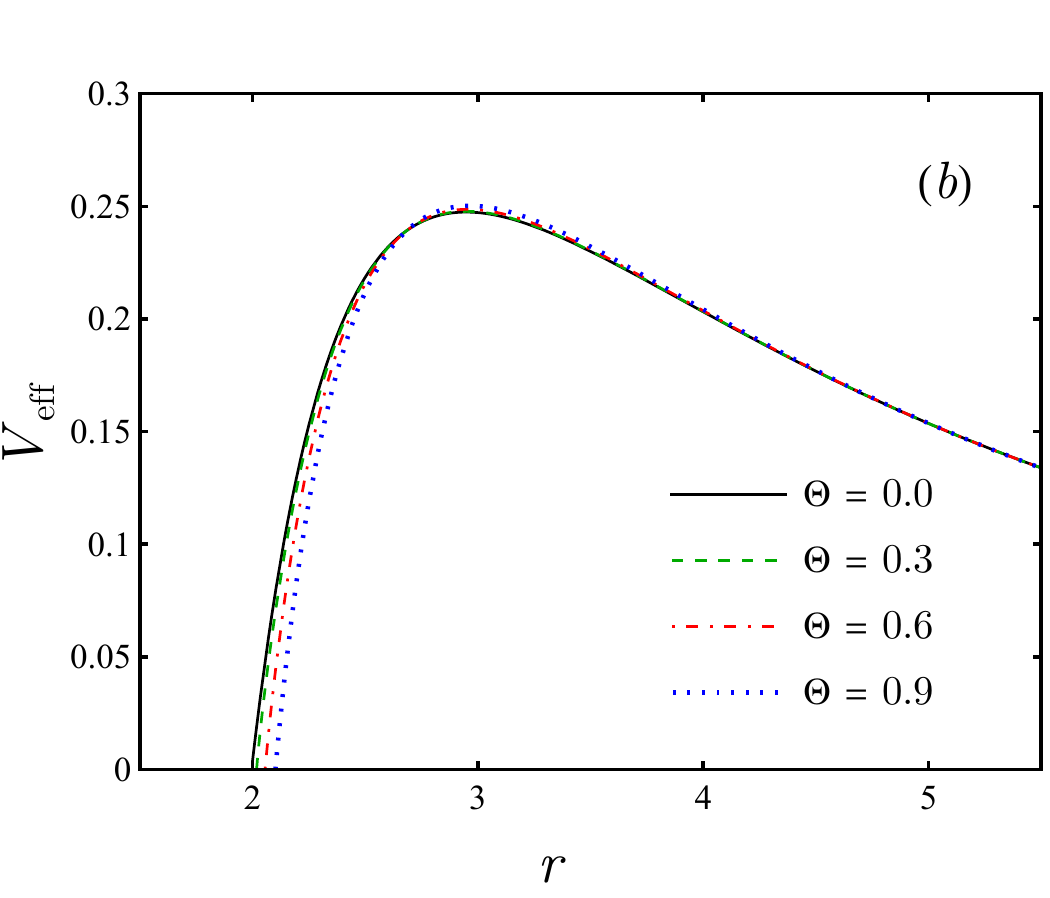}
		\hfil
		\includegraphics[width=52mm]{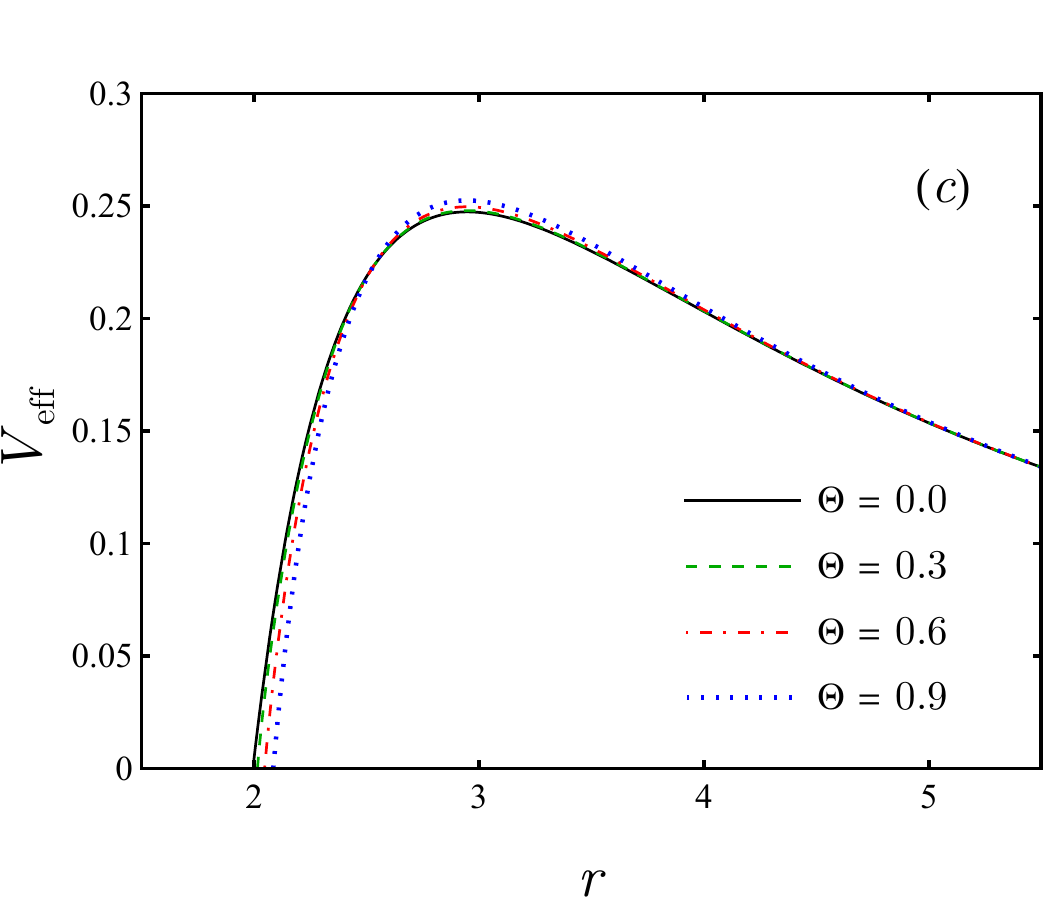}\\
		\caption{{Effective potential for scalar field with $M = 1$, $l = 0.1$ in left panel $(a)$ for $\ell = 1  (m=\pm 1) $, in middle panel $(b)$ for $\ell = 2$ $(m=\pm 1)$ and in the right panel $(c)$ for $\ell = 2$ $(m=\pm 2)$ concerning different values of $\Theta$.}}
		\label{fig:Veffrstar}
	\end{figure}

The coefficients \( a_{lm}^j, b_{lm}^j, c_{lm}^j, d_{lm}^j \) are calculated based on the specific values of the non--commutative parameter \( \Theta \), the multipole number $\ell (m = \pm 1)$, as detailed in Ref. \cite{zhao2023quasinormal}. Notably, the effective potential, when extended to the non--commutative formalism, depends not only on the multipole \( \ell \) but also on the azimuthal number \( m \). In addition, these coefficients are the same for $\pm$ of each azimuthal number $m$. For example, the explicit form of \( V_{\rm{eff}} \) for \( \ell = 1 ( m = \pm 1) \) is given by 
{
\begin{align}\nonumber
    &V_\text{eff}=f(r) \left(\frac{2}{r^2}+\frac{f'(r)}{r}\right)+\frac{\Theta^2}{1024 r^4 \sqrt{f(r)}}\Big[+2 r \sqrt{f(r)} \left(\frac{128 r^2 f'(r)^2}{f(r)}+192\right) f'(r)\\ \nonumber
    &+32 f(r)^{7/2}+2 f(r)^{3/2} \left(\frac{r^2 f'(r) \left(\frac{96 r^2 f'(r)^3}{f(r)^{3/2}}-\frac{32 f'(r)}{\sqrt{f(r)}}+2624 r \Xi\right)}{2 \sqrt{f(r)}}-312\right)\\ \nonumber
    &+32rf(r)^2\left[-\frac{81 f'(r)}{2 \sqrt{f(r)}}+\frac{29 r^2 f'(r)^3}{2 f(r)^{3/2}}-16 r \Xi\frac{82 r^3 f'(r)^2 \Xi}{f(r)}+\right.\\ \nonumber
    &+\left.28 r^2 \left(\frac{f^{(3)}(r)}{2 \sqrt{f(r)}}+\frac{3 f'(r)^3}{8 f(r)^{5/2}}-\frac{3 f'(r) f''(r)}{4 f(r)^{3/2}}\right)\right]+f(r)^{5/2}\Big[176-\frac{264 r^2 f'(r)^2}{f(r)}\\ \nonumber
    &+3200 r^4 \Xi^2+\frac{r f'(r) \left(9024 r^2 \Xi+6016 r^3 \left(\frac{f^{(3)}(r)}{2 \sqrt{f(r)}}+\frac{3 f'(r)^3}{8 f(r)^{5/2}}-\frac{3 f'(r) f''(r)}{4 f(r)^{3/2}}\right)\right)}{2 \sqrt{f(r)}}\Big]\\ \nonumber
    &+rf(r)^3\left(-\frac{64 f'(r)}{\sqrt{f(r)}}+32 r \Xi+1856 r^2 \left(\frac{f^{(3)}(r)}{2 \sqrt{f(r)}}+\frac{3 f'(r)^3}{8 f(r)^{5/2}}-\frac{3 f'(r) f''(r)}{4 f(r)^{3/2}}\right)\right.\\ \label{VeffL1M1}
    &\left.\left.+768 r^3 \left(\frac{f^{(4)}(r)}{2 \sqrt{f(r)}}-\frac{3 f''(r)^2}{4 f(r)^{3/2}}-\frac{15 f'(r)^4}{16 f(r)^{7/2}}-\frac{f^{(3)}(r) f'(r)}{f(r)^{3/2}}+\frac{9 f'(r)^2 f''(r)}{4 f(r)^{5/2}}\right)\right)\right)\Big]
\end{align}

where $\Xi=\frac{f''(r)}{2 \sqrt{f(r)}}-\frac{f'(r)^2}{4 f(r)^{3/2}}$ and the prime denotes the derivation with respect to the coordinate $r$. We can notice that when $\Theta=0$, the standard Hayward effective potential for the scalar field with $\ell=1$ is recovered.} 
In Fig. \ref{fig:VeffSH}, the effective potentials for Schwarzschild and Hayward black holes are shown. Visibly, the Hayward spacetime, in the non--commutative framework has a higher effective potential than the Schwarzschild case. To explore the effect of Hayward parameter value $l$ on the effective potential, we illustrate the effective potential, for specific values of mass $M$, $\ell$, and $m$ in Fig.\ref{fig:Veffrstar}, left panel displays the effective potential for varying values of $\Theta$ when $\ell = 1$ and $m=\pm 1$, while panels middle and right panels correspond to $\ell = 2(m=\pm 1)$ and $\ell = 2 (m=\pm 2)$, respectively.  

Although the influence of the NC parameter on the system is subtle on the potential, as $\Theta$ increases, the peak value of the potential also rises, suggesting that the effective potential forms a stronger barrier against the propagation of the field.

	
\section{Quasinormal modes}

Black holes respond to perturbations through damped oscillations known as quasinormal modes (QNMs), which are characterized by complex frequencies. The real component determines the oscillation rate, while the imaginary part governs the damping behavior. These modes reflect how disturbances behave near the event horizon and reveal properties of the surrounding geometry. Within the setting of non--commutative geometry, QNMs provide a way to examine how the non--commutative parameter affects scalar field (for instance) behavior and its interaction with the effective potential.

To extract the QNM frequencies, one must solve the wave equation under boundary conditions that restrict solutions to purely ingoing waves at the event horizon and outgoing waves at infinity. Given the intricate nature of the equation, exact solutions are rarely achievable. The importance of QNMs in describing the temporal and spatial features of perturbations has led to a variety of approximation methods, both analytical and numerical \cite{leaver1986solutions,ferrari1984new, blome1984quasi, cardoso2001quasinormal,heidari2023investigation}. In what follows, the WKB method is applied to assess how the non--commutative parameter alters the quasinormal spectrum.

The WKB approximation is used to determine QNMs through the following relation
\begin{equation}\label{omegawkb}
\frac{{i(\omega _n^2 - V_{0})}}{{\sqrt { - 2V''_0} }} + \sum\limits_{j = 2}^3 {{\Omega _j} = n + \frac{1}{2}},
\end{equation}
where \( V_0 \) represents the effective potential at its maximum, and \( V_0'' \) is its second derivative with respect to \( r^* \). The terms \( \Omega_j \) account for the WKB corrections \cite{konoplya2011quasinormal}.


\begin{table}[]
\caption{Quasinormal modes of Hayward black hole in non--commutative framework, calculated by WKB method, for $M = 1$. }
\label{Tab:AllQNMS}
\begin{tabular}{|cc|c|c|c|c|c|}
\hline
\multicolumn{2}{|c|}{} &
  $\Theta = 0.000$ &
  $\Theta = 0.025$ &
  $\Theta = 0.050$ &
  $\Theta = 0.075$ &
  $\Theta = 0.100$ \\ \hline
\multicolumn{1}{|c|}{\multirow{2}{*}{\begin{tabular}[c]{@{}c@{}}$l=1$,\\$m=\pm 1$\end{tabular}}} &
  $n = 0$ &
 {0.29132-0.09781$i$} & {0.29133-0.09782$i$} & {0.29134-0.09785$i$} & {0.29137-0.09788$i$} & {0.29140-0.09794$i$} \\ \cline{2-7} 
\multicolumn{1}{|c|}{} &
  $n = 1$ &
 {0.26245-0.30685$i$} & {0.26247-0.30687$i$} & {0.26251-0.30692$i$} & {0.26257-0.30701$i$} & {0.26266-0.30713$i$}  \\ \hline
\multicolumn{1}{|c|}{\multirow{3}{*}{\begin{tabular}[c]{@{}c@{}}$l=2$ \\ $m=\pm 2$\end{tabular}}} &
  $n = 0$ &
  {0.48357-0.09664$i$} &
 { 0.48358-0.09665$i$} &
  {0.48359-0.09667$i$} &
  {0.48362-0.09671$i$} &
  {0.48365-0.09676$i$} \\ \cline{2-7} 
\multicolumn{1}{|c|}{} &
  $n = 1$ &
 {0.46362-0.29531$i$} &
 {0.46363-0.29534$i$} &
 {0.46366-0.29540$i$} &
 {0.46370-0.29550$i$} &
 {0.46376-0.29565$i$} \\ \cline{2-7} 
\multicolumn{1}{|c|}{} &
  $n = 2$ &
 {0.43217-0.50257$i$} &
 {0.43219-0.50260$i$} &
 {0.43223-0.50270$i$} &
 {0.43231-0.50286$i$} &
 {0.43242-0.50309$i$} \\ \hline
\multicolumn{1}{|c|}{\multirow{3}{*}{\begin{tabular}[c]{@{}c@{}}$l=2$\\ $m=\pm1$\end{tabular}}} &
  $n = 0$ &
  {0.48357-0.09664$i$} &
  {0.48358-0.09665$i$} &
  {0.48359-0.09668$i$} &
  {0.48360-0.09672$i$} &
  {0.48363-0.09678$i$} \\ \cline{2-7} 
\multicolumn{1}{|c|}{} &
  $n = 1$ &
 {0.46362-0.29531$i$} &
 {0.46363-0.29534$i$} &
 {0.46366-0.29541$i$} &
 {0.46370-0.29553$i$} &
 {0.46376-0.29570$i$} \\ \cline{2-7} 
\multicolumn{1}{|c|}{} &
  $n = 2$ &
 {0.43217-0.50257$i$ }&
{ 0.43219-0.50261$i$ }&
 {0.43224-0.50272$i$ }&
 {0.43233-0.50290$i$} &
 {0.43245-0.50316$i$} \\ \hline
\end{tabular}
\end{table}

\begin{figure}
	\centering
	\includegraphics[width=80mm]{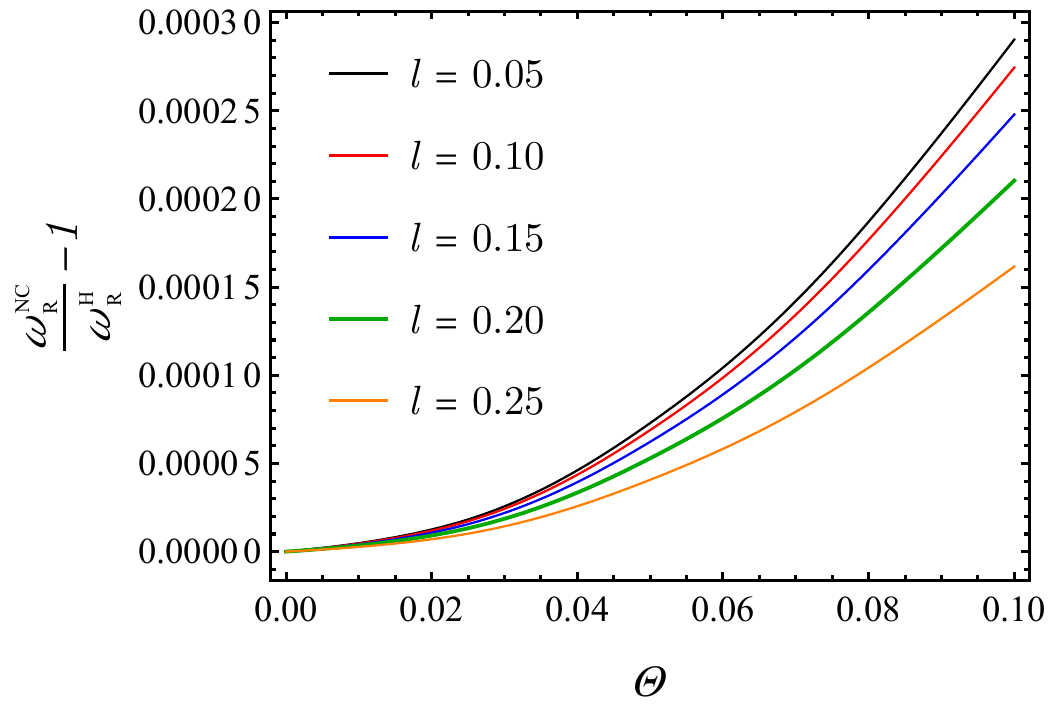}
	\hfil
    \includegraphics[width=80mm]{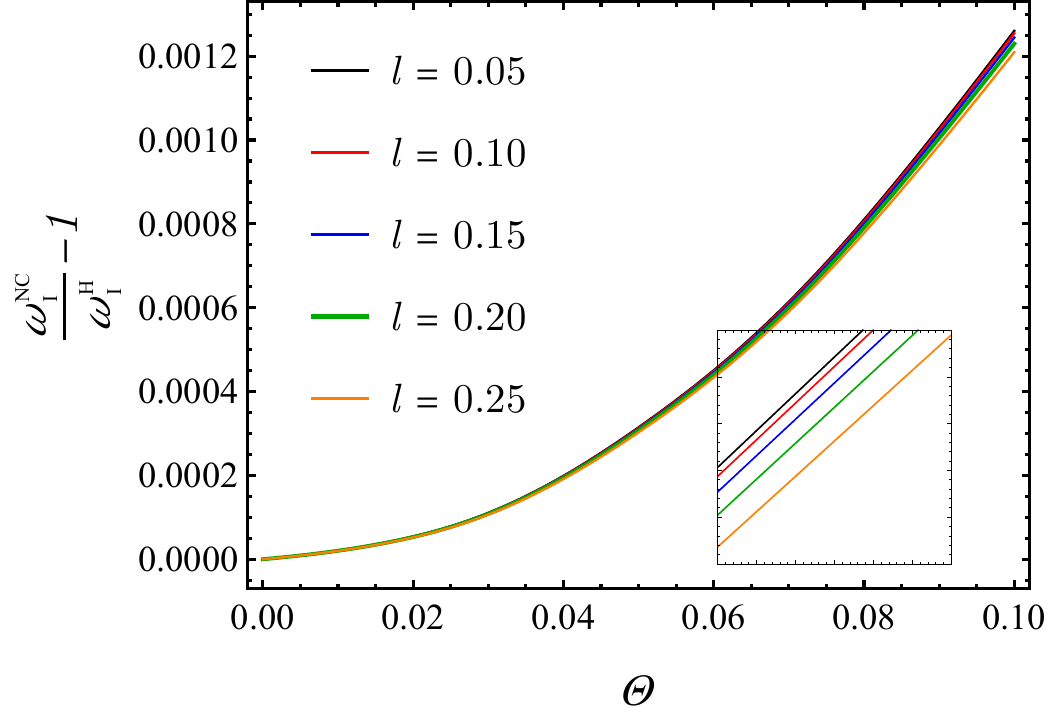}\\
	
	\caption{{Real and imaginary deviation of QNMs obtained by WKB method with respect to variation of $\Theta$ when $M = 1$, $\ell = 1 (m = \pm 1)$ and overtone $n = 0$.}}
	\label{fig:L1M1}
    \end{figure}
\begin{figure}
	\centering
	\includegraphics[width=80mm]{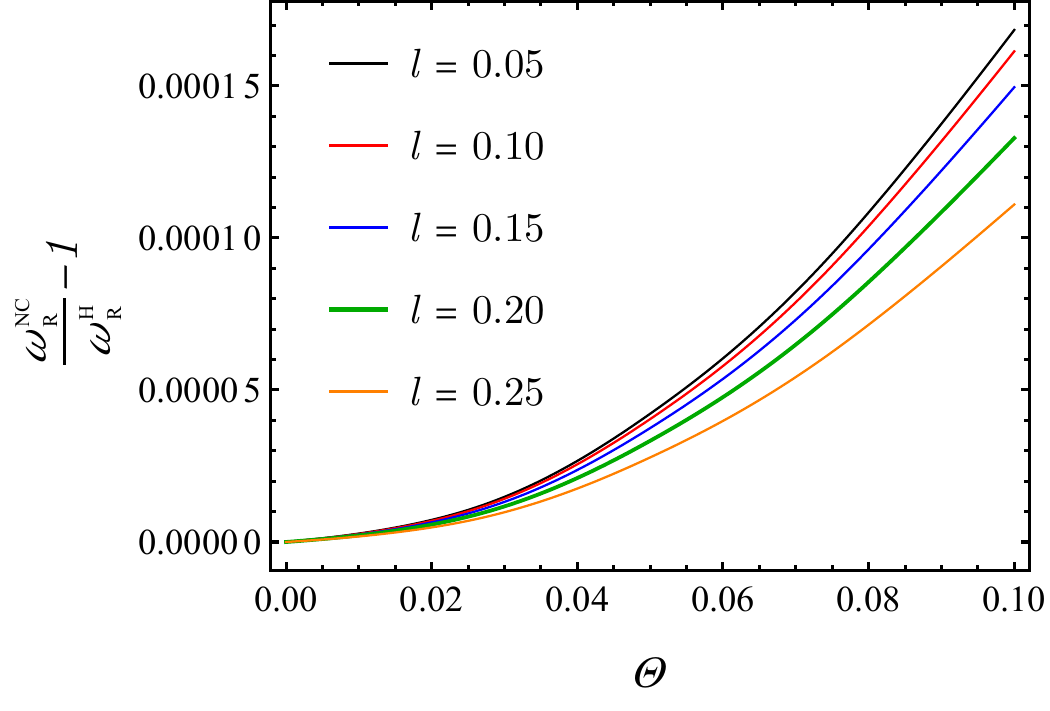}
	\hfil
    \includegraphics[width=80mm]{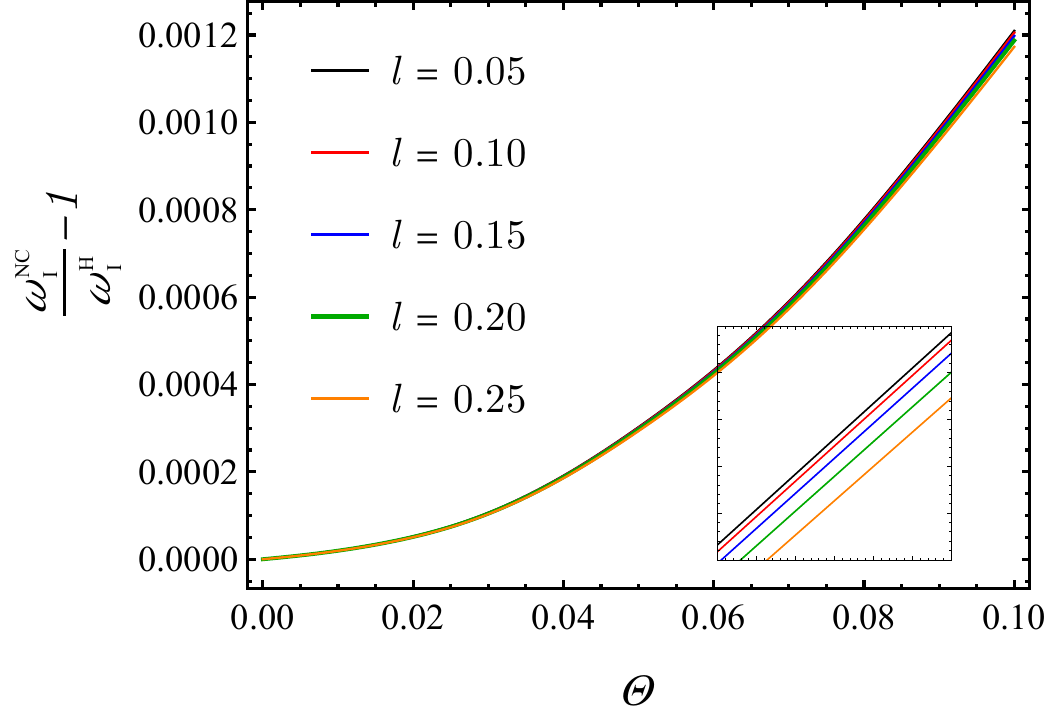}\\
	
	\caption{{Real and imaginary deviation of QNMs obtained by WKB method concerning variation of $\Theta$ when $M = 1$, $\ell = 2 (m = \pm 2)$ and overtone $n = 0$.}}
	\label{fig:L2M2}
    \end{figure}
\begin{figure}
	\centering
	\includegraphics[width=80mm]{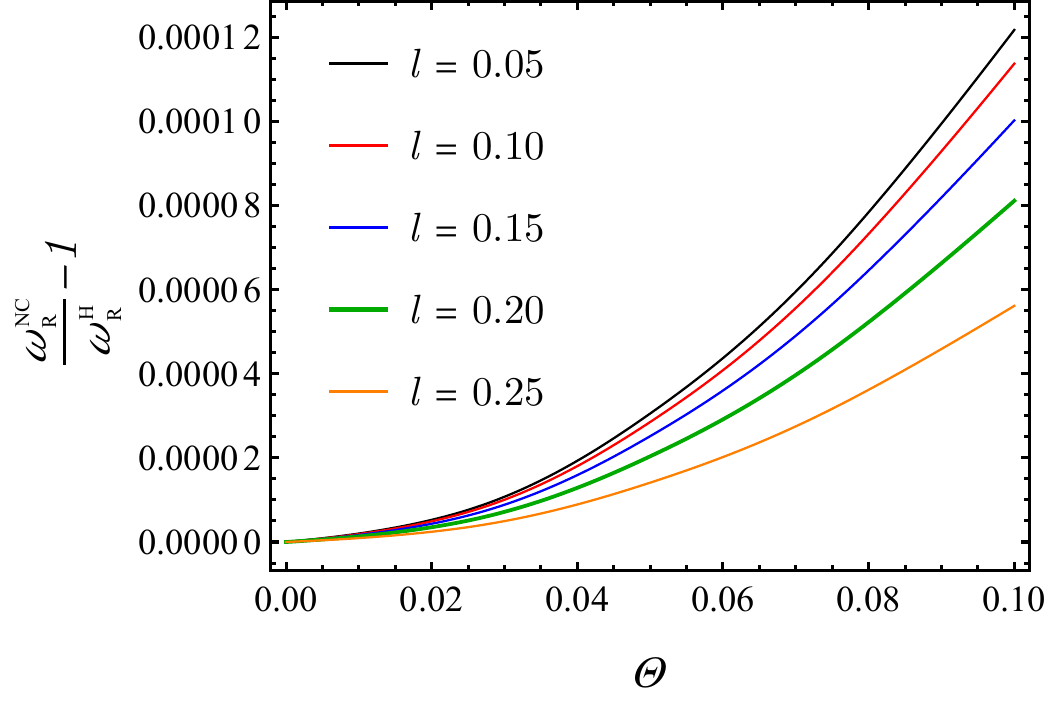}
	\hfil
    \includegraphics[width=80mm]{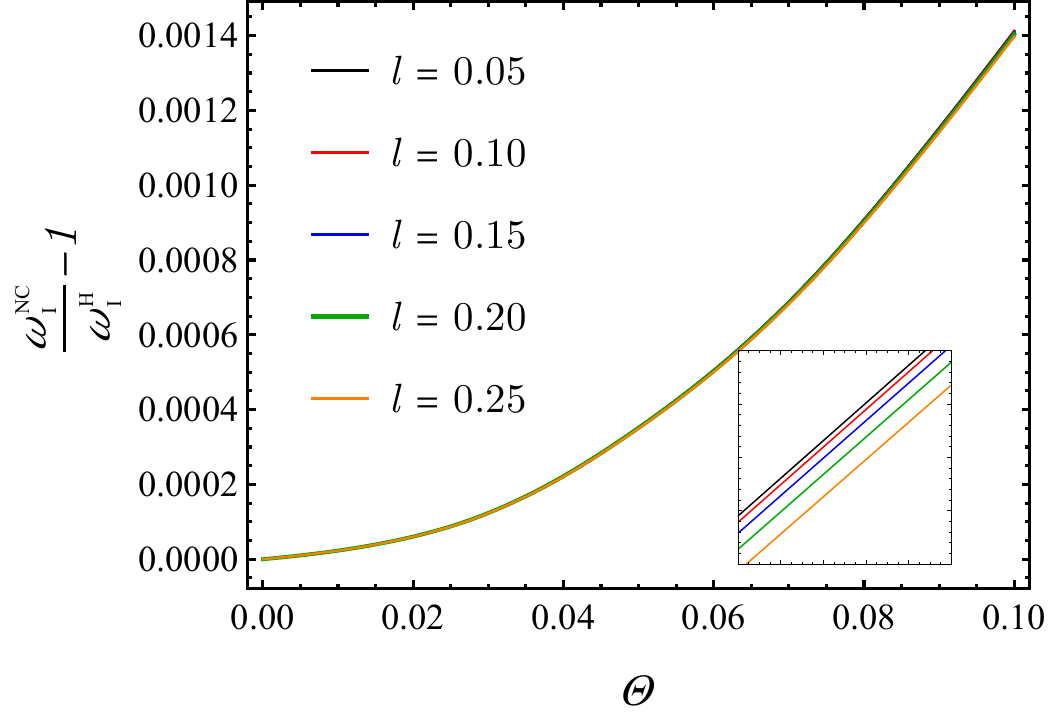}
	
	\caption{{Real and imaginary deviation of QNMs obtained by WKB method with respect to variation of $\Theta$ when $M = 1$, $\ell = 2 (m = \pm 1)$ and overtone $n = 0$.}}
	\label{fig:L2M1}
    \end{figure}
\begin{figure}
	\centering
	\includegraphics[width=80mm]{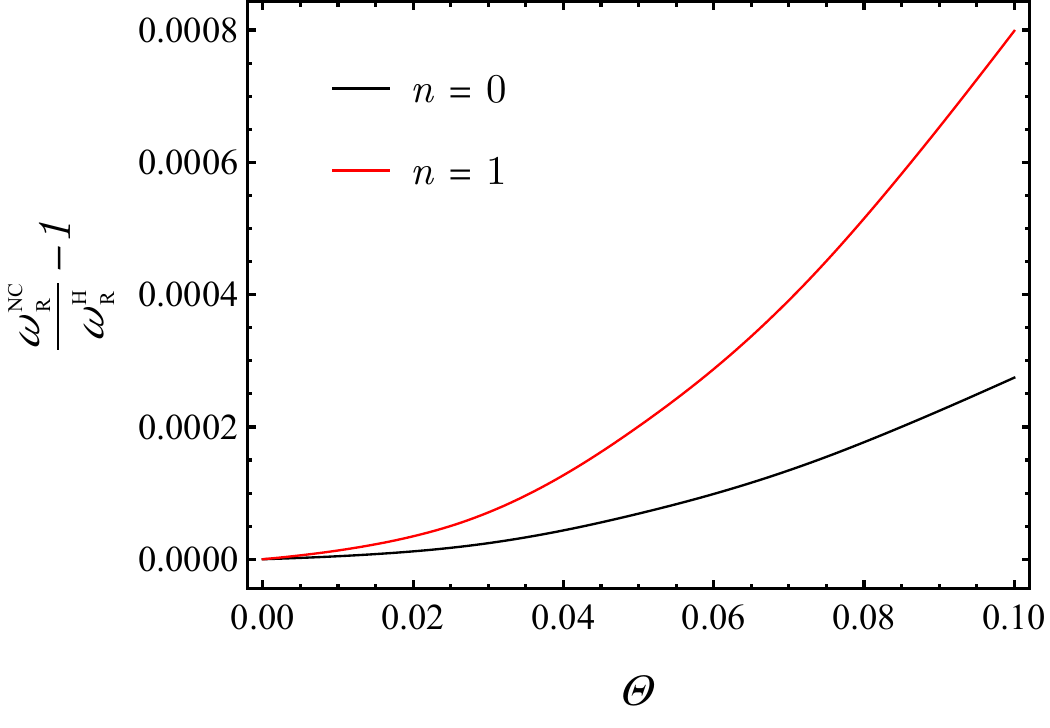}
	\hfil
    \includegraphics[width=80mm]{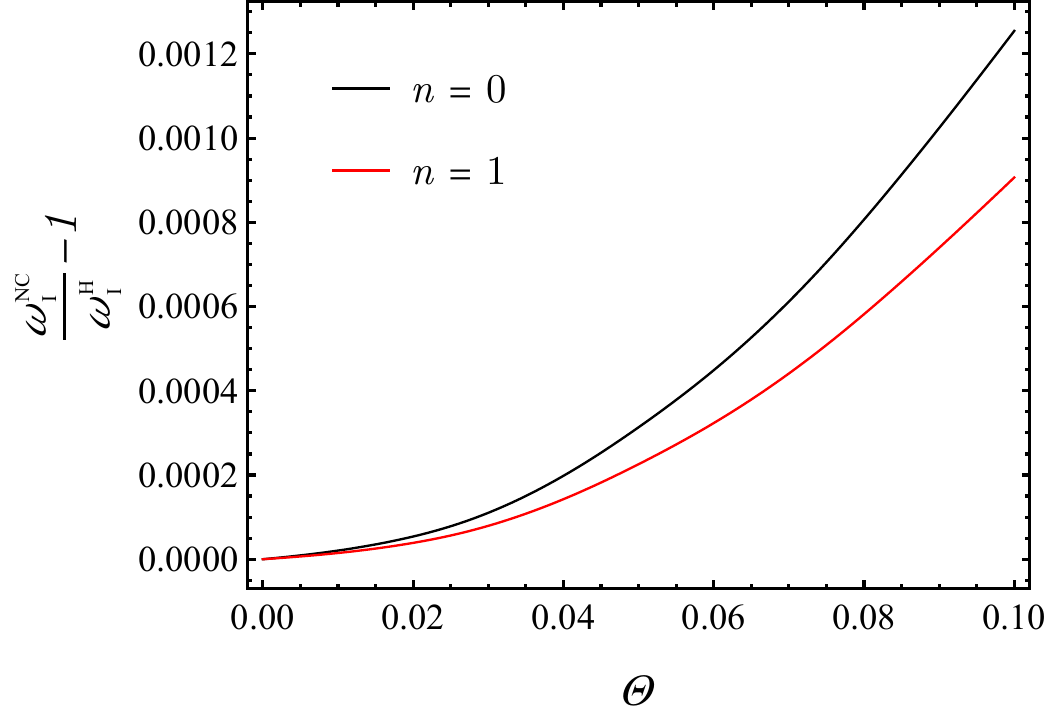}
	
	\caption{{QNMs obtained by WKB method for $M = 1$ and $l = 0.1$ with respect to variation of $\Theta$ when $\ell = 1 (m = \pm 1)$ and overtones $n = 0, 1$.}}
	\label{fig:L1M1N01}
    \end{figure}
\begin{figure}
	\centering
	\includegraphics[width=80mm]{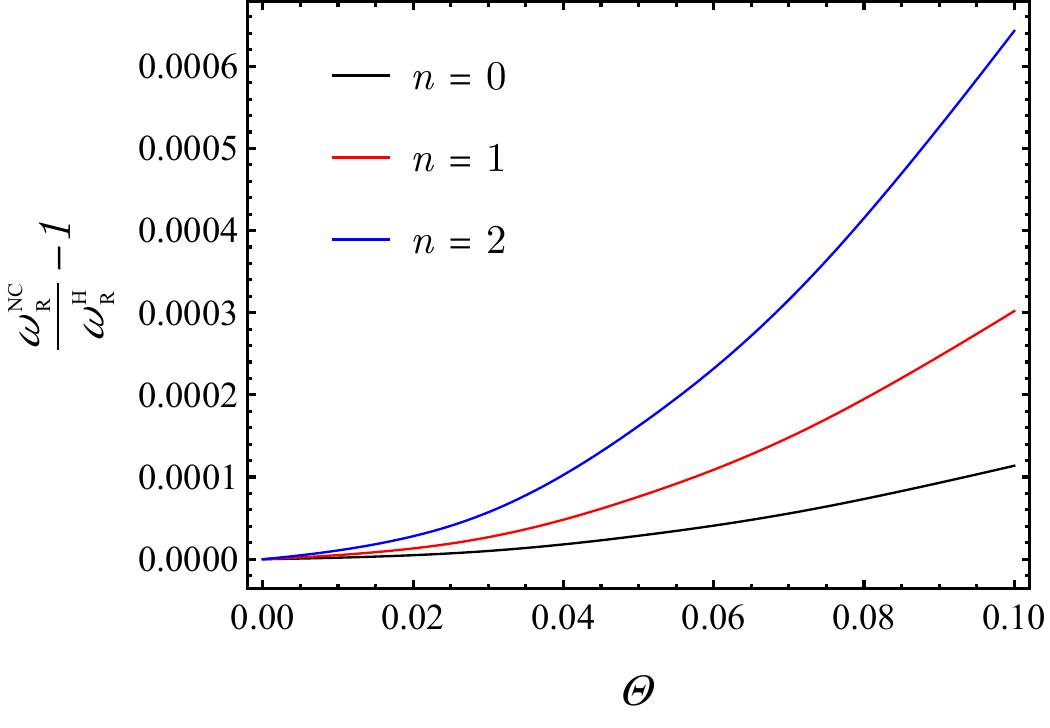}
	\hfil
    \includegraphics[width=80mm]{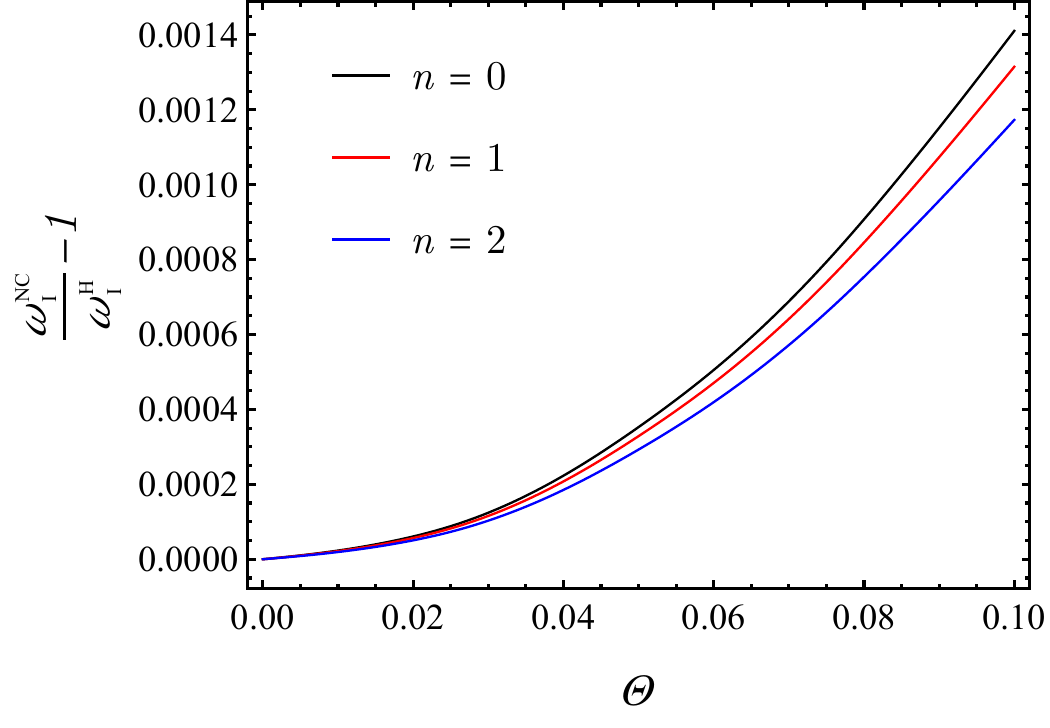}
	
	\caption{{QNMs obtained by WKB method for $M = 1$ and $l = 0.1$ with respect to variation of $\Theta$ when $\ell = 2 (m = \pm 1)$ and overtones $n = 0, 1$ and $2$.}}
	\label{fig:L2M1N}
\end{figure}

Applying the above--mentioned method, QNMs are explored and presented in Tab. \ref{Tab:AllQNMS}. The results show that the real and imaginary parts of QNMs delicately depend on $\Theta$ and $\ell$, therefore for better analysis, a normalized deviation is defined to quantify the difference between the QNMs of the non--commutative black hole and the standard Hayward black hole as
\begin{equation}
    \delta = \frac{\omega^{\rm {NC}}}{\omega^{\rm {H}}}-1.
\end{equation}
The above deviation has been investigated for both real and imaginary terms of QNMs.
The results are summarized in Fig. \ref{fig:L1M1}-\ref{fig:L2M1N} .

The analysis considers two sets of multipole numbers, \( \ell = 1,2 \), along with their associated monopoles, satisfying \( n \leq \ell \), for \( M = 1 \). Notably, when \( \Theta = 0 \), the solution reduces to the standard Hayward black hole, as expected. \\
The results show that $\delta$ in both real and absolute imaginary values
increases with $\Theta$.

For instance, the deviation in the real part of the QNMs for $\ell=1$, $m = \pm 1$ and $n = 0$ (Fig. \ref{fig:L1M1}) shows a gradual increase with $\Theta$, implying a higher propagation frequency. Moreover, the deviation in the real part is more pronounced for lower values of $l$.
Although the absolute value of the imaginary component of the frequency exhibits a similar trend, suggesting that larger values of \( \Theta \) correspond to shorter damping timescales for the black hole. These behaviors are consistent across all modes in Fig. \ref{fig:L2M2}--\ref{fig:L2M1} and the behavior of both real and imaginary terms, are similar for $\ell = 1 (m=\pm1)$, $\ell = 2 (m=\pm 2)$ and $\ell = 2 (m=\pm 1)$. \\

Furthermore, the deviation for various related monopoles for $n \leq \ell$, are represented in Fig.\ref{fig:L1M1N01}--\ref{fig:L2M1N}, respectively. Both real and imaginary parts of the QNMs's, have similar behaviour as their corresponding multipole number $\ell$. The real part, as a propagation frequency, goes higher with NC parameter, but the effect is more pronounced for the overtones with higher numbers. The imaginary part, governing the damping rate, also increases with $\Theta$, however, it is more significant for the lower overtone. This suggests that the damping rate is more sensitive to non--commutative effects for the fundamental mode.




\section{Time--domain solution}

Examining scalar perturbations within the time domain is fundamental to understanding the role of the quasinormal spectrum in time--dependent scattering processes. Given the complexity of the effective potential, a precise numerical approach is necessary to properly capture its behavior. To address this, the characteristic integration method developed by Gundlach et al. \cite{Gundlach:1993tp} is applied, providing a reliable tool for studying quasinormal modes in dynamical scattering and their implications for black hole physics.

The approach outlined in Refs. \cite{Baruah:2023rhd, Bolokhov:2024ixe, Guo:2023nkd, Yang:2024rms, Gundlach:1993tp, Skvortsova:2024wly, Shao:2023qlt} employs light--cone coordinates, defined as $u = t - r^{*}$ and $v = t + r^{*}$. Utilizing these coordinates streamlines the wave equation, facilitating a more effective analysis. Within this formulation, the wave equation can be reformulated as
\ie
\left(4 \frac{\partial^{2}}{\partial u \partial v} + V(u,v)\right) \Tilde{\psi} (u,v) = 0 \label{timedomain}.
\fe

An effective numerical strategy for solving the equation involves a discretization scheme that combines the finite-difference method with other computational techniques to improve accuracy and stability
\ie
\Tilde{\psi}(N) = -\Tilde{\psi}(S) + \Tilde{\psi}(W) + \Tilde{\psi}(E) - \frac{h^{2}}{8}V(S)[\Tilde{\psi}(W) + \Tilde{\psi}(E)] + \mathcal{O}(h^{4}).
\fe

The coordinate points are designated as $S = (u, v)$, $W = (u + h, v)$, $E = (u, v + h)$, and $N = (u + h, v + h)$, where $h$ represents the grid spacing parameter. The null surfaces defined by $u = u_{0}$ and $v = v_{0}$ play a fundamental role in establishing reference points for initializing the system. In this study, the initial conditions along the null surface $u = u_{0}$ are described by a Gaussian profile centered at $v = v_{c}$ with a width parameter $\sigma$
\ie
\Tilde{\psi}(u=u_{0},v) = A e^{-(v-v_{0})^{2}}/2\sigma^{2}, \,\,\,\,\,\, \Tilde{\psi}(u,v_{0}) = \Tilde{\psi}_{0}.
\fe
The initial condition is defined as $\Tilde{\psi}(u, v_{0}) = \Tilde{\psi}_{0}$ at $v = v_{0}$, with $\Tilde{\psi}_{0}$ set to zero for convenience, as this does not affect generality. The integration process advances along surfaces of constant $u$ while incrementing $v$, following the specified null data. To simplify the analysis, a scalar test field is considered with $M = 1$. The initial configuration adopts a Gaussian profile centered at $v_{0} = 0$, characterized by a width of $\sigma = 1$ and centered at $\Tilde{\psi}_{0} = 0$. The computational region spans $u \in [0, 1000]$ and $v \in [0, 1000]$, with the grid factors $h=0.1$.

\begin{figure}
    \centering
    \includegraphics[scale=0.51]{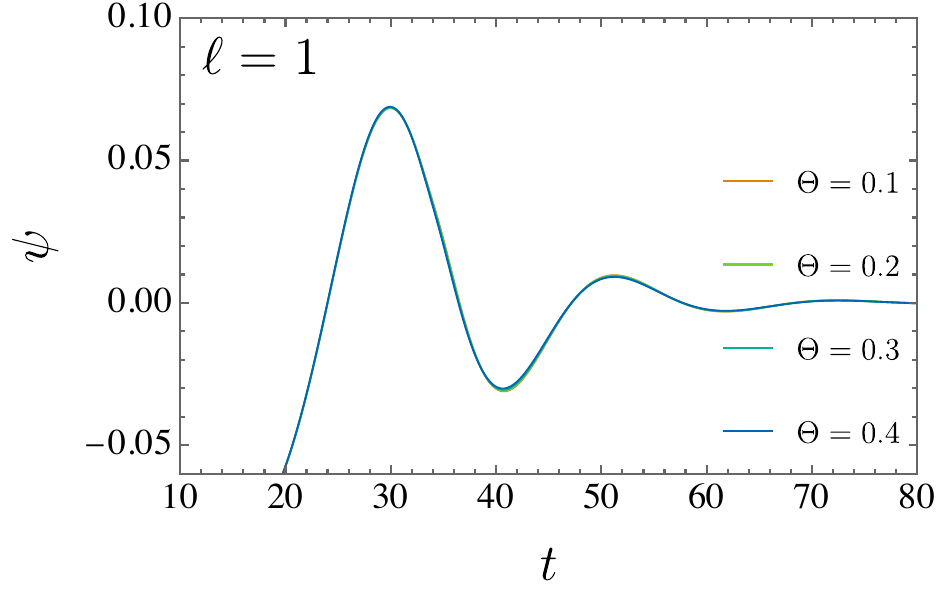}
     \includegraphics[scale=0.51]{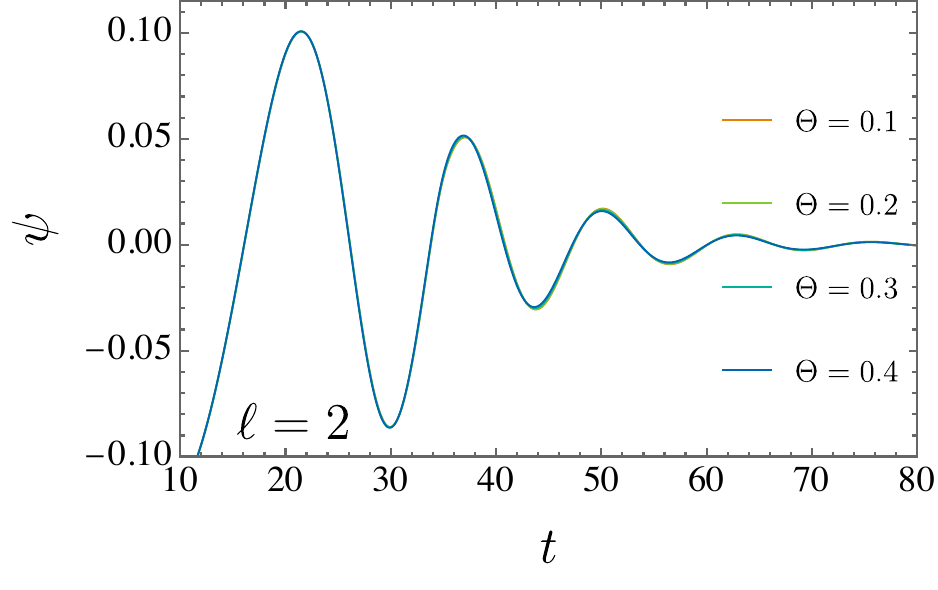}
      \includegraphics[scale=0.51]{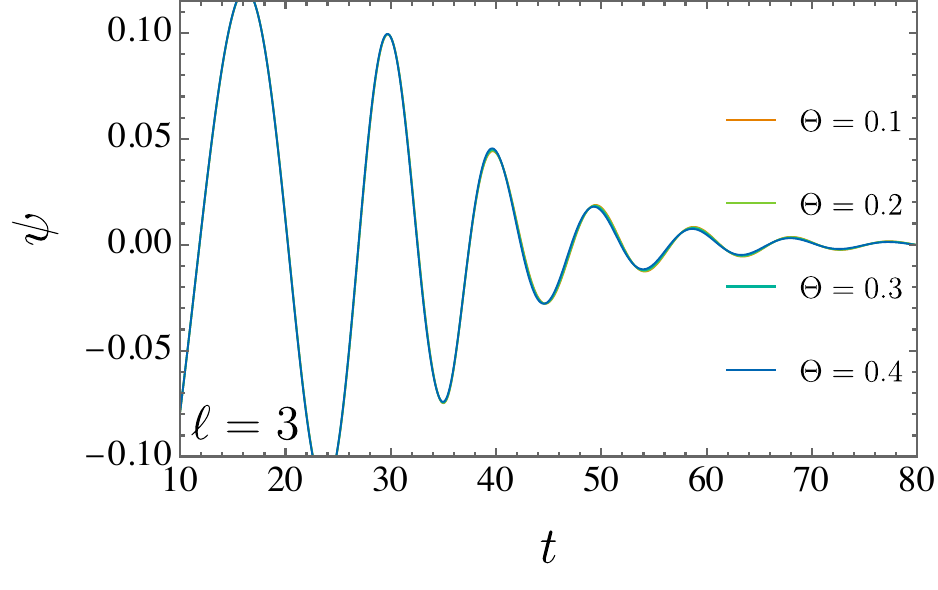}
    \caption{{The function $\Tilde{\psi}$ is shown as a function of time $t$ for $\ell = 1$ (top left panel), $\ell = 2$ (top right panel), and $\ell = 3$ (bottom panel), with $m$ fixed at $+1$. The plots consider various values of $\Theta$—namely, $0.1$, $0.2$, $0.3$, and $0.4$—while holding $l = 0.1$ constant.}}
    \label{time-domain-wave}
\end{figure}

\begin{figure}
    \centering
    \includegraphics[scale=0.51]{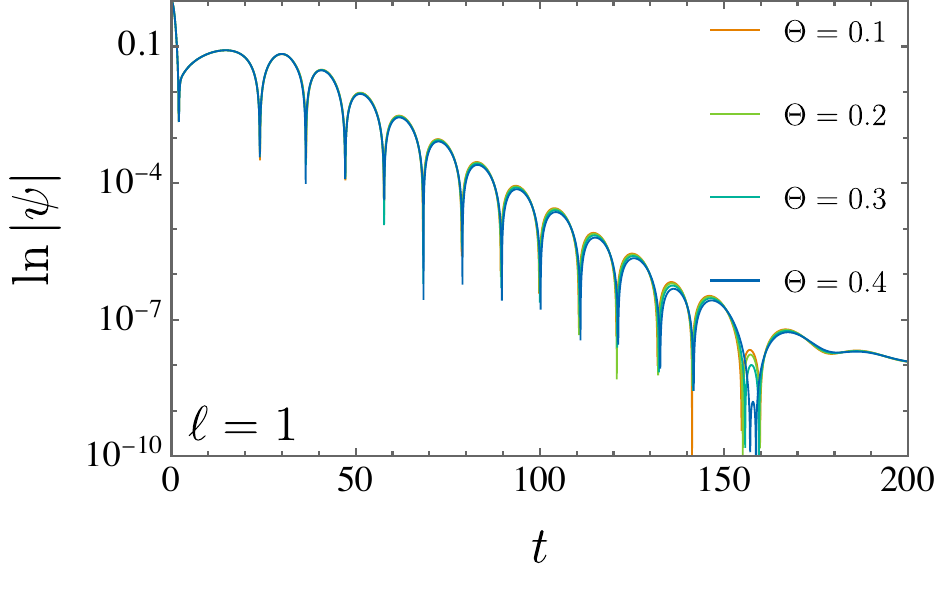}
     \includegraphics[scale=0.51]{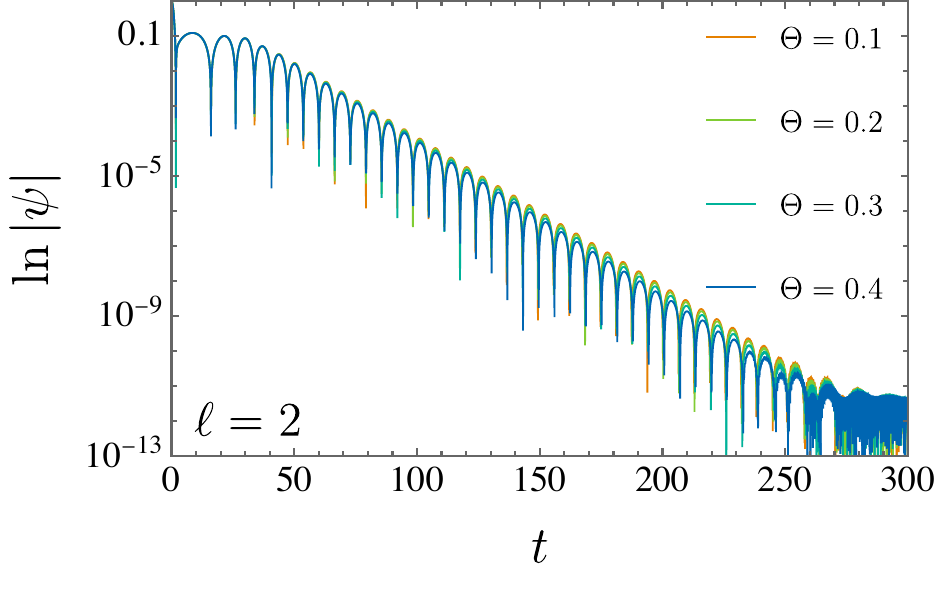}
      \includegraphics[scale=0.51]{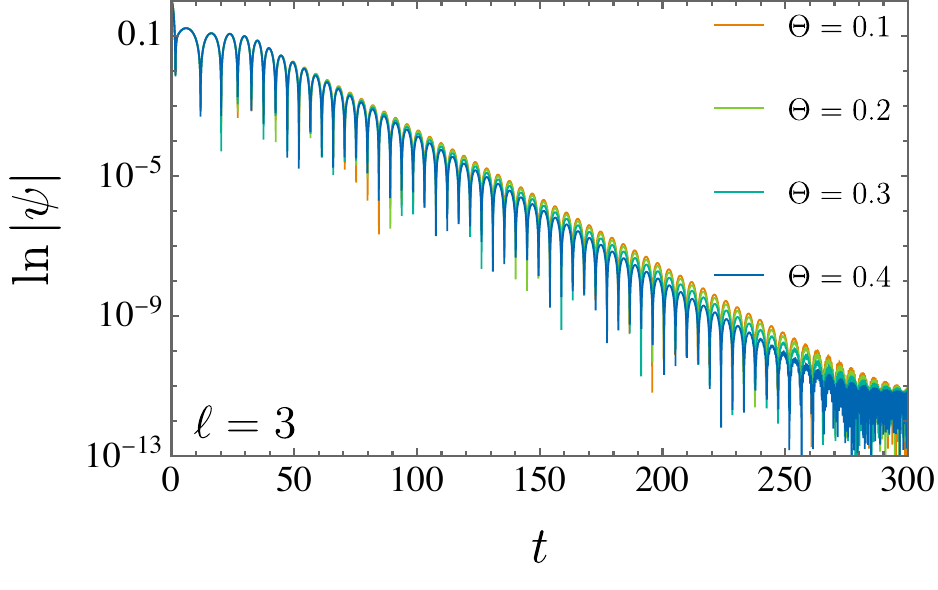}
    \caption{ { The quantity $\ln|\Tilde{\psi}|$ is plotted against time $t$ for $\ell = 1$ (top left panel), $\ell = 2$ (top right panel), and $\ell = 3$ (bottom panel), with $m$ fixed at $+1$. The analysis considers different values of $\Theta$—specifically, $0.1$, $0.2$, $0.3$, and $0.4$—while keeping $l = 0.1$ unchanged.}}
    \label{time-domain-wave-abs}
\end{figure}

\begin{figure}
    \centering
    \includegraphics[scale=0.51]{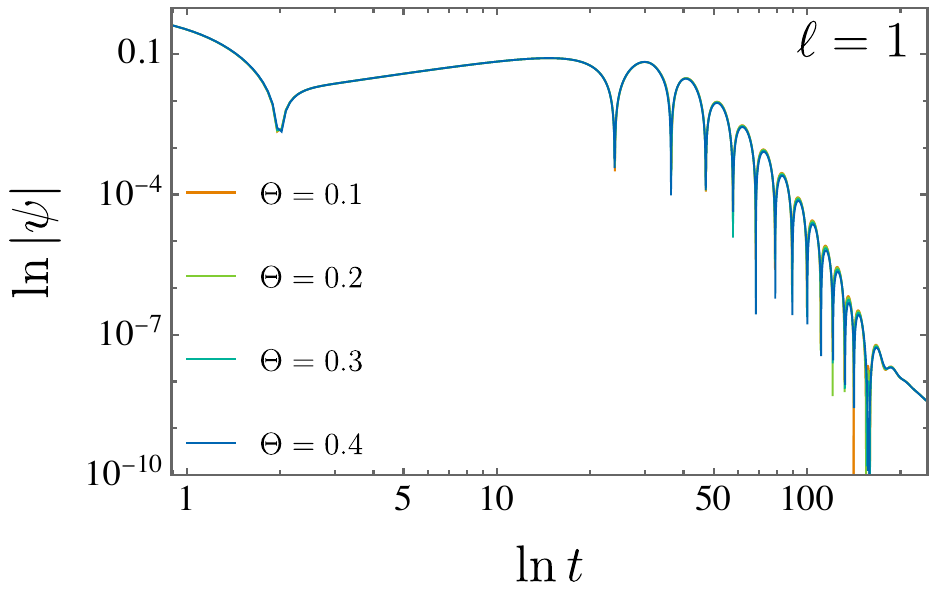}
     \includegraphics[scale=0.51]{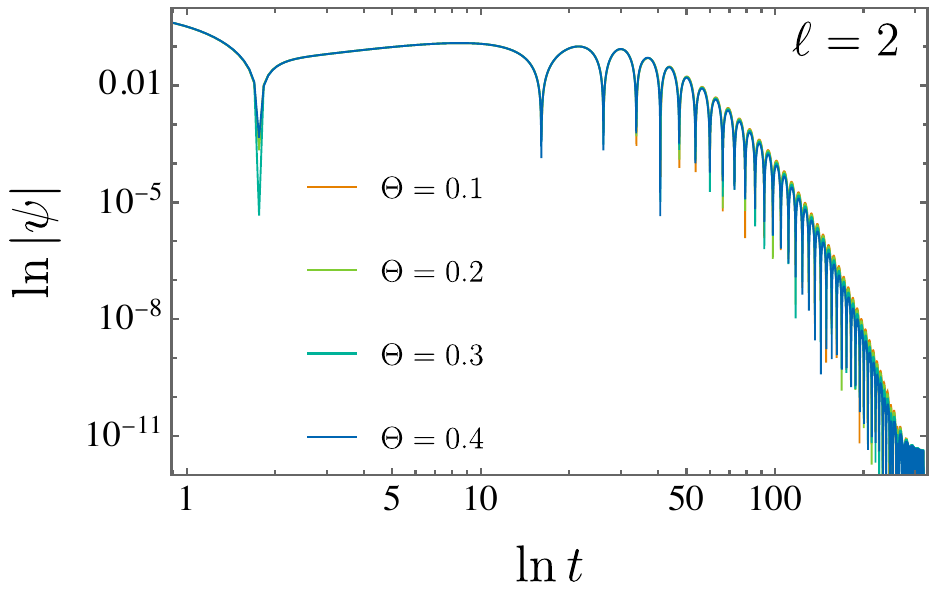}
     \includegraphics[scale=0.51]{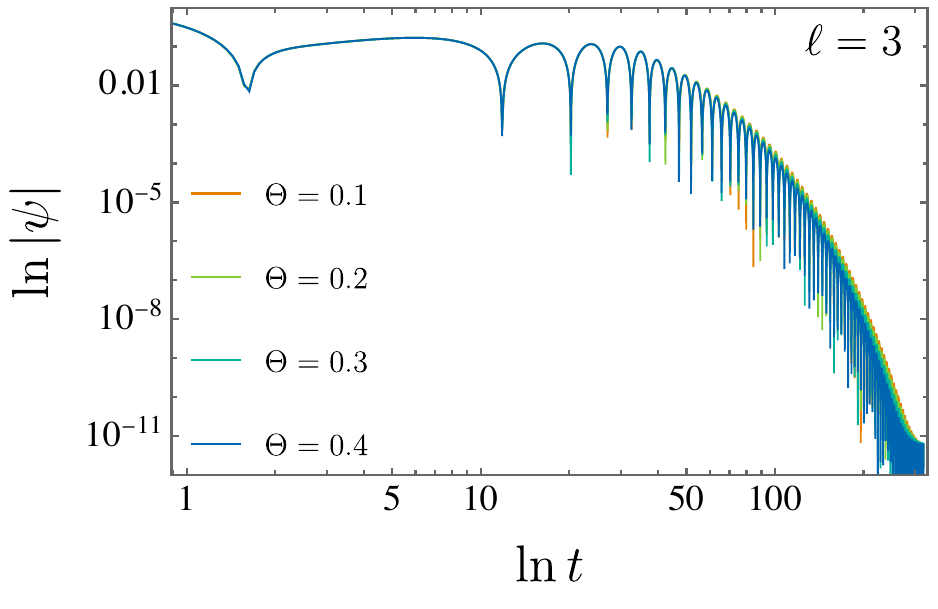}
    \caption{{ The logarithmic quantity $\ln|\Tilde{\psi}|$ is plotted as a function of $\ln t$ for $\ell = 1$ (top left panel), $\ell = 2$ (top right panel), and $\ell = 3$ (bottom panel), with $m$ fixed at $+1$. The analysis explores several values of $\Theta$—namely, $0.1$, $0.2$, $0.3$, and $0.4$—while keeping $l = 0.1$ constant.}}
    \label{time-domain-wave-abs-loglog}
\end{figure}

In Fig. \ref{time-domain-wave}{, We compute the function $\Tilde{\psi}$ and plot its evolution over time $t$ for distinct values of $\Theta$—specifically, $0.1$, $0.2$, $0.3$ and $0.4$—while keeping $l = 0.1$ fixed. The analysis is performed for $\ell = 1$ (top left panel), $\ell = 2$ (top right panel), and $\ell = 3$ (bottom panel), with $m$ set to $+1$ throughout.}  These plots reveal that as time progresses, the waveform tends to stabilize at more attainable values. Additionally, Fig. \ref{time-domain-wave-abs} illustrate{s} the $\ln$ absolute values of $\Tilde{\psi}$ as a function of time {for the same configuration emplyed to $\Tilde{\psi}$}. The straight--line segments in these plots indicate exponential decay, reflecting the typical profile of a quasinormal mode, where oscillations diminish exponentially over time. Finally, Fig. \ref{time-domain-wave-abs-loglog} present{s} a log--log plot of the $\ln$ absolute value of $\Tilde{\psi}$ against time. {As in the previous plots, the same initial conditions (for $\ell, m, l$ and $\Theta$) have been adopted here.} Overall, the inclusion of the non--commutative parameter in the black hole model results in a more strongly damped configuration for gravitational waves. Such a conclusion is corroborated in the quasinormal section as well (see Tab. \ref{Tab:AllQNMS}).

\section{Null Geodesics}

This part of the study is dedicated to examining the geodesic structure. As discussed earlier, determining the Christoffel symbols plays a crucial role in this analysis. Therefore, we begin by expressing
\ie
\frac{\mathrm{d}^{2}x^{\mu}}{\mathrm{d}\tau^{2}} + \Gamma\indices{^\mu_\alpha_\beta}\frac{\mathrm{d}x^{\alpha}}{\mathrm{d}\tau}\frac{\mathrm{d}x^{\beta}}{\mathrm{d}\tau} = 0. \label{geogeo}
\fe

\begin{figure}
    \centering
    \includegraphics[scale=0.62]{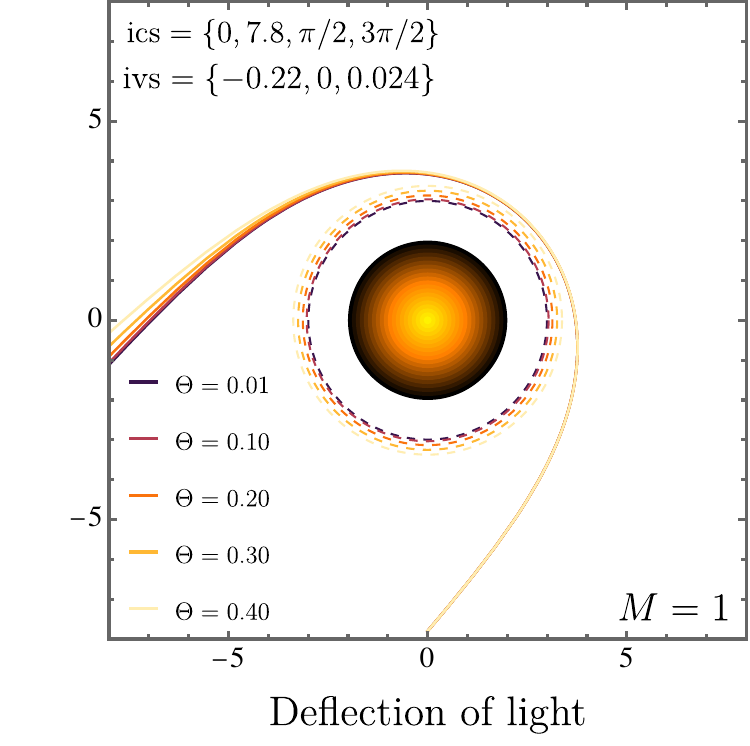}
    \includegraphics[scale=0.62]{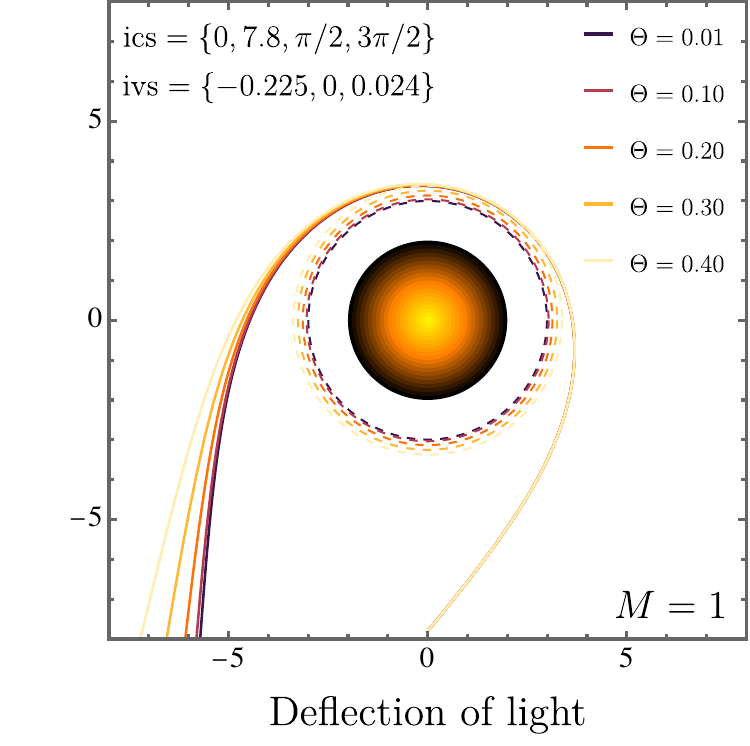}
    \includegraphics[scale=0.62]{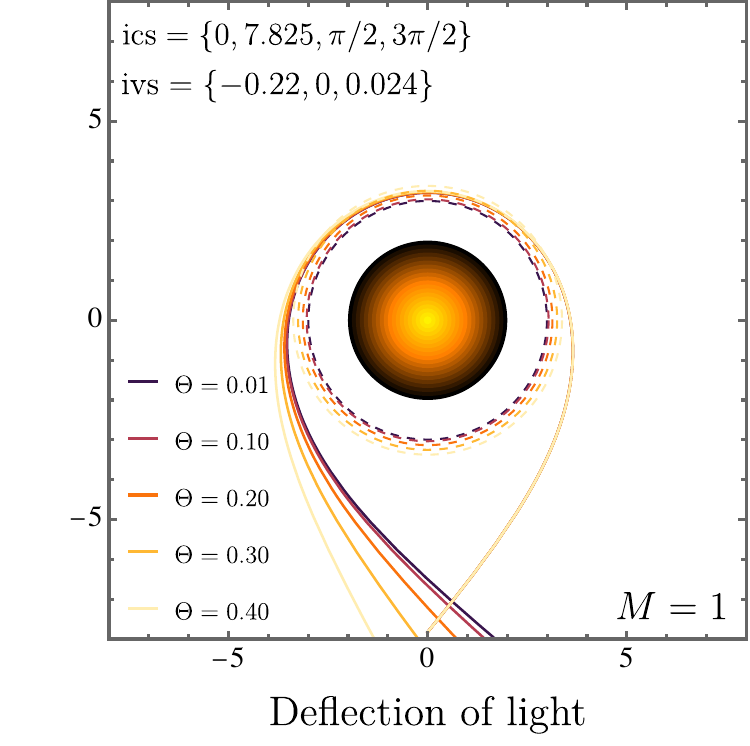}
    \includegraphics[scale=0.62]{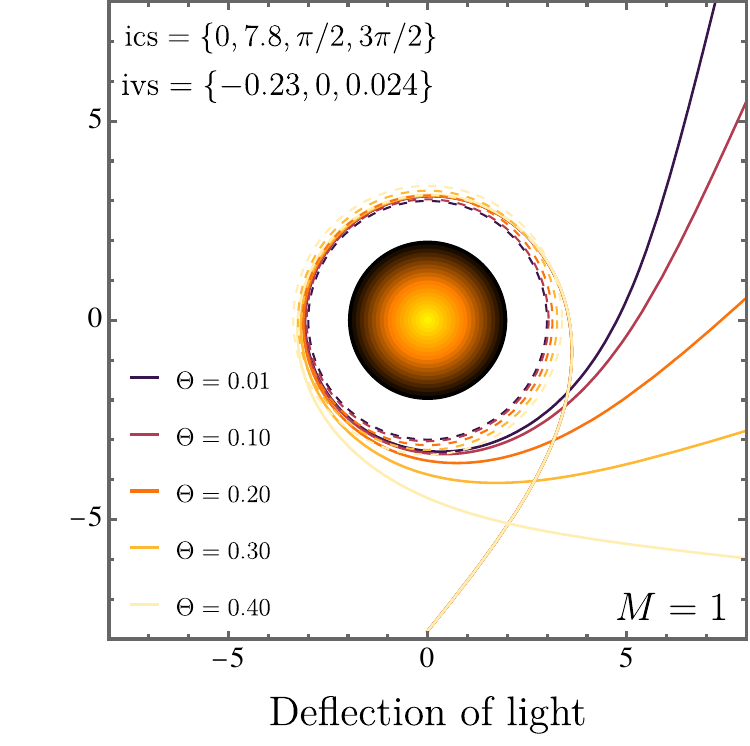}
    \caption{{The representation of light trajectories as the non--commutative parameter $\Theta$ increases, with fixed values of $l = 0.1$ and $M=1$, considering a range of initial conditions for the numerical inputs.}}
    \label{defleeeded}
\end{figure}

In this context, $\tau$ represents a general affine parameter. This formulation leads to a system of four coupled differential equations, each describing motion along a specific coordinate direction. Due to their considerable length, spanning approximately $20$ pages, their explicit expressions are omitted in this paper. 
{Fig. \ref{defleeeded} illustrates the behavior of light trajectories for different values of the non--commutative parameter $\Theta$, while keeping $l = 0.1$ and $M = 1$ fixed. The values of $\Theta$ vary from $0.01$ to $0.4$ in increments. In general lines, as $\Theta$ increases, the light rays exhibit a more ``open'' trajectory, indicating a reduced deflection. This analysis was performed using a range of numerical initial conditions, as specifically indicated within the plots themselves. The colored disk denotes the event horizon, while the dashed curves illustrate the outermost critical orbits.  }

 \subsection{Photonic radius and shadows}\label{sec:null}

A key area of research in black hole physics is the study of the shadows cast \cite{zeng2022shadows, HAMIL2023101293, anacleto2023absorption, yan2023shadows}. Interest in these phenomena has increased significantly, especially after the groundbreaking observation of the $Sgr A^*$ and $M87^*$ black holes shadow by the Event Horizon Telescope (EHT) \cite{ball2019first, gralla2021can, akiyama2019first}. To investigate the photonic and the shadow radius, we begin with a general diagonal metric given by 
\begin{equation}
g^{(\Theta,l)}_{\mu \nu }\mathrm{d}{x^\mu }\mathrm{d}{x^\nu } = g_{tt}^{(\Theta,l)} \mathrm{d}{t^2} + g_{rr}^{(\Theta,l)}\mathrm{d}{r^2} + g_{\theta\theta}^{(\Theta,l)}\mathrm{d}{\theta ^2} + g_{\varphi\varphi}^{(\Theta,l)}\mathrm{d}{\varphi ^2}.
\end{equation}

Using this metric, the corresponding Lagrangian is given by

\begin{equation}
\label{113}
\mathcal{L} = \frac{1}{2}\Big[  g_{tt}^{(\Theta,l)}{{\dot t}^2} + g_{rr}^{(\Theta,l)}{{\dot r}^2} + g_{\theta\theta}^{(\Theta,l)}{{\dot \theta }^2} + g_{\varphi\varphi}^{(\Theta,l)} {{\dot \varphi }^2}\Big].
\end{equation}

For simplicity, we assume that geodesics lie in the equatorial plane ($\theta=\frac{\pi}{2}$).
Also, we consider two conserved quantities, the energy $E$ and angular momentum $L$ as

\begin{equation}\label{constant}
E = |g_{tt}^{(\Theta,l)}| \dot t \quad\mathrm{and}\quad L = g_{\varphi\varphi}^{(\Theta,l)}\dot \varphi .
\end{equation}
which are related to the impact parameter as $b=\frac{L}{E}$. Since light follows null geodesics as $\mathcal{L}=0$, the following condition must hold
\begin{equation}\label{light}
g_{tt}^{(\Theta,l)}{{\dot t}^2} + g_{rr}^{(\Theta,l)}{{\dot r}^2} + g_{\varphi\varphi}^{(\Theta,l)}{{\dot \varphi }^2} = 0.
\end{equation}
 Substituting Eq. \eqref{constant} into Eq. \eqref{light}, the radial equation governing the motion of photons in the equatorial plane can be expressed as

\begin{equation}\label{rdot}
\frac{{{{\dot r}^2}}}{{{{\dot \varphi }^2}}} = \frac{{g_{{\varphi\varphi}}^{(\Theta,l)}}}{{g_{rr}^{(\Theta,l)}}}\left({-}\frac{{g_{\varphi\varphi}^{(\Theta,l)}}}{{g_{tt}^{(\Theta,l)}}}\frac{{{E^2}}}{{{L^2}}} - 1\right) .
\end{equation}

{Additionally, it is important to emphasize that
\ie
\frac{\mathrm{d}r}{\mathrm{d}\lambda} = \frac{\mathrm{d}r}{\mathrm{d}\varphi} \frac{\mathrm{d}\varphi}{\mathrm{d}\lambda}  = \frac{\mathrm{d}r}{\mathrm{d}\varphi}\frac{L}{g_{\varphi\varphi}^{(\Theta,l)}}, 
\fe
where
\ie
\Dot{r}^{2} = \left( \frac{\mathrm{d}r}{\mathrm{d}\lambda} \right)^{2} =\left( \frac{\mathrm{d}r}{\mathrm{d}\varphi} \right)^{2} \frac{L^{2}}{\left( g_{\varphi\varphi}^{(\Theta,l)}\right)^{2}}.
\fe

In this way, the effective potential $\mathrm{V}(\Theta,l)$ may properly be defined as
\ie
\label{potential}
\mathrm{V}(\Theta,l) = \frac{{g_{\varphi\varphi}^{(\Theta,l)}}}{{g_{rr}^{(\Theta,l)}}}\left(-\frac{{g_{\varphi\varphi}^{(\Theta,l)}}}{{g_{tt}^{(\Theta,l)}}}\frac{{{E^2}}}{{{L^2}}} - 1\right)\frac{L^{2}}{\left( g_{\varphi\varphi}^{(\Theta,l)}\right)^{2}}  .
\fe}

{To determine the photon spheres (also referred to as critical orbits), one must consider the following condition: }
\begin{equation}
\frac{\mathrm{d} \,{\mathrm{V}(\Theta,l)}}{\mathrm{d}r} = 0 .
\end{equation}

\begin{table}[!ht]
    \centering
    \caption{{The values of the outermost critical orbit are presented for various combinations of $\Theta$ and $l$, with the black hole mass fixed at $M = 1$. }}
    \begin{tabular}{|c|c|c|c|c|c|}
    \hline
         {$r_{ph1}$} & ${l = 0.02}$ & ${l = 0.04}$ & ${l = 0.06}$ & ${l = 0.08}$& ${l = 0.10
        }$\\ \hline\hline
        $\Theta = 0.0$ & 2.99982 & 2.99929 & 2.99840 & 2.99715 &2.99554 \\ \hline
        $\Theta= 0.2$ & {3.13641} & {3.13602} & {3.13537} & {3.13446} & {3.13329}\\ \hline
        $\Theta = 0.4$& {3.37380} & {3.37372} & {3.37358} & {3.37339} & {3.37315} \\ \hline
        $\Theta = 0.6$ & {3.59937} & {3.59962} & {3.60004} & {3.60062} & {3.60137}\\ \hline
        $\Theta = 0.8$ & {3.79557} & {3.79612} & {3.79705} & {3.79834} & {3.80001} \\ \hline
        $\Theta = 0.99$ & {3.95476} & {3.95557} & {3.95693} & {3.95883} & {3.96128}\\ \hline
    \end{tabular}
    \label{Tab:rphoton}
\end{table}

{Remarkably, this procedure yields three real and positive values. However, only two of them—$r_{ph1}$ (the outermost) and $r_{ph2}$ (the innermost)—are located outside the event horizon\footnote{{As shown in Tab. \ref{Tab:rphoton2}, only a few configurations of $\Theta$ (big values) and $l$ yield photon spheres ($r_{ph2}$) within the black hole event horizon $r_{h}$.} }. Due to their considerable length, the explicit expressions for these radii are not presented here. Instead, Tab. \ref{Tab:rphoton} provides a quantitative analysis of the photon sphere radius $r_{ph1}$ for a black hole with mass $M = 1$, evaluated for varying values of the non--commutative and Hayward parameters. Notice that, for fixed values of $\Theta$, increasing $l$ results in a decrease of $r_{ph1}$. On the other hand, when $l$ is held constant and $\Theta$ is varied, the corresponding radius of the photon sphere increases. 

In addition, Tab. \ref{Tab:rphoton2} displays the quantitative values of the innermost photon sphere radius, $r_{ph2}$. Unlike $r_{ph1}$, increasing $l$ for fixed values of $\Theta$ leads to a larger critical radius. On the other hand, when $l$ is kept constant and $\Theta$ is varied, the changes in $r_{ph2}$ are minimal. It is important to highlight, based on this Tab., which contrasts to our case, the pure Hayward black hole does not exhibit two photon spheres located outside the event horizon.

\begin{table}[!ht]
    \centering
    \caption{{The values of the innermost critical orbit are presented for various combinations of $\Theta$ and $l$, with the black hole mass fixed at $M = 1$. }}
    \begin{tabular}{|c|c|c|c|c|c|}
    \hline
         {$r_{ph2}$} & ${l = 0.02}$ & ${l = 0.04}$ & ${l = 0.06}$ & ${l = 0.08}$& ${l = 0.10
        }$\\ \hline\hline
        $\Theta = 0.0$ & {0.0000} & {0.0000} & {0.0000} & {0.0000} & {0.0000} \\ \hline
        $\Theta= 0.2$ & {2.0004} & {2.00161} & {2.00363} & {2.00648} & {2.01019}\\ \hline
        $\Theta = 0.4$& {2.0004} & {2.00161} & {2.00363} & {2.00648} & {2.0102} \\ \hline
        $\Theta = 0.6$ & {2.0004} & {2.00161} & {2.00363} & {2.00648} & {2.0102}\\ \hline
        $\Theta = 0.8$ & {\text{inside $r_{h}$}} & {\text{inside $r_{h}$}} & {2.00363} & {2.00648} & {2.01021} \\ \hline
        $\Theta = 0.99$ & {\text{inside $r_{h}$}} & {\text{inside $r_{h}$}} & {\text{inside $r_{h}$}} & {\text{inside $r_{h}$}} & {2.01021}\\ \hline
    \end{tabular}
    \label{Tab:rphoton2}
\end{table}

}

\begin{figure}[ht]
		\centering
		\includegraphics[width=81mm]{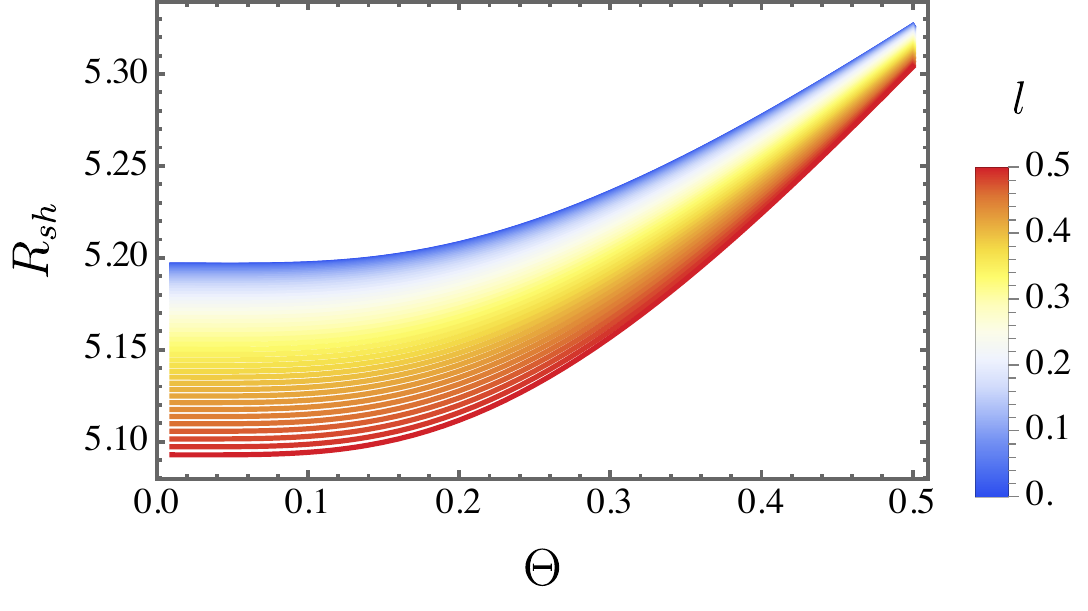}
        \includegraphics[width=81mm]{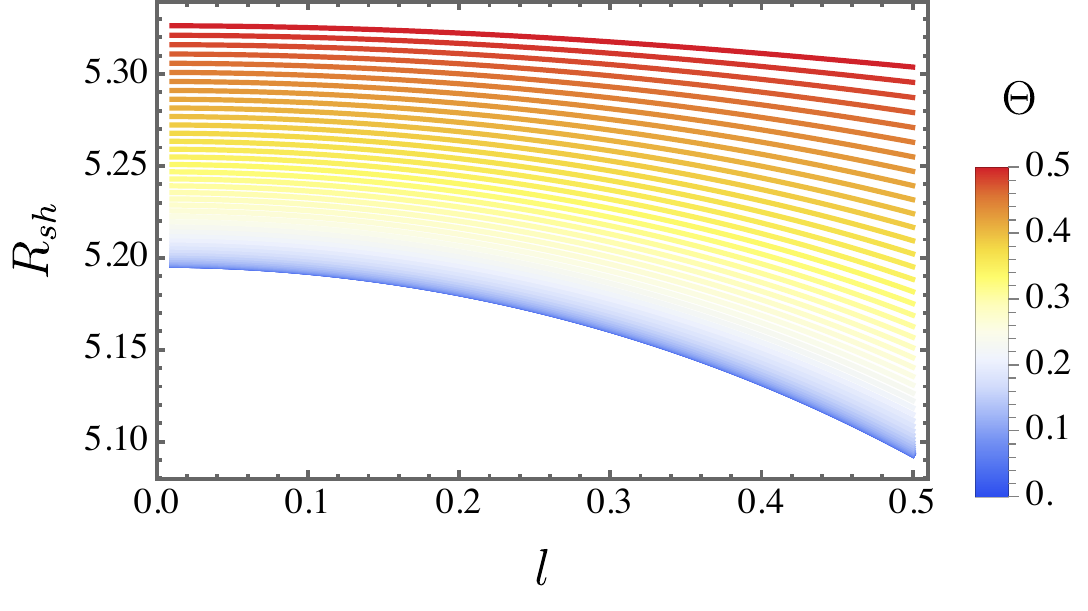}
		\caption{ {The left panel displays $R_{sh}$ as a function of $\Theta$ for different values of $l$, while the right panel presents shadow radii versus $l$ for various choices of $\Theta$. All analyses are performed with the black hole mass fixed at $M = 1$.}}
		\label{fig:Shadow}
	\end{figure}

{Furthermore,} following the methodologies outlined in Refs. \cite{perlick2015influence, konoplya2019shadow, touati2022geodesic}, the expression for the black hole shadow in a spherically symmetric spacetime read the following equation
\ie
\label{shadow}
\begin{split}
R_{sh} & = {  \sqrt {\frac{{{g_{\varphi\varphi}^{(\Theta,l)}(r_{ph1})}}}{{{|g_{tt}^{(\Theta,l)}(r_{ph1})|}}}} } \\
& {\approx \, \, \frac{\Theta ^2 \left(24 M^2-12 M r_{ph1} + r_{ph1}^2\right)}{8 \sqrt{\frac{r_{ph1}^3}{r_{ph1} - 2 M}} (r_{ph1} - 2 M)^2} +\sqrt{\frac{r_{ph1}^3}{r_{ph1} - 2 M}} } \\
& {  + \left( \frac{3 \Theta ^2 M^2 \sqrt{\frac{r_{ph1}^3}{r_{ph1} - 2 M}} \left(360 M^2-336 M r_{ph1}+77 r_{ph1}^2\right)}{4 r_{ph1}^6 (r_{ph1}-2 M)^2}-\frac{2 M^2}{\sqrt{\frac{r_{ph1}^3}{r_{ph1}-2 M}} (r_{ph1}-2 M)^2} \right)l^{2}, }
\end{split}
\fe
{where we have considered $\theta = \pi/2$. }{ Concerning the shadow radii, our analysis will focus exclusively on the configuration corresponding to the outermost photon sphere. In this manner, the corresponding values are presented in Tab. \ref{tabshadows}.}
\begin{table}[!ht]
    \centering
    \caption{{The values for the shadow radii $R_{sh}$ are presented for various combinations of $\Theta$ and $l$, with the black hole mass fixed at $M = 1$. }}
    \begin{tabular}{|c|c|c|c|c|c|}
    \hline
         {$R_{sh}$} & ${l = 0.02}$ & ${l = 0.04}$ & ${l = 0.06}$ & ${l = 0.08}$& ${l = 0.10
        }$\\ \hline\hline
        $\Theta = 0.0$ & {5.19600} & {5.19554} & {5.19477} & {5.19368} & {5.19229} \\ \hline
        $\Theta= 0.2$ & {5.20759} & {5.20717} & {5.20646} & {5.20546} & {5.20418}\\ \hline
        $\Theta = 0.4$& {5.27683} & {5.27658} & {5.27617} & {5.27559} & {5.27485} \\ \hline
        $\Theta = 0.6$ & {5.37941} & {5.37943} & {5.37945} & {5.37949} & {5.37954}\\ \hline
        $\Theta = 0.7$ & {5.43349} & {5.43365} & {5.43391} & {5.43427} & {5.43475} \\ \hline
    \end{tabular}
    \label{tabshadows}
\end{table}
For better visualization, we {also} display an analysis of the shadows of our black hole for a range of $\Theta$ and $l$ values in Fig. \ref{fig:Shadow}.  
    
{ The left panel displays the variation of the shadow radius as a function of the Hayward parameter $l$, with the black hole mass fixed at $M = 1$ and the non--commutative parameter set to $\Theta = 0.01$. As $l$ increases, the shadow radius decreases. In the right panel, the plot shows how the shadow radius changes with respect to $\Theta$, while keeping $M = 1$ and fixing the Hayward parameter at $l = 0.01$. In this case, the shadow radius also decreases as $\Theta$ increases.}

In the following section, we will explore the constraints on non--commutativity concerning the EHT observations.


\subsection{{Constraints on black hole shadow radii from EHT observations}}

{

Taking into account the horizon--scale data from the Event Horizon Telescope for $Sgr A^*$, one combines the mass--to--distance ratio estimates independently provided by Keck and VLTI. After applying a $2\sigma$ confidence interval \cite{heidari2023gravitational}, this procedure yields two distinct bounds for the shadow radius, as reported in Refs. \cite{vagnozzi2022horizon,akiyama2022firstSgrA}
\ie
\label{adaconst1asas}
4.55 < \frac{R_{sh}}{M} < 5.22,
\fe
and
\ie
\label{adaconst1asas2}
4.21 < \frac{R_{sh}}{M} < 5.56.
\fe
Constraints on the parameter $\Theta$ and $l$ are extracted by confronting theoretical predictions with data from the Event Horizon Telescope. The dependence of the shadow radius on them, expressed in units of the black hole mass $M$, is illustrated in Figs. \ref{adaconstraints} and \ref{adaconstraints2}. The parameter space boundaries are determined by the points at which the plotted curves intersect these observational bands. All resulting constraints are listed in Tabs. \ref{adatab:constr} and \ref{adatab:constr2}.

In general, the constraints on $\Theta$ for different values of $l$ are as follows: for $l = 0.1$, we find $0.26 \lesssim \Theta \lesssim 0.94$; for $l = 0.2$, $0.29 \lesssim \Theta \lesssim 0.92$; for $l = 0.3$, $0.33 \lesssim \Theta \lesssim 0.88$; and for $l = 0.4$, $0.36 \lesssim \Theta \lesssim 0.90$. On the other hand, we also observe the following: for $\Theta = 0.1$ and $\Theta = 0.2$, there are no values of $l$ that satisfy the bounds established in Eqs. (\ref{adaconst1asas}) and (\ref{adaconst1asas2}). However, for $\Theta = 0.3$, the constraint becomes $0.23 \lesssim l$, while for $\Theta = 0.4$, we obtain $0.52 \lesssim l$.

\begin{figure}
    \centering
     \includegraphics[scale=0.6]{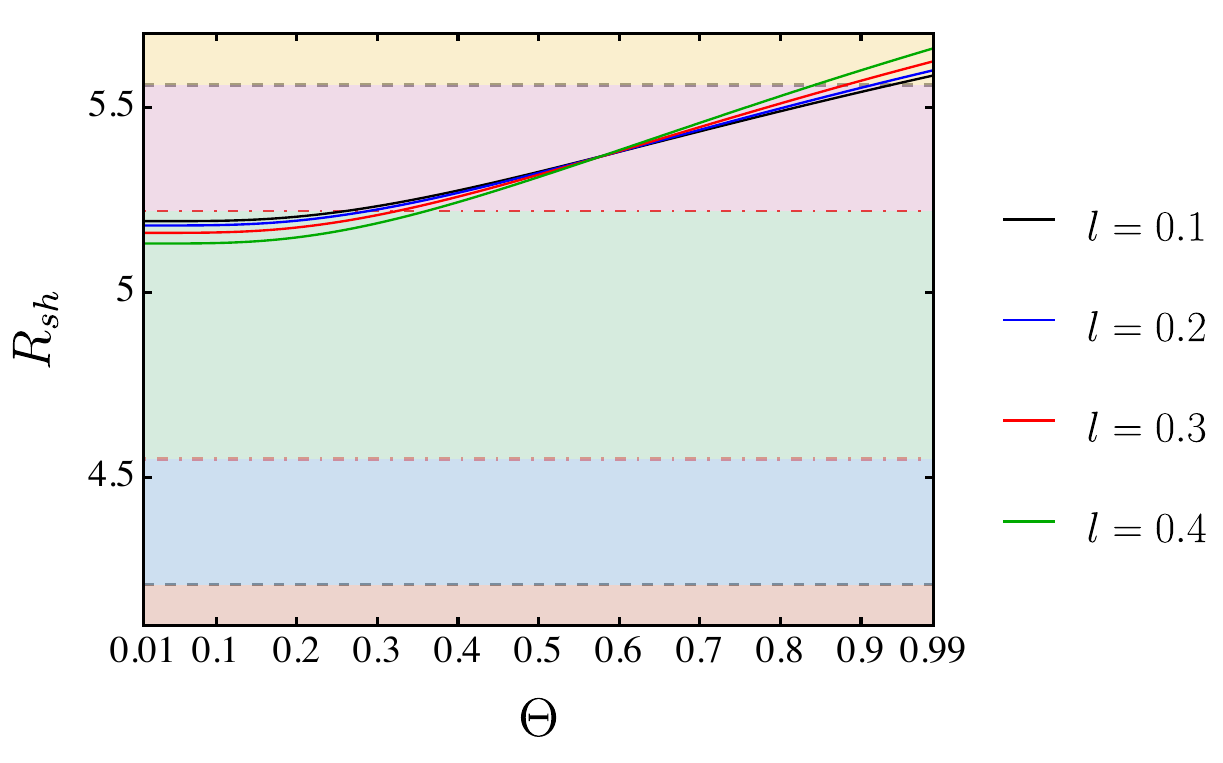}
    \caption{{The shadow radius is plotted as a function of $\Theta$, considering the experimental constraints from $SgrA^{*}$, for various values of the parameter $l$ \cite{vagnozzi2022horizon,akiyama2022firstSgrA}.}}
    \label{adaconstraints}
\end{figure}

\begin{figure}
    \centering
     \includegraphics[scale=0.6]{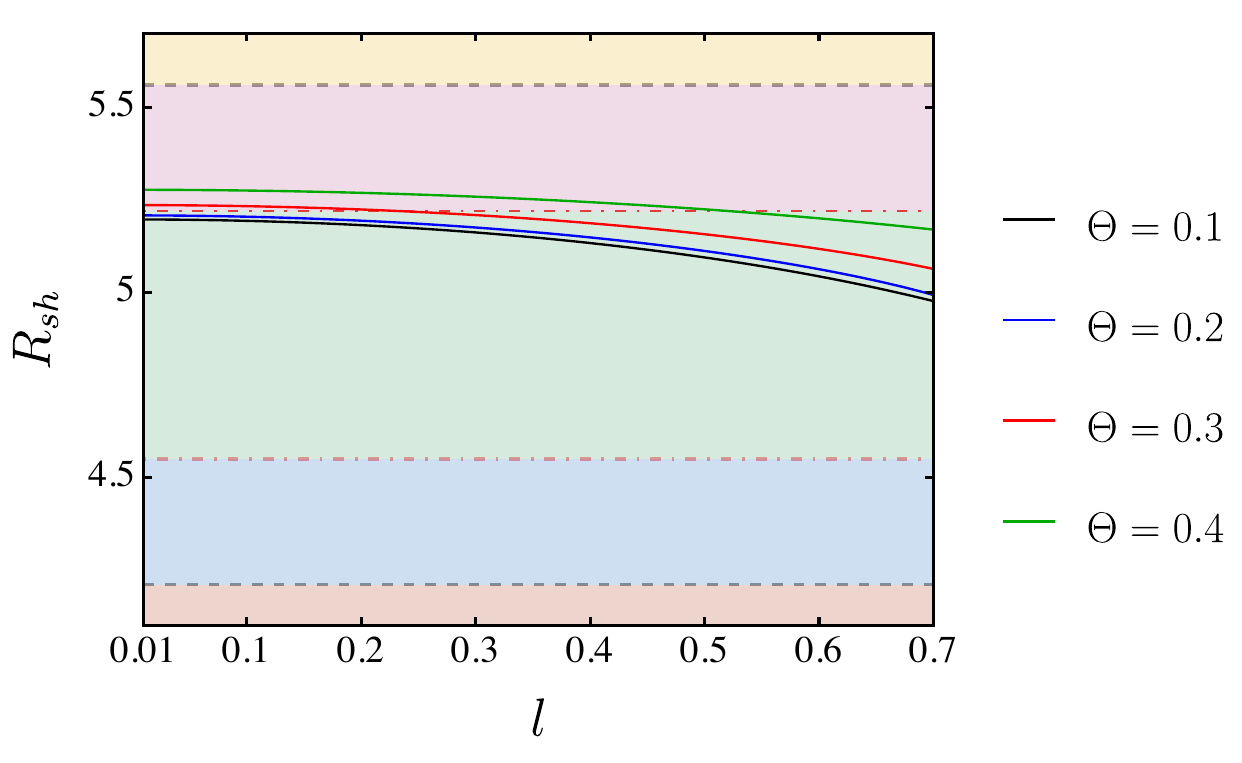}
    \caption{{The shadow radius is plotted as a function of $l$, taking into account the experimental constraints from $SgrA^{*}$, for different values of the parameter $\Theta$ \cite{vagnozzi2022horizon,akiyama2022firstSgrA}.}}
    \label{adaconstraints2}
\end{figure}

\begin{table}[h!]
\centering
\caption{{Bounds for $\Theta$ based on the observational data of EHT concering $SgrA^{*}$ \cite{vagnozzi2022horizon,akiyama2022firstSgrA}.}}
\label{adatab:constr2}
\begin{tabular}{lc}
\hline\hline
\textbf{Parameter} & Bounds  \\
\hline
\quad  $l =0.1$  & \makecell{$ 0.26  \lesssim  \Theta \lesssim 0.94$} \\
\quad  $l = 0.2$     & \makecell{$0.29 \lesssim \Theta \lesssim 0.92$}  \\
\quad  $l = 0.3$   & \makecell{$0.33 \lesssim \Theta \lesssim 0.88$}  \\
\quad $l = 0.4 $  & \makecell{$0.36 \lesssim \Theta \lesssim  0.90  $}  \\
\hline\hline
\end{tabular}
\end{table}

\begin{table}[h!]
\centering
\caption{{Bounds for $l$ based on the observational data of EHT concering $SgrA^{*}$ \cite{vagnozzi2022horizon,akiyama2022firstSgrA}.}}
\label{adatab:constr}
\begin{tabular}{lc}
\hline\hline
\textbf{Parameter} & Bounds  \\
\hline
\quad  $\Theta = 0.1$  & \makecell{-----------} \\
\quad  $\Theta = 0.2$     & \makecell{-----------}  \\
\quad  $\Theta = 0.3$   & \makecell{$0.23 \lesssim l$}  \\
\quad $\Theta = 0.4$  & \makecell{$0.52 \lesssim l$}  \\
\hline\hline
\end{tabular}
\end{table}

}


\section{Gravitational lensing}

This section examines the Gaussian curvature $\mathcal{K}(r,l,\Theta)$, which serves as a key factor in evaluating the stability of the photon sphere. Using this curvature, the deflection angle in gravitational lensing is derived through the Gauss--Bonnet theorem, following the method outlined in Ref. \cite{Gibbons:2008rj}. As demonstrated, the photon spheres analyzed here are inherently unstable, as indicated by $\mathcal{K}(r,l,\Theta) < 0$. Additionally, increasing $\Theta$ leads to more pronounced deflection angles, $\hat{\alpha}(b,l,\Theta)$.


\subsection{Stability of the photon sphere}

The stability of photon spheres in the vicinity of black holes is largely governed by the geometric and topological characteristics of optical spacetime, where conjugate points play a significant role. When photon trajectories experience perturbations, their response depends on whether the photon sphere is stable or unstable. In the case of instability, slight deviations lead photons to either be absorbed by the black hole or escape to infinity. Conversely, stable photon spheres enable photons to remain in bounded orbits nearby \cite{qiao2022curvatures,qiao2022geometric}.

The presence or absence of conjugate points within the spacetime manifold plays a crucial role in this behavior. Photon spheres classified as stable contain conjugate points, whereas unstable ones lack them. The Cartan–Hadamard theorem establishes a connection between Gaussian curvature, $\mathcal{K}(r)$, and these points, providing a method for assessing the stability of critical orbits \cite{qiao2024existence}. Following this approach, the null geodesics, which satisfy the condition $\mathrm{d}s^2=0$, can be expressed as \cite{araujo2024effects,heidari2024absorption}:
\ie
\mathrm{d}t^2=\gamma_{ij}\mathrm{d}x^i \mathrm{d}x^j = -\frac{g^{(\Theta,l)}_{rr}(r)}{g^{(\Theta,l)}_{tt}(r)}\mathrm{d}r^2  -\frac{\Bar{g}^{(\Theta,l)}_{\varphi\varphi}(r)}{g^{(\Theta,l)}_{tt}(r)}\mathrm{d}\varphi^2   ,
\fe
where $i$ and $j$ range from $1$ to $3$, $\gamma_{ij}$ represents the optical metric, and $\Bar{g}^{(\Theta,l)}_{\varphi\varphi} (r) \equiv g^{(\Theta,l)}_{\varphi\varphi}(r,\theta = \pi/2)$. Additionally, the Gaussian curvature is given by \cite{qiao2024existence}
\ie
\label{dffdsf}
\mathcal{K}(r,l,\Theta) = \frac{R}{2} =  \frac{g^{(\Theta,l)}_{tt}(r)}{\sqrt{g^{(\Theta,l)}_{rr}(r) \,  \Bar{g}^{(\Theta,l)}_{\varphi\varphi}(r)}}  \frac{\partial}{\partial r} \left[  \frac{g^{(\Theta,l)}_{tt}(r)}{2 \sqrt{g^{(\Theta,l)}_{rr}(r) \, \Bar{g}^{(\Theta,l)}_{\varphi\varphi}(r) }}   \frac{\partial}{\partial r} \left(   \frac{\Bar{g}^{(\Theta,l)}_{\varphi\varphi}(r)}{g^{(\Theta,l)}_{tt}(r)}    \right)    \right],
\fe
where $R$ represents the Ricci scalar in two dimensions. For sufficiently small values of $l$ and $\Theta$, its explicit expression is
\ie
\begin{split}
\label{gaussiancurvature}
\mathcal{K}(r,l,\Theta)  & { \, \,  \approx  \, \frac{3 M^2}{r^4}-\frac{2 M}{r^3} -\frac{72 l^2 M^3}{r^7}+\frac{40 l^2 M^2}{r^6} -\frac{47136 l^2 \Theta ^2 M^6}{r^{10} (r-2 M)^2}+\frac{88456 l^2 \Theta ^2 M^5}{r^9 (r-2 M)^2}   } \\
& {  -\frac{61574 l^2 \Theta ^2 M^4}{r^8 (r-2 M)^2}+\frac{18770 l^2 \Theta ^2 M^3}{r^7 (r-2 M)^2}-\frac{2102 l^2 \Theta ^2 M^2}{r^6 (r-2 M)^2}+\frac{624 \Theta ^2 M^4}{8 M r^7-4 r^8}            } \\
& { -\frac{848 \Theta ^2 M^3 r}{8 M r^7-4 r^8}+\frac{374 \Theta ^2 M^2 r^2}{8 M r^7-4 r^8}+\frac{\Theta ^2 r^4}{8 M r^7-4 r^8}-\frac{54 \Theta ^2 M r^3}{8 M r^7-4 r^8}   .}
\end{split}
\fe
{ The first two terms of the Gaussian curvature presented above correspond to the Schwarzschild black hole; the third and fourth terms represent contributions from the Hayward solution; and the remaining terms encode the corrections introduced by non--commutativity, as we can naturally deduce.

}

\begin{figure}
    \centering
    \includegraphics[scale=0.6]{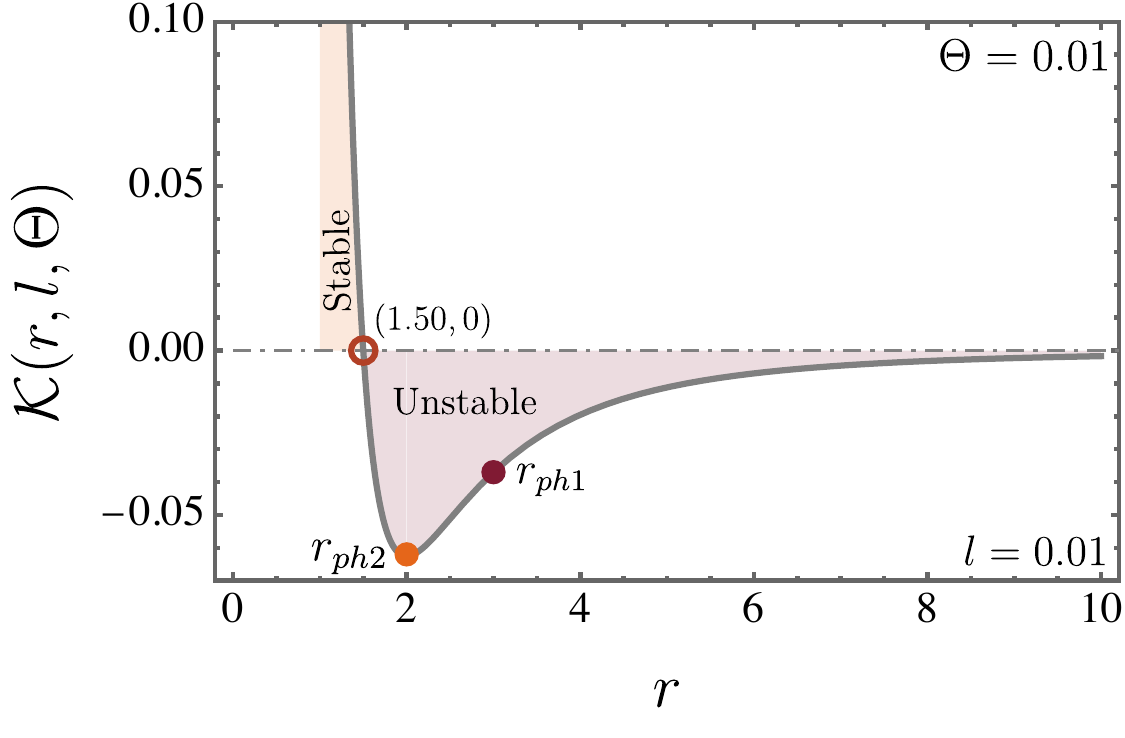}
    \caption{The Gaussian curvature $\mathcal{K}(r,l,\Theta)$ is presented, with the stable and unstable regions distinctly highlighted. Here, it is considered $M = 0.85$, $l = 0.001$, and $\Theta = 0.01$.}
    \label{gauss}
\end{figure}

As explored in Refs. \cite{qiao2022curvatures, qiao2022geometric, qiao2024existence}, the stability of photon spheres is determined by the sign of $\mathcal{K}(r,l,\Theta)$. A negative curvature, $\mathcal{K}(r,l,\Theta) < 0$, indicates instability, while a positive curvature, $\mathcal{K}(r,l,\Theta) > 0$, corresponds to stability. To illustrate this behavior, Fig. \ref{gauss} displays the Gaussian curvature $\mathcal{K}(r,l,\Theta)$ as a function of $r$, highlighting distinct regions associated with stable and unstable photon spheres. The analysis is carried out using the parameter values {$M = 1.0$}, {$l = 0.01$}, and $\Theta = 0.01$; {the transition point between stable and unstable configurations occurs at $(1.50, 0)$, as marked by an wine--colored circle}. {For these specific parameter values, the photon spheres are  $r_{ph1} \approx 3.00038$ (wine dot) and $r_{ph} \approx 2.0001$ (orange dot). In other words, this indicates that both of them turn out to be unstable photon orbits.}

\subsection{The weak deflection limit in the Gauss--Bonnet framework}

Building on the Gaussian curvature derived in Eq. (\ref{gaussiancurvature}), the next step is to evaluate the deflection angle in the weak--field limit using the Gauss--Bonnet theorem \cite{Gibbons:2008rj}. To achieve this, the surface area on the equatorial plane is determined and expressed as
\ie
\mathrm{d}S = \sqrt{\gamma} \, \mathrm{d} r \mathrm{d}\varphi = \sqrt{\frac{g^{(\Theta,l)}_{rr}}{g^{(\Theta,l)}_{tt}}  \frac{g^{(\Theta,l)}_{\varphi\varphi}}{g^{(\Theta,l)}_{tt}} } \, \mathrm{d} r \mathrm{d}\varphi,
\fe
allowing the deflection angle to be determined through the following expression:
\ie
\begin{split}
& \hat{\alpha} (b,l,\Theta) =  - \int \int_{D} \mathcal{K} \mathrm{d}S = - \int^{\pi}_{0} \int^{\infty}_{\frac{b}{\sin \varphi}} \mathcal{K} \mathrm{d}S \\
& \simeq  \, \frac{4 M}{b} + \frac{3 \pi  M^2}{4 b^2} -\frac{15 \pi  l^2 M^2}{16 b^4} { -\frac{23 \Theta ^2 M}{18 b^3} +\frac{5225 \pi  l^2 \Theta ^2 M^2}{384 b^6}+\frac{243 \pi  \Theta ^2 M^2}{1024 b^4}}    .
\end{split}
\fe

\begin{figure}
    \centering
    \includegraphics[scale=0.635]{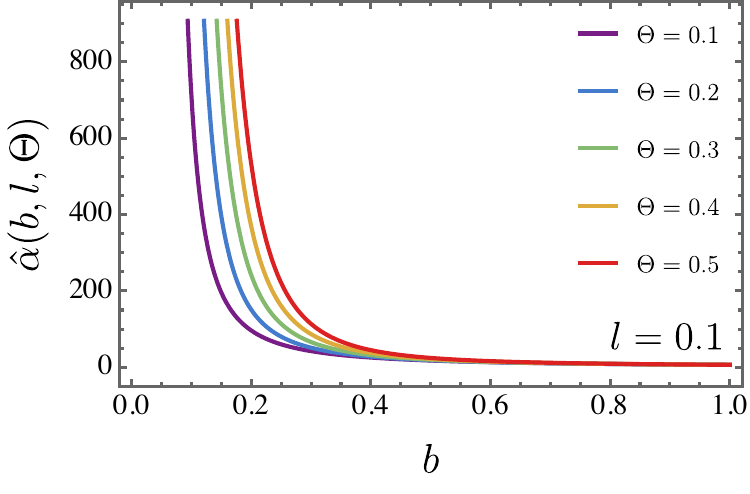}
    \includegraphics[scale=0.65]{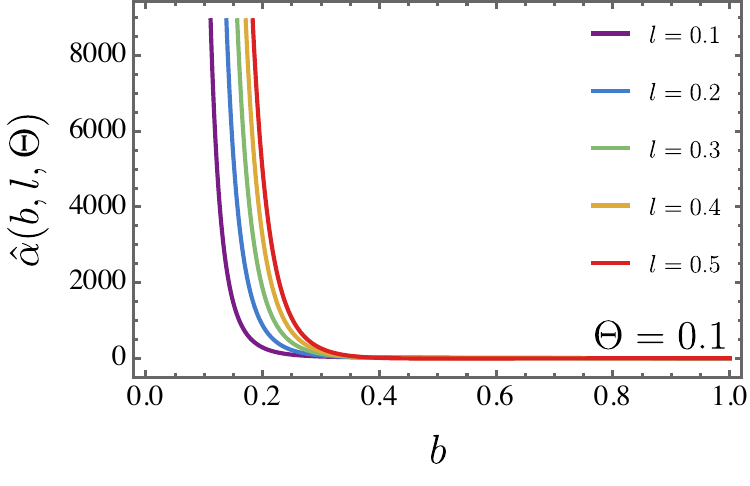}
    \caption{The deflection angle $\hat{\alpha}(b,l,\Theta)$ is plotted as a function of the impact parameter $b$ for different values of $\Theta$, while maintaining $l$ fixed.}
    \label{angldefc}
\end{figure}

Fig. \ref{angldefc} illustrates the behavior of the deflection angle $\hat{\alpha}(b,l,\Theta)$. {Notice that the first two terms correspond to the deflection angle in the Schwarzschild case, while the third term is associated with the Hayward black hole. The remaining terms represent the contributions arising from the non--commutativity introduced in this work, as straightforwardly expected.}

For a given impact parameter, such as $b = 0.2$, an increase in $\Theta$ leads to a larger magnitude of $\hat{\alpha}(b,l,\Theta)$. A similar trend is observed when the parameter $l$ is increased, resulting in a greater deflection angle.

\subsection{Lensing observables by EHT}

By analyzing the Event Horizon Telescope (EHT) observations of the black hole shadows associated with $Sgr A^*$ and $M87^*$ \cite{akiyama201987,akiyama2019M871,akiyama2022firstSgr,akiyama2022firstSgrA}, we explore potential constraints on the non--commutative parameter $\Theta$ within the framework of a Hayward black hole in non--commutative geometry. The angular diameter of the shadow, denoted as $\theta_\text{sh}$, serves as a critical observable for exploring the modified theories in quantum gravity \cite{afrin2023tests}. Through a comparison of theoretical shadow diameter predictions with the empirical data provided by the EHT, we evaluate whether the observational results impose any limitations on $\Theta$.

For a distant observer situated at a distance $D_O$ from the black hole, the angular diameter $\theta_{\text{sh}}$ of the black hole shadow is measured as follows \cite{kumar2020rotating}
\begin{equation}
\theta_{\text{sh}} = \frac{2b_c}{D_O},
\end{equation}
where $b_c$ represents the critical impact parameter. 

For the $Sgr A^*$ scenario, we adopt the mass and distance values reported by the EHT collaboration: $M = 4 \times 10^6 M_{\odot}$ and $D_O = 8.15 \, \text{kpc}$. The observed angular diameter of the shadow was found to lie within the range $\theta_{\text{sh}} =48.7 \pm 7 \, \mu\text{as}$ \cite{akiyama2022firstSgr,akiyama2022firstSgrA}. In the case of $M87^*$, we assume the mass and distance values as : $M = (6.5 \pm 0.7) \times 10^9 M_{\odot}$ and $D_O = 16.8 \, \text{Mpc}$. The angular diameter of the $M87^*$ black hole shadow was determined to be $\theta_{\text{sh}} = 42 \pm 3 \, \mu\text{as}$ \cite{akiyama201987,akiyama2019M871}.

\begin{figure}
    \centering
    \includegraphics[scale=0.39]{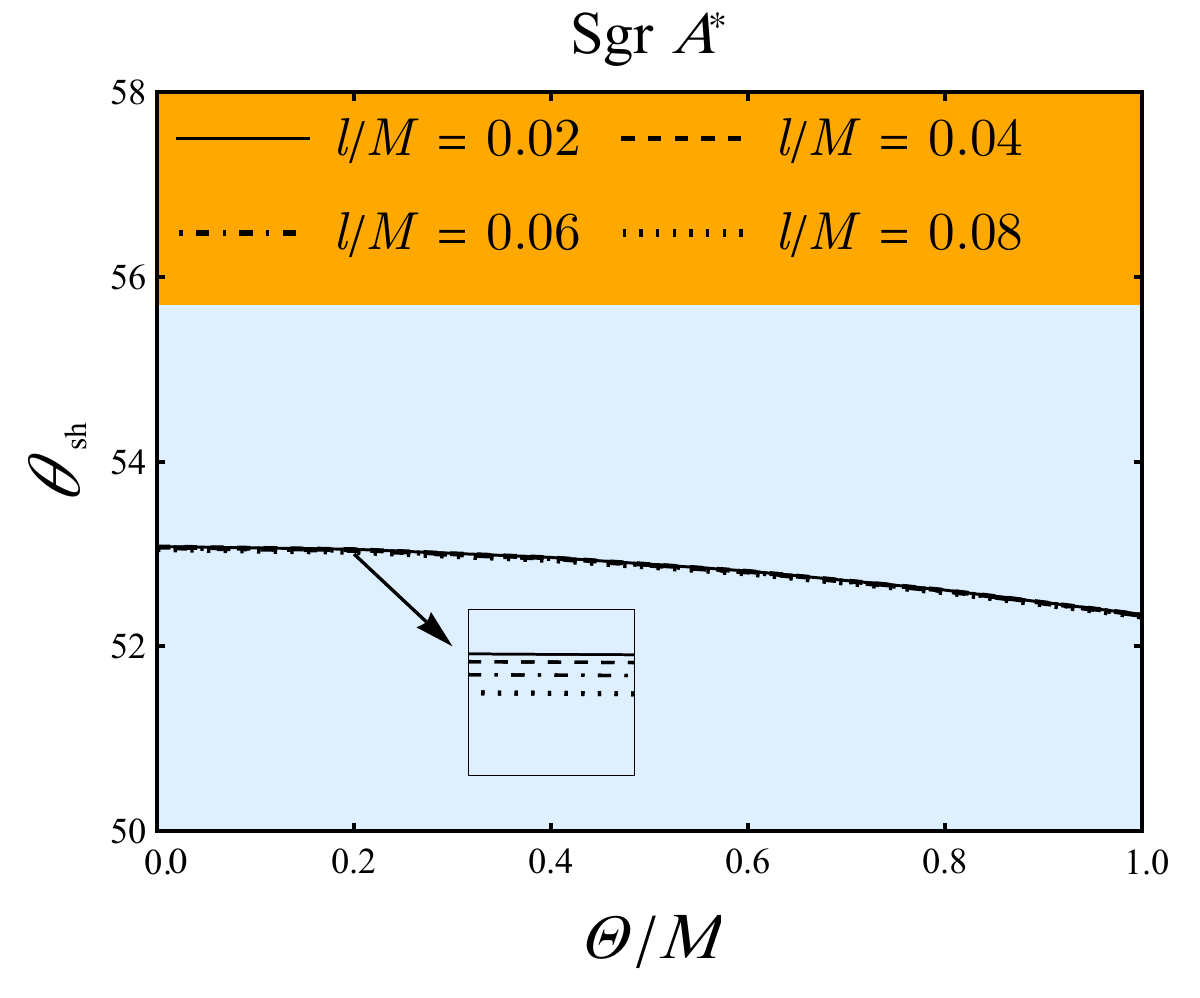}
    \includegraphics[scale=0.41]{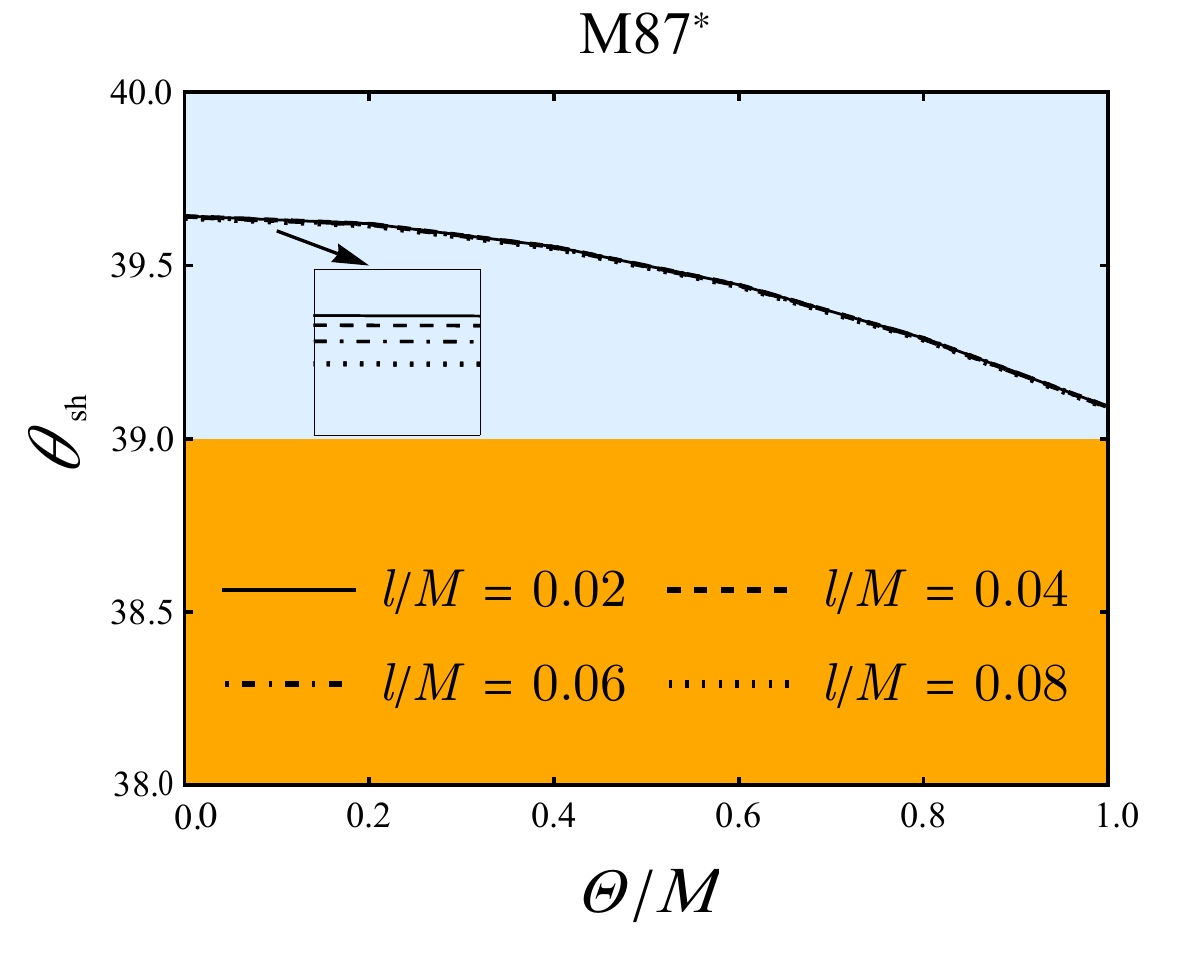}
    \caption{Angular diameter $\theta_{\textit{sh}}(\mu as)$ based on $M$ and $D_{O}$ of $Sgr A^*$ and $M87^*$ in relation to NC parameter $\Theta$.  The regions of angular shadow that are permitted and excluded based on EHT observations are shown by the orange and blue areas, respectively }
    \label{RCons}
\end{figure}

Fig. \ref{RCons} analyzes the black hole angular shadow behavior for  $Sgr A^*$ and $M87^*$, respectively, within the non--commutative geometry framework. Both plots show the angular shadow diameter as a function of the non--commutative parameter $\Theta$ for different values of $l/M$.
In both plots, the angular shadow diameter decreases monotonically with increasing  $\Theta$. The shadow radius is slightly sensitive to $l/M$. For smaller $l/M$ values, the angular shadow diameter is larger, while for larger $l/M$, it decreases. By considering the Hayward black hole in the non--commutativity framework, the calculated angular diameters fall within the observed range $\theta_{\text{sh}} \in (41.5, 55.7) \mu\text{as}$ and $\theta_{\text{sh}} \in (39, 45)\mu \text{as}$ for $Sgr A^*$ and $M87^*$, respectively.


\section{Estimation of bounds based on the solar system tests}

We revisit the results from Sec. \ref{sec:null}, now considering a Lagrangian of the form $L(x,\dot{x})=-\eta/2$, where $\eta=0$ corresponds to massless particles, while $\eta=1$ describes massive particles parametrized by the proper time $\lambda$. The Lagrangian for motion in the equatorial plane is given by
\begin{equation}
{A}(r){{\dot t}^2} - B(r){{\dot r}^2} - D(r){{\dot \varphi }^2} = \eta\, ,
\end{equation}
where we refer to $A(r)\doteq -g_{tt}^{(\Theta,l)}$, $B(r)\doteq g_{rr}^{(\Theta,l)}$, $D(r)\doteq g_{\varphi\varphi}^{(\Theta,l)}$.
The above expression, together with the conserved quantities \eqref{constant}, lead to
\begin{equation}\label{massive}
    \left[\frac{\mathrm{d}}{\mathrm{d}\varphi}\left(\frac{1}{r}\right)\right]^2=r^{-4}D^2(r)\left[\frac{E^2}{A(r)B(r)L^2}-\frac{1}{B(r)L^2}\left(\eta+\frac{L^2}{D(r)}\right)\right]\, .
\end{equation}

Defining a new variable as $u = L^2/(Mr)$ and differentiating Eq. \eqref{massive} with respect to $\varphi$, we obtain the leading contributions in $\Theta$ and $l$ expressed as:
{\begin{align}\label{eq:u-massive}
&\frac{\mathrm{d}^2 u}{\mathrm{d}\varphi^2} = \eta - u + \frac{3 M^2 u^2}{L^2} \\
&\Theta^2 \Bigg(\frac{L^4 \left(E^2-\eta \right) u}{2 \left(L^2-2 M^2 u\right)^3} + \frac{3 L^2 M^2 \left(3 \eta -4 E^2\right) u^2}{2 \left(L^2-2 M^2 u\right)^3} - \frac{M^2 u^3 \left(4 M^2 \left(7 \eta -13 E^2\right)+L^2\right)}{2 \left(L^2-2 M^2 u\right)^3} \\
&+ \frac{M^4 u^4 \left(M^2 \left(16 \eta -47 E^2\right)+3 L^2\right)}{L^2 \left(L^2-2 M^2 u\right)^3} - \frac{3 M^6 u^5}{L^2 \left(L^2-2 M^2 u\right)^3} + \frac{30 E^2 M^8 u^5}{L^4 \left(L^2-2 M^2 u\right)^3} \\
&- \frac{14 M^8 u^6}{L^4 \left(L^2-2 M^2 u\right)^3} - \frac{8 \eta M^{10} u^6}{L^6 \left(L^2-2 M^2 u\right)^3} + \frac{36 M^{10} u^7}{L^6 \left(L^2-2 M^2 u\right)^3} - \frac{24 M^{12} u^8}{L^8 \left(L^2-2 M^2 u\right)^3} \\
&+ l^2 \left( -\frac{6 M^{10} u^8 \left(4 M^2 \left(137 E^2-63 \eta \right)+15 L^2\right)}{L^{10} \left(L^2-2 M^2 u\right)^3} + \frac{12 M^8 u^7 \left(M^2 \left(391 E^2-91 \eta \right)+L^2\right)}{L^8 \left(L^2-2 M^2 u\right)^3} \right. \\
&\left. - \frac{14 M^8 \left(157 E^2-25 \eta \right) u^6}{L^6 \left(L^2-2 M^2 u\right)^3} + \frac{42 M^6 \left(8 E^2-\eta \right) u^5}{L^4 \left(L^2-2 M^2 u\right)^3} + \frac{144 M^{16} u^{11}}{L^{14} \left(L^2-2 M^2 u\right)^3} \right. \\
&\left. - \frac{312 M^{14} u^{10}}{L^{12} \left(L^2-2 M^2 u\right)^3} + \frac{28 M^{12} u^9 \left(9 L^2-28 \eta M^2\right)}{L^{12} \left(L^2-2 M^2 u\right)^3} \right) \Bigg) \nonumber\\
&- \frac{4 l^2 M^4 u^3 \left(2 \eta  L^2+3 M^2 u^2\right)}{L^8} \nonumber.
\end{align}}

The first two terms on the right--hand side represent the Newtonian contribution. The nonlinear term proportional to $M^2/L^2$, when expressed in SI units, takes the form $G^2M^2/(c^4L^2)$, which is significantly smaller than the Newtonian terms, introducing a relativistic correction. With this in mind, a perturbative approach is applied to constrain the parameters of the theory. This requires considering perturbations in three parameters: $\Theta$, $l$, and up to sixth--order in $M/L$. Under these conditions, Eq. \eqref{eq:u-massive} simplifies considerably to
{\begin{align}\label{eq:u}
     u''(\varphi)=& \, \eta -\frac{u \left(\Theta ^2 \left(E^2-\eta \right)-2 L^2\right)}{2 L^2}+ \frac{3 M^2 u^2 \left(\Theta ^2 \left(\eta -2 E^2\right)+2 L^2\right)}{2 L^4}  \nonumber\\ 
     &+u^3 \left(\frac{M^4 \left(2 \Theta ^2  \left(2 E^2+\eta \right)-16 l^2 \eta \right)}{2 L^{6}}-\frac{\Theta ^2 M^2}{2 L^4}\right)+ \frac{5 E^2 \Theta ^2 M^6 u^4}{L^8} \nonumber\\
     &+\frac{3 M^6 u^5 \left(\Theta ^2 L^2-4 l^2 L^2\right)}{L^{10}}\, .
\end{align}}

This equation reveals several noteworthy qualitative aspects. First, the non--commutative contribution $\Theta$ is coupled to the test particle's energy, a feature widely explored in quantum gravity phenomenology, further reinforced by this model \cite{Addazi:2021xuf,AlvesBatista:2023wqm}. Second, the Hayward contribution only appears in terms proportional to $M^4$, making it less significant than the leading non--commutative correction. Moreover, for massless particles ($\eta=0$), corrections from the Hayward parameter would only emerge at order $M^6/L^{10}$, rendering this case less favorable for observational constraints. {Besides that, we observe that the first correction is due to the non--commutative effect and corrects the Newtonian trajectory. Also, this is something that is known from the quantum gravity community \cite{Addazi:2021xuf} and can be a source of different constraints on $\Theta$, for instance from in-vacuo dispersion and threshold effects.}


\subsection{Perihelion precession of Mercury}

A fundamental test of General Relativity involves accurately predicting Mercury’s perihelion precession, quantified by angular shifts per century. To examine this, Mercury is treated as a test particle moving in the static spacetime generated by the Sun. Based on the previous discussion, we focus on the massive case with $\eta=1$. {For $\Theta=0$,} the first correction to GR in Eq. \eqref{eq:u}, proportional to $M^2$, is of the order $E^2M^2u^2\Theta^2/L^4$. However, for Mercury, this term is significantly smaller than the next-order correction, $M^4\Theta^2u^3/L^6$. This suppression arises because the energy $E$ is relatively small at the large distances considered in this context, as gravitational energy decreases with distance. {These corrections are also smaller than the Newtonian one, proportional to $u$.} Consequently, the $l$ corrections proportional to $u^2$, $u^4$, and $u^5$ are negligible and can be disregarded.

{Let us analyze the Hayward contribution.} By introducing the redefinition $m = M/L$ and denoting $\epsilon^2$ as a general factor multiplying $u^3$, the equation {for $\Theta=0$} takes the form
\begin{equation}\label{eq:hay_merc}
u''(\varphi)=1 - u + 3m^2u^2 +  \epsilon^2u^3\, .
\end{equation}

The parameter $m$ is treated as a perturbative quantity, introducing the first correction to GR, while the parameter $\epsilon$ accounts for modifications arising from the Hayward term. Based on this, the solution is expanded in terms of these parameters as $u = u_0 + m^2 u_m + \epsilon^2 u_{\epsilon}$, where $u_0$ corresponds to the Newtonian expression given by
\begin{equation}
    u_0=1+e \cos(\varphi)\, .
\end{equation}

Substituting this perturbative expansion into Eq. \eqref{eq:hay_merc}, the resulting expression takes the form
\begin{equation}\label{eq:pert_hay_merc}
    m^2 \left(3 (e \cos (\varphi )+1)^2-u_m''-u_m\right)+\epsilon^2 \left((e \cos (\varphi )+1)^3-u_{\epsilon}''-u_{\epsilon}\right)=0\, ,
\end{equation}
where terms of order $m^2\epsilon^2$ have been neglected. The term proportional to $m^2$ corresponds to the General Relativity correction, given by
\begin{equation}
    u_m=3m^2\left[\left(1+\frac{e^2}{2}\right)-\frac{e^2}{6}\cos\left(2\varphi\right)+e\varphi\sin\left(\varphi\right)\right]\, .
\end{equation}

The first term within the square brackets remains constant, while the second term oscillates around zero. Since neither contribute to the perihelion shift, they are omitted from the final expression. The remaining term, being proportional to $\varphi$, accumulates over multiple orbits, leading to a measurable effect.  

The term proportional to $\epsilon^2$ in Eq. \eqref{eq:pert_hay_merc} vanishes, introducing new corrections. Its solution contains several trigonometric terms that oscillate around zero, but only one contributes proportionally to $\varphi$, given by $\frac{3}{8}e\left(4+e^2 \right)\varphi  \sin (\varphi )$. Incorporating the Newtonian, GR, and new contributions and reinstating the original model parameters, we obtain a useful expression for $u(\varphi)$
\begin{equation}
    u(\varphi)=1+e \cos (\varphi )+\frac{3M^2}{L^2}\left(1+\epsilon^2\frac{(4+e^2)L^2}{8M^2}\right)e\varphi\sin(\varphi)\, .
\end{equation}

This expression can be further simplified by noting that the terms proportional to $\varphi\sin(\varphi)$ contribute minimally. This observation enables us to combine the last two terms using trigonometric identities, leading to
\begin{equation}
    u(\varphi)\approx 1+e\cos\left[\left(1-\frac{3M^2}{L^2}\left(1+\epsilon^2\frac{(4+e^2)L^2}{8M^2}\right)\right)\varphi\right]\doteq 1+e\cos\left[\left(1-\frac{3\widetilde{M}^2}{L^2}\right)\varphi\right]\, .
\end{equation}

This can be interpreted as a redefinition of the mass term, expressed as $\widetilde{M}^2 = M^2 \left[ 1 + \epsilon^2 (4 + e^2) L^2 / (8M^2) \right]$, modifying the standard GR result. From this relation, the perihelion shift induced by the new term is given by:
\begin{equation}
    \Delta\Phi=6\pi\frac{\widetilde{M}^2}{L^2}=6\pi\frac{M^2}{L^2}\left(1+\epsilon^2\frac{(4+e^2)L^2}{8M^2}\right)\, .
\end{equation}

This leads to a dimensionless correction to the General Relativity result, given by $\delta_{\text{Perih}} = \epsilon^2(4+e^2)L^2/(8M^2)$. Since Mercury's orbital period is approximately 88 days, the number of orbits per century is estimated as $100 \times 365.25 / 88 \approx 415$. By multiplying this by $\Delta\Phi$, the precession per century is obtained. General Relativity predicts a perihelion shift of $\Delta \Phi_{\text{GR}} = 42.9814''$/century, while the observed value is $\Delta \Phi_{\text{Exp}} = (42.9794 \pm 0.0030)''$/century \cite{Casana:2017jkc,Yang:2023wtu}. Using this, experimental constraints can be placed on the parameter $l$.

The angular momentum is related to the eccentricity $e$ and the semi-major axis $a$ by $L^2 = M a (1 - e^2)$, while the energy per unit mass is given by $E = -M/(2a)$ \cite{Goldstein:2002}. Using natural units, the values for the Sun’s mass, Mercury’s semi-major axis, and its eccentricity are $M = M_{\odot} = 9.138 \times 10^{37}$, $a = 3.583 \times 10^{45}$, and $e = 0.2056$. This results in an angular momentum of $L = 5.600 \times 10^{41}$, justifying the perturbative treatment due to the presence of terms of the form $M^2/L^2$. Additionally, the energy term is approximately $E^2 = 1.627 \times 10^{-16}$, making energy--dependent contributions negligible.

{In this case, $\epsilon^2=-8 l^2 M^4/L^6$. Therefore, the bound on $l$ is found to be $-1.730 \times 10^{16} \, \text{m}^2 \leq l^2 \leq 8.650 \times 10^{16} \, \text{m}^2$.} Negative values for the squared parameters are considered for the sake of generality.

{For the case of the non--commutative correction, we consider $l=0$, which gives the dominant equation $u''(\varphi)=1-u(1+\Theta^2/(2L^2))+3M^2u^2/L^2$. Again, if we redefine $m=M/L$ and $\xi=\Theta^2/(2L^2)$, we have
\begin{equation}
 u''(\varphi)=1-(1+\xi^2)u+3m^2u^2\, .  
\end{equation}}

{The equation can be treated perturbatively in $m$ and $\xi$, leading to a cumulative solution 
\begin{equation}
    u(\varphi)=1+e \cos (\varphi )+\frac{3M^2}{L^2}\left(1-\frac{\Theta^2}{12M^2}\right)e\varphi\sin(\varphi)\approx 1+e\cos\left[\left(1-\frac{3\widetilde{M}^2}{L^2}\right)\varphi\right]\, .
\end{equation}}

{The correction of the mass is $\widetilde{M}^2=M^2[1-\Theta^2/(12M^2)]$, which leads to the following bounds $-2976.57\, \text{m}^2\leq\Theta^2\leq 595.315\, \text{m}^2$.}

\subsection{Deflection of light}

When a light ray grazes the surface of a massive object, it gets deflected before reaching the detector. This parallax affects the apparent location of a light emitter in the sky. The deflection of light is carried out by the analysis of null geodesics, which consists in taking $\eta=0$ in \eqref{eq:u}. Now, we redefine the function $u$ as simply $1/r$. This gives the following expression
{\begin{align}\label{eq:light}
    u''(\varphi)=&-\frac{(2L^2-E^2\Theta^2)}{2L^2}u+\frac{3 M \left(-E^2 \Theta ^2+ L^2\right)}{ L^2}u^2 +\frac{(4E^2M^2-L^2) \Theta ^2}{2L^2}u^3+\frac{5M^3E^2}{L^2}\Theta ^2u^4
    \nonumber \\&-\frac{3 M^2  \left(\Theta ^2 L^2-4 l^2 \left(L^2-28 E^2 \Theta ^2\right)\right)}{L^2}u^5\, .
\end{align}}

In this case, the factor $L/E$ is the impact parameter $b$.  
{We begin by analyzing the non-commutative case. We have also universal corrections in $\Theta$ that do not depend on the mass $M$.
This way, with a redefinition of the mass parameter, we have the dominant contributions
{\begin{equation}\label{eq:redef_light_theta}
     u''(\varphi)+\left(1-\frac{\Theta^2}{2b^2}\right)u=3\widetilde{M}u^2\, ,
\end{equation}
where $\widetilde{M}=M\left(1-\Theta^2/b^2\right)$.} The Newtonian case consists in taking the left hand side of the above equation as zero. This gives a Newtonian correction \cite{Yang:2023wtu}
\begin{equation}
    u_0=b^{-1}\sin\left(\left(1-\frac{\Theta^2}{4b^2}\right)\varphi \right)\, ,
\end{equation}
where $b$ is the impact parameter and the incident angle is $\varphi_0=0$, and this solution describes a straight line. Inserting this solution into \eqref{eq:redef_light_theta}, we have the perturbed solution given by
\begin{equation}
    u(\varphi)=\frac{1}{b}\sin\left(\left(1-\frac{\Theta^2}{4b^2}\right)\varphi\right)+\frac{\widetilde{M}}{b^2(1-\Theta^2/(2b^2))}\left[1+\cos^2\left(\left(1-\frac{\Theta^2}{4b^2}\right)\varphi\right)\right]\, .
\end{equation}}

{The light ray asymptotically obeys $u\rightarrow 0$ (when $r\rightarrow\infty)$. Therefore, by solving the equation $u=0$, we find the incident and exit angles $\varphi_{\text{in}}$ and $\varphi_{\text{ex}}$. Considering a perturbation in $\varphi$ and the other parameters of the model, we find $\varphi_{\text{in}}=-2\bar{M}/b$ and $ \varphi_{\text{ex}}=\pi+2\bar{M}/b$, where $\bar{M}=M\left(1-\frac{\Theta^2}{4b^2}\right)$. The angle of deflection therefore is $\delta=-2\varphi_{u\rightarrow0}$
\begin{equation}
    \delta_{\Theta}=\frac{4\bar{M}}{b}=4\frac{M}{b}\left(1-\frac{\Theta^2}{4b^2}\right)\, .
\end{equation}}

For light that grazes the surface of the sun, we have $b\approx R_{\odot}=4.305\times 10^{43}$. Besides that, the mass of the sun is $M=M_{\odot}=9.138\times 10^{37}$. The dimensionless correction due to the non--commutativity parameter is {$1-\Theta^2/(4b^2)$}. Notice that this correction does not depend on the mass of the star directly, but on its impact parameter. Therefore, for a light ray that grazes the surface of a star, it depends only on its radius. This universal behavior is a prediction of this kind of non--commutativity model.

The deflections predicted by GR is $\delta_{\text{GR}}=4M/b=1.7516687''$. And the observational value is $\delta_{\text{Exp}}=\frac{1}{2}(1+\gamma)1.7516687''$, where $\gamma=0.99992\pm 0.00012$ \cite{dsasdas}. {For this reason, we compare the dimensionless correction $1-\Theta^2/(4b^2)$ with $(1+\gamma)/2$. This gives the 
 $-3.872\times 10^{13}\, \text{m}^2\leq\Theta^2\leq 1.936\times 10^{14}\, \text{m}^2$.}
 
 The analysis of the Hayward case requires that we set $\Theta=0$ in Eq.\eqref{eq:light}. In this case, the radial equation reads
\begin{equation}\label{eq:light-hay}
    u''(\varphi)=-u+3Mu^2-12M^2l^2u^5\, .
\end{equation}

Now we proceed by redefining the quantity $Ml=\alpha$. Using the same technique of the previous subsection, we approximate the solution as $u(\varphi)\approx u_0+M u_M+\alpha^2u_{\alpha}^2$. By substituting this approximation in Eq.\eqref{eq:light-hay}, we find 
\begin{equation}
    M\left(\frac{3\sin^2(\varphi)}{b^2}-u_M-u_M''\right)-\alpha^2\left(\frac{12\sin^2(\varphi)}{b^5}+u_{\alpha}+u_{\alpha}''\right)=0\, ,
\end{equation}
where we again discarded terms proportional to $\alpha^2M$. The solution for $u_M$ gives the GR correction $u_M=\frac{M}{b^2}\left(1+\cos^2(\varphi)\right)$. The other solution is\footnote{We set the initial conditions such that the solution must depend on the impact parameter.}
\begin{equation}
    u_{\alpha}=\frac{60\varphi\cos(\varphi)-(47+14\cos(2\varphi)-\cos(4\varphi))\sin(\varphi)}{b^5}\, .
\end{equation}

For small angles $\varphi$ and considering that in infinity, $u\rightarrow 0$, we find the deflection angle $\delta_l=-2\varphi_{u\rightarrow0}$ as
\begin{equation}
    \delta_l=\frac{4M}{b}\left(1-\frac{15M^2}{4b^4}l^2\right)\, ,
\end{equation}
which gives the dimensionless correction $15M^2l^2/(4b^4)$. Using the experimental result above, we find  $-5.729\times 10^{23}\, \text{m}^2\leq l^2\leq 2.863\times 10^{24}\, \text{m}^2$.

The ratio $M^2/b^4$ is much smaller than $1/b^2$. In fact, they are $10^{12}$ orders of magnitude apart. Therefore the Hayward correction is much fainter that the non--commutative one. Also in this case, we considered positive and negative bounds on the quadratic parameters, for the sake of generality.


\subsection{Time delay of light}

This measurable effect, known as the Shapiro delay \cite{Shapiro:1964uw}, refers to the additional time taken by radar signals to travel from Earth to planets in the solar system and return, accounting for the curvature of spacetime caused by the Sun. To determine this delay, we analyze the null geodesic equation \eqref{massive}. By utilizing the null curve condition alongside the expressions for energy and angular momentum given in \eqref{constant}, the radial coordinate can be expressed as a function of time
\begin{equation}
  \left(  \frac{\mathrm{d}r}{\mathrm{d}t}\right)^2=\frac{A(r)D(r)-\frac{L^2}{E^2}A^2(r)}{B(r)D(r)}\, .
\end{equation}

The angular momentum and energy can be rewritten in terms of the closest approach distance $b$ of the light ray to the Sun \cite{Wang:2024fiz}, which is determined by imposing the condition $\dot{r} = 0$. This yields the relation $L^2/E^2 = D(r_{\text{min}})/A(r_{\text{min}})$. Consequently, the time of flight can be expressed as a function of the radial coordinate as follows
\begin{equation}\label{eq:shapiro_main}
    \mathrm{d} t=\pm \frac{1}{A(r)}\frac{1}{\sqrt{\frac{1}{A(r)B(r)}-\frac{D(r_{\text{min}})/A(r_{\text{min}})}{B(r)D(r)}}}\, .
\end{equation}

Next, we examine the Hayward and non-commutative cases separately, beginning with the non-commutative scenario by setting $l = 0$. Since our focus is on deviations from the flat spacetime case, we retain only the leading-order corrections in $M$ and $\Theta^2$. Under this approximation, integrating Eq. \eqref{eq:shapiro_main} yields
{\begin{align}
    t=\sqrt{r^2-r_{\text{min}}^2}+M\left(\sqrt{\frac{r-r_{\text{min}}}{r+r_{\text{min}}}}+2\ln\left(\frac{r+\sqrt{r^2-r_{\text{min}}^2}}{r_{\text{min}}}\right)\right)\\
    +\frac{\Theta ^2 \left(2 (r_{\text{min}}-M) \arctan\left(\frac{\sqrt{r^2-r_{\text{min}}^2}-r}{r_{\text{min}}}\right)-\frac{M (19 r_{\text{min}}+26 r) \sqrt{r^2-r_{\text{min}}^2}}{r (r_{\text{min}}+r)}\right)}{8 r_{\text{min}}^2}\, .\nonumber
\end{align}}

For the regime where $r_{\text{min}} \ll r$, the leading contributions from General Relativity and non--commutative corrections are
{\begin{equation}\label{eq:sh_t_nc}
    t(r)=r+M+2M\ln\left(\frac{2r}{r_{\text{min}}}\right)-\frac{13M}{4r_{\text{min}}^2}\Theta^2\, .
\end{equation}}

The travel time from the emitter to the Sun is denoted as $t(r_E)$, while the time from the Sun to the receiver is $t(r_R)$, where $t(r)$ is determined by Eq. \eqref{eq:sh_t_nc}. Here, $r_E$ and $r_R$ represent the radial coordinates of the emitter and receiver, respectively. The total round-trip time, from emission to reflection at the receiver and back to the emitter, is given by $T = 2t(r_E) + 2t(r_R)$. In this scenario, this results in
{\begin{equation}
T_{\Theta}=2(r_E+r_R)+4M\left[1+\ln\left(\frac{4r_Rr_E}{r_{\text{min}}^2}\right)-\frac{13\Theta^2}{4r_{\text{min}}^2}\right]=T_{\text{flat}}+\delta T\, .
\end{equation}}

The Shapiro delay is defined as the difference between the time required for the signal to travel in the presence of relativistic corrections, which depend on the mass $M$, and the corresponding travel time in flat spacetime, given by $T_{\text{flat}} = 2(r_E + r_R)$. In the parametrized post-Newtonian (PPN) framework, this time delay is expressed as
\begin{equation}
    \delta T = 4M\left(1+\frac{1+\gamma}{2}\ln \left(\frac{4r_Rr_E}{r_{\text{min}}^2}\right)\right)\, .
\end{equation}

Using data from the Cassini probe measurements \cite{Bertotti:2003rm,Will:2014kxa}, the most precise constraint on $\gamma$ is found to be $|\gamma - 1| < 2.3 \times 10^{-5}$. The mean Earth-Sun distance, expressed in natural units, is one astronomical unit (AU), given by $r_E = 1\, \text{AU} = 2.457 \times 10^{45}$. At the time of measurement, the Cassini probe was located at a distance of $r_R = 8.46\, \text{AU}$ from the Sun, while the closest approach of the light signal to the Sun was $r_{\text{min}} = 1.6 R_{\odot}$, where the solar radius is $R_{\odot} = 4.305 \times 10^{43}$. These values allow us to impose a constraint on $\Theta$, yielding {$-5.001 \times 10^{14}\, \text{m}^2 \leq \Theta^2 \leq 5.001 \times 10^{14}\, \text{m}^2$.}  

For the Hayward scenario, we set $\Theta = 0$. The dominant contributions obtained from integrating Eq. \eqref{eq:shapiro_main} are
\begin{align}
   t=\sqrt{r^2-r_{\text{min}}^2}+M\left(\sqrt{\frac{r-r_{\text{min}}}{r+r_{\text{min}}}}+2\ln\left(\frac{r+\sqrt{r^2-r_{\text{min}}^2}}{r_{\text{min}}}\right)\right)\nonumber\\
  -M^2\left(\frac{(4r+5r_{\text{min}})}{2r_{\text{min}}(r+r_{\text{min}})^2}\sqrt{r^2-r_{\text{min}}^2}-\frac{15 \arctan \left(\frac{r-\sqrt{r^2-r_{\text{min}}^2}}{r_{\text{min}}}\right)}{r_{\text{min}}}\right)\nonumber\\
  +l^2 M^2\left(\frac{10  \arctan\left(\frac{r-\sqrt{r^2-r_{\text{min}}^2}}{r_{\text{min}}}\right)}{r_{\text{min}}^3}-\frac{3  \sqrt{r^2-r_{\text{min}}^2}}{r_{\text{min}}^2 r^2}\right)\, .
\end{align}

For $b\ll r$, the dominant terms are the following 
\begin{equation}
     t=r+M+2M\ln\left(\frac{2r}{r_{\text{min}}}\right)-\frac{2M^2}{r_{\text{min}}}-\frac{3 M^2l^2}{r_{\text{min}}^2r}\, .
\end{equation}

The Shapiro time due to the Hayward contribution is then
\begin{equation} T_l=2(r_E+r_R)+4M\left[1+\ln\left(\frac{4r_Rr_E}{r_{\text{min}}^2}\right)-\frac{3M^2l^2}{2r_\text{min}^2}\left(\frac{1}{r_E}+\frac{1}{r_R}\right)\right]\, .
\end{equation}

This effect is weaker compared to the contributions from non--commutative corrections. Using Cassini tracking data once again, the constraints on the Hayward parameter are found to be $-9.437 \times 10^{26}\, \text{m}^2 \leq l^2 \leq 1.562 \times 10^{19}\, \text{m}^2$, where the solar mass $M$ was used in the calculations. Considering the bounds derived for both $\Theta$ and $l$, negative values for the squared parameters were included for the sake of generality. A summary of all constraints obtained in this section is presented in Tab. \ref{tab:constr}. Finally, it is important to mention that, although we consider the positive quantity $\Theta^2$, the negative branch of the bound corresponds to the formal replacement $\Theta^2 \rightarrow -\Theta^2$, which is included here for the sake of generality.

\begin{table}[h!]
\centering
\caption{Bounds for $\Theta^2$ and $l^2$ derived from Solar System tests.}
\label{tab:constr}
\begin{tabular}{lc}
\hline\hline
\textbf{Solar System Test} & Constraints \((\text{m}^2)\) \\
\hline
{\bf{Mercury precession}}   & \makecell{\(-2976.57\leq\Theta^2\leq 595.315\) \\ \(-1.730 \times 10^{16} \leq l^2 \leq 8.650 \times 10^{16} \)} \\
{\bf{Light deflection}}     & \makecell{\(-3.872\times 10^{13}\leq\Theta^2\leq 1.936\times 10^{14}\) \\ \(-5.729\times 10^{23}\leq l^2\leq 2.863\times 10^{24}\)}  \\
{\bf{Shapiro time delay}}   & \makecell{\(-5.001 \times 10^{14} \leq \Theta^2 \leq 5.001 \times 10^{14}\) \\ \(-9.437\times 10^{26}\leq l^2\leq 1.562\times 10^{19}\)}  \\
\hline\hline
\end{tabular}
\end{table}


\section{Conclusion}

A black hole solution incorporating non--commutative corrections into a Hayward--like metric was proposed in this work. {This solution was obtained within the framework of non-commutative gauge theory, incorporating the Moyal product defined by $\partial_{r} \wedge \partial_{\theta}$, in accordance with the recent approach presented in Ref.~\cite{Juric:2025kjl}.}

The analysis began with an investigation of the metric’s fundamental properties, including the structure of the event horizon $r_h$ and the Kretschmann scalar $\mathcal{K}$. {The results revealed that $\Theta$ introduced no corrections to the event horizon.} In contrast to the Hayward solution, the modified black hole studied here showed a regular behavior ({featuring a dependence on the angle $\theta$}), as revealed by the analysis of the Kretschmann scalar. {It remains to verify is such anisotropic behavior leaves signatures on other observables. Where measurements done in the northern or southern hemispheres may lead to different predictions.}

Thermodynamic aspects were then explored by evaluating the Hawking temperature, entropy, and heat capacity. {The temperature profile indicated the absence of a physical remnant mass when it was considered $T^{(\Theta,l)} \to 0$.} For a fixed $l$, increasing $\Theta$ raised $T^{(\Theta,l)}$. The heat capacity featured regions of both positive and negative values, indicating transitions between thermodynamic stability and instability.

The emission of quantum radiation was also examined, with a focus on Hawking radiation. The particle creation densities for fermionic, $n_{\psi}$, and bosonic, $n$, modes were determined, allowing a comparison of emission rates. The results indicated that bosons were more abundantly emitted than fermions for a given frequency $\omega$ {(in a low regime scenario)}.

A perturbative evaluation of the effective potential allowed us to study the quasinormal modes and the time domain solution in response to scalar perturbations. In essence, it was demonstrated that, by increasing $\Theta$ and $l$, it led to more damped oscillations.

Through the calculation of null geodesics, we could address the photon sphere and black hole shadows. Comparisons with Event Horizon Telescope (EHT) observations enabled constraints on the shadow radii. The stability of critical orbits was assessed via the Gaussian curvature, followed by a gravitational lensing analysis using the Gauss--Bonnet theorem. The deflection angle $\hat{a}(\Theta,l)$ was found to increase with both $\Theta$ and $l$ for a fixed impact parameter $b$, while the critical orbit remained unstable.

Lastly, constraints on the parameters $\Theta$ and $l$  were established based on observational tests in the solar system, including Mercury’s perihelion precession, light deflection, and the Shapiro time delay.

As a further perspective, it is worth investigating {the emission rate, the evaporation lifetime,} the absorption cross section and greybody factors of the black hole analyzed in this manuscript, following the methodology adopted in \cite{12aa2025particle,12araujo2024particle,12araujo2025does}. Moreover, inspired by a recently proposed technique \cite{Juric:2025kjl}, we plan to examine other configurations --- such as Bardeen-- and Frolov--like black holes (and others) — within the framework of non--commutative gauge theory, along with their corresponding gravitational signatures. These and other ideas are now under development.

\section*{Acknowledgments}
\hspace{0.5cm} A. A. Araújo Filho is supported by Conselho Nacional de Desenvolvimento Cient\'{\i}fico e Tecnol\'{o}gico (CNPq) and Fundação de Apoio à Pesquisa do Estado da Paraíba (FAPESQ), project No. 150891/2023-7. I. P. L. was partially supported by the National Council for Scientific and Technological Development - CNPq grant 312547/2023-4.  I. P .L. would like to acknowledge the contribution of the COST Action BridgeQG (CA23130), supported by COST (European Cooperation in Science and Technology). The authors would like to thank C. Molina for his assistance with the calculations in the time--domain section.


	\bibliography{main}
	\bibliographystyle{unsrt}
	
\end{document}